\documentclass{article}
\usepackage{graphicx} 

\usepackage{mathtools}

\usepackage{color,xcolor,ucs}
\usepackage{mathtools}   \usepackage{tikz} 
\usepackage{ amssymb }
\usepackage{extarrows} 
\usepackage{pgf,tikz}
\usepackage{float}
\usetikzlibrary{positioning}
\usetikzlibrary{shapes.geometric}
\usetikzlibrary{shapes.misc}
\usetikzlibrary{arrows}
\usepackage{caption}
\usepackage{mathrsfs}
\usetikzlibrary{arrows,shapes,automata,backgrounds,petri,positioning}
\usetikzlibrary{decorations.pathmorphing}
\usetikzlibrary{decorations.shapes}
\usetikzlibrary{decorations.text}
\usetikzlibrary{decorations.fractals}
\usetikzlibrary{decorations.footprints}
\usetikzlibrary{shadows}
\usetikzlibrary{calc}
\usetikzlibrary{spy}
\usepackage{amsmath}
\usepackage{array}
\usepackage{ amssymb }
\usepackage{braket}
\usepackage{qcircuit}
\usepackage{soul}
\usepackage{braket} 
\usepackage{relsize}

\usepackage{amsmath}
\usepackage{ amssymb }
\usepackage{braket}
\usepackage{qcircuit}
\usepackage{soul}
\usepackage{braket}

\begin{document}

\title{Probability distributions over CSS codes: two-universality, QKD hashing, collision bounds, security}

\author{Pete Rigas}

\maketitle

\begin{abstract}
We characterize novel probability distributions for CSS codes. Such classes of error correcting codes, originally introduced by Calderbank, Shor, and Steane, are of great significance in advancing the fidelity of Quantum computation, with implications for future near term applications. Within the context of Quantum key distribution, such codes, as examined by Ostrev in arXiv: 2109.06709 along with two-universal hashing protocols, have greatly simplified Quantum phases of computation for unconditional security. To further examine novel applications of two-universal hashing protocols, particularly through the structure of parity check matrices, we demonstrate how being able to efficiently compute functions of the parity check matrices relates to marginals of a suitably defined probability measure supported over random matrices. The security of the two-universal QKD hashing protocol will be shown to depend upon the computation of purified states of random matrices, which relates to probabilistic collision bounds between two hashing functions. Central to our approach are the introduction of novel real, simulator, and ideal, isometries, hence allowing for efficient computations of functions of the two parity check matrices. As a result of being able to perform such computations involving parity check matrices, the security of the two-universal hashing protocol is a factor of $2^{ \frac{5}{2} ( 5 - \frac{3}{2} ) + \mathrm{log}_2 \sqrt{C}}$ less secure, for some strictly positive constant $C$.
\end{abstract}



\section{Introduction}

\subsection{Overview}

\noindent Quantum key distribution (QKD) is an object that is of great interest in the field of Quantum computing, with potential industrial applications pertaining to number theory, algebraic geometry, and group theory, {\color{blue}[53]}, transition dynamics, {\color{blue}[54]}, entanglement lower bounds, {\color{blue}[52]}, generalized XOR games, {\color{blue}[45},{\color{blue}51]}, human cooperation, {\color{blue}[44]}, probability theory, {\color{blue}[43]}, energy applications, {\color{blue}[42]}, amongst several other possibilities {\color{blue}[36}, {\color{blue}37}, {\color{blue}41]}. While progress on the reduction of error rates has been recently achieved, {\color{blue}[33]}, potentially several years of additional work need to still be carried out for realizing Quantum advantage. Currently, several proposed sources of experimental advantage relate to density-functional theory, {\color{blue}[34]}, classification of correlations, {\color{blue}[32]}, differences between classical and Quantum annealing, {\color{blue}[31]}, variational quantum eigensolvers, {\color{blue}[29},{\color{blue}47]}, multipartite games, {\color{blue}[28]}, data-driven optimization, {\color{blue}[27]}, and nonlinear transformations, {\color{blue}[26]}, all of which fit under the general premises of Quantum information theory, {\color{blue}[30]}. 

While previous works of the author, initially beginning with variational quantum algorithms (VQAs), {\color{blue}[47]}, examined prospective Quantum advantage in polynomial time, other more theoretically intricate, and complicated, consequences remain of interest to explore. From previous work of Ostrev, {\color{blue}[38]}, the author determined how error bounds, and optimal values, can be leveraged for characterizations of exact, and approximate, optimality {\color{blue}[46]}. However, beyond the dual $\mathrm{XOR}^{*}$ game and $\mathrm{FFL}$ game, multiplayer game-theoretic interpretations can be obtained from the two player game-theoretic setting {\color{blue}[48]}. Towards the direction of QKD, in work of the author following {\color{blue}[46},{\color{blue}48]}, converse bit transmission rates were obtained {\color{blue}[49]}. Upper bounds on the bit transmission rate, which determines the total number of qubits that can be transmitted over independent uses of a Quantum channel, are achieved \textit{without} the use of QKD. The \textit{secrecy} capacity, which is related to the capacity at which bits are transmitted over single uses of a noisy Quantum channel, is described in {\color{blue}[35]}.

Generally speaking, the question as to whether a Quantum communication protocol does, or does not, make use of QKD is of great importance to consider. Despite the fact that protocols that are \textit{independent} of secret keys sampled from QKD have certain advantages, perhaps some of the most well known problems in Quantum information theory can be phrased in terms of the security, and related aspects, of QKD. To this end, we consider the \textit{two-universal} hashing protocol developed in {\color{blue}[40]}, which we now dedicate some efforts to describe. Such a group of hashing protocols not only is designed for determining bit flip, and phase flip, errors, but is also independent of sampling \textit{random} positions of the generated secret key. Furthermore, the \textit{two-universal} hashing protocol is dependent upon Quantum CSS codes, which simplifies Quantum steps of the protocol as much as possible.

Explicitly, the description of eight steps for the \textit{two-universal} hasing QKD protocol provided in {\color{blue}[40]}, as well as its accompanying security, depend upon the computation of two functions that Alice and Bob use to output the desired secret key. These two functions, while obviously being dependent upon the $n$ qubit state that either player initially receives, also depend upon the parity check matrices of the CSS code. Such matrices are straightforwardly obtained by taking the appropriate rows of some random, invertible matrix, over the field of two elements, and from the transpose of its inverse. As previously mentioned, given the fact that the \textit{two-universal} QKD hashing protocol detects the number of bit flip, and phase flip, errors, it remains of interest, as mentioned by Ostrev, to explore novel probability distributions of CSS codes.

Exploring such a distribution over CSS codes would entail: (1) determining how the block size, ie the number of secret bit transmitted, could be used to construct a secret key; (2) exploring whether the two functions described above, which are each dependent upon the parity check matrices, can be computed more efficiently; (3) establishing further perspectives between \textit{dependent}, and \textit{independent}, QKD communication protocols. One can begin to anticipate why the aforementioned three reasons are helpful for probability distributions over CSS codes; depending upon the parity check matrices which are further processed within the hashing protocol, Alice and Bob determine whether there are bit flip, and phase flip, errors. Moreover, from computations provided in {\color{blue}[40]}, the two functions dependent upon the parity check matrices, over the set of possible errors within the Hamming ball, are used in several computations, including: approximating the probability that bit flips, and phase flip, errors are contained within the Hamming ball with radius $r$ (which is taken to be equal to the number of such errors); formulating tensor products involving a purification of the random matrix that is sampled over the field with two elements; concluding that, with sufficiently good probability, that Alice and Bob can output a secret key drawn from the QKD.

Regardless of computations provided in {\color{blue}[40]}, an interesting research question relates to formulating additional aspects of two-universal hashing protocols. That is, if we restrict ourselves to the field of two elements, would there exist probability distributions over CSS code so that the parity check matrices allow for more direct, and explicit, computations of the functions used to output the secret key? Adopting similar notation to that provided in {\color{blue}[40]}, denote $g_1$ and $g_2$ as the two functions which Alice and Bob, respectively, compute in the steps of the two-universal hashing protocol before outputting the secret key. Even if Alice and Bob are able to compute $g_1$ and $g_2$, with support over the Hamming ball, such computations are intricately related to the parity check matrices. As a result, determining whether there are any favorable properties of parity check matrices, and hence of the \textit{two-universal} hashing protocol, that would enable such a task to be completed in polynomial time are of interest. As established in previous remarks, similar computations, and aspects of Quantum communication protocols that are independent of secret keys, provide great intuition with regard to the characteristics of probability distributions over CSS codes that will be examined in this work {\color{blue}[1},{\color{blue}2},{\color{blue}3},{\color{blue}4},{\color{blue}5},{\color{blue}6},{\color{blue}8},{\color{blue}9},{\color{blue}10},{\color{blue}11},{\color{blue}12},{\color{blue}13},{\color{blue}14},{\color{blue}15},{\color{blue}16},{\color{blue}17},{\color{blue}18]}.

Along such lines, we conclude the overview section by providing several characteristics of the functions $g_1$, and $g_2$, which Alice and Bob compute before outputting the desired secret key. As functions of the two parity check matrices, which in themselves are related to an instance of a random matrix over the field with two elements, one must consider: (1) decompositions of outer products, drawn uniformly at random from the field of two elements, corresponding to bit flip and phase flip errors; (2) incorporating the purified state of a random matrix whose entries are drawn uniformly at random from the field of two elements; (3) incorporating the parity check matrices which Alice and Bob use to compute $g_1$ and $g_2$. In formulating such observations between Probability theory and Quantum information theory, the existence of a measure over CSS codes which makes it easier to compute $g_1$ and $g_2$ is desirable. Related tasks of computational interest are provided in {\color{blue}[7},{\color{blue}19},{\color{blue}20},{\color{blue}21},{\color{blue}22},{\color{blue}23},{\color{blue}24},{\color{blue}25},{\color{blue}26},{\color{blue}27},{\color{blue}28]}.

Computing tensor products of the above indicated form, particularly which involve tensor products after purification, impacts the parity check matrices in the following manner. When performing computations with respect to the tensor product of the purified state, the parity check matrices of the CSS code determines the computational complexity of approximating $g_1$ and $g_2$. Given the prospective, though in some cases very tentative, expected Quantum advantage, the existence of suitable Probability measures over CSS codes determines whether error correction can be executed in the first place. To approximate functions associated with the parity check matrices, we: (1) provide motivation for the difficulties associated with computing such functions; (2) providing additional steps of the \textit{two-universal} QKD hashing protocol associated with computing functions of the parity check matrices; (3) establish characteristics of novel probability distributions over CSS codes. In comparison to probability distributions that have been identified for CSS codes, {\color{blue}[40]}, those identified in forthcoming arguments are obtained by: explicitly providing expressions for the parity check matrices; determining the occurrence of bit flip, and phase flip, errors; outputting the Quantum states associated with bit flip, and phase flip, errors; expressing, as a function of the two parity check matrices, the shared secret key, drawn uniformly at random from the QKD, between Alice and Bob.

\subsection{This paper's contributions}

\noindent This paper provides several characterizations of probability distributions over CSS codes. Motivated by one possible direction of research provided in {\color{blue}[40]}, we determine how the structure of parity check matrices is related to functions used for outputting the shared secret key. In comparison to previous work of the author on Quantum communication protocols, {\color{blue}[46},{\color{blue}48},{\color{blue}49]}, making use of the secret key translates to \textit{two-universal} hashing protocols. More broadly, characterizing steps of QKD hashing protocols can provide instances of \textit{unconditional} security. Although aspects of \textit{unconditional} security have been alluded to in {\color{blue}[46},{\color{blue}48},{\color{blue}49]}, it is obtained by completely different means. Nevertheless, examining instances of \textit{unconditional}, or \textit{conditional}, security from secret keys, or bit codewords, shared between Alice and Bob is intriguing. 

In comparison to previous computations that have been provided by the author, {\color{blue}[46},{\color{blue}48},{\color{blue}49]}, characterizations of exact, and approximate, optimality need not rely upon QKD. Regardless, having knowledge of the structure of optimal states, EPR pairs, optimal values, and related quantities relates to the security associated with \textit{two-universal} hashing. For future applications, whether theoretical or experimental, it is imperative to formalize what the expected Classical-Quantum gap in performance is.

\subsection{Paper organization}

\noindent Before being providing expressions for functions of parity check matrices, and the parity check matrices themselves, we provide an overview of the objects from {\color{blue}[40]}. Such objects relate to the distribution of random matrices over the field of two elements; several probabilities related to sampling invertible random matrices over the field of two elements; computing where bit flip, and phase flip, errors can occur; probabilities associated with determining whether functions, from one finite set to another, are related to the number of suitable \textit{two-universal} functions. While several properties of \textit{two-universal} functions are related to those identified for probability distributions over CSS codes that have been studied in {\color{blue}[40]}, the parity check matrices significantly impact computations associated with approximating the output of the secret key shared between Alice and Bob. For probability distributions over CSS codes, and accompanying parity check matrices examined in this work, to approximate bit flip, and phase flip, errors, one computes: the outer product associated with a random matrix over the field of two elements; the purified state associated with a random matrix over the field with two elements; incorporating aspects from the parity check matrices into outer products of random matrices and the output state obtained after purification. While computations provided in {\color{blue}[40]} indeed make use of previously described outer products and purified states, incorporating constraints to the parity check matrices, along with the CSS probability distribution, can strongly influence the secret key distribution obtained via \textit{two-universal} hashing.

\subsection{Statement of main results}

\noindent For the results below, denote: $\rho_{ABE}$ as the state encoded by Alice (resp. Bob) which is distributed to Bob (resp. Alice) and to Eve, $\rho_{W_A W_B C E}$ as the state corresponding to a classical transcript of the message communication between Alice and Bob, $P$ as the projection operator,

\begin{align*}
   P \equiv    P \big( \vec{g} , x  \big) =      2^{-n} \underset{1 \leq j \leq m}{\prod}  \big[ \textbf{I} + \big( - 1 \big)^{x_j} g_j  \big]       , 
\end{align*}

\noindent for $\vec{g} \in G_n$, where $G_n$ denotes an element of the Pauli group, on $n$ qubits,

\begin{align*}
  G_n \equiv \big\{ \omega \sigma^u_1 \sigma^u_3 : \omega \in \big\{ \pm 1 , \pm i \big\} , u , v \in \textbf{F}^{1 \times n}_2 \big\}   , 
\end{align*}

\noindent over the Pauli matrix basis,

\begin{align*}
 \sigma_1 =    \begin{bmatrix}
   0 & 1   \\ 1 & 0 
    \end{bmatrix} ,   \sigma_2 = \begin{bmatrix}
    0 & - i  \\ i & 0 
    \end{bmatrix} ,   \sigma_3 = \begin{bmatrix}
    1 & 0  \\ 0 & -1 
    \end{bmatrix} ,
\end{align*}

\noindent and the field of two elements, $\textbf{F}_2$. Over this field, denote $L$ as the matrix drawn uniformly at random, where each entry of the matrix is chosen from $\textbf{F}_2$. We demonstrate that computations of the following form can be used to conclude that a particular security threshold exists for the \textit{two-universal} hashing protocol, from:

\begin{itemize}
    \item[$\bullet$] Specifying decompositions for outer products which depend upon elements from the base field,

    \item[$\bullet$] Expressing, through a trace operation, the output of Eve's register throughout the hashing protocol,

    \item[$\bullet$] Determining the number of bit flip, and phase flip, errors, associated with the outer product decomposition, 

    \item[$\bullet$] Concluding that a desired security threshold is obtained from a purification routine applied to random matrices supported over the base field.
\end{itemize}

\noindent Straightforwardly, the action associated with the multiplication operation of $G_n$, for some $g,g^{\prime} \in G_n$, satisfies,

\begin{align*}
 g g^{\prime} =  \big( - 1 \big)^{\mathcal{F}  ( g ) \mathcal{S} \mathcal{F} ( g^{\prime} )^{\mathrm{T}} }  g^{\prime} g , 
\end{align*}

\noindent for the group homomorphism,

\begin{align*}
  \mathcal{F} : G_n \longrightarrow \textbf{F}^{1 \times 2n }_2 : \omega \sigma^u_1 \sigma^u_3 \mapsto \begin{bmatrix}
u & v  \end{bmatrix} ,
\end{align*}

\noindent given some $n>0$, in addition to the block matrix,

\begin{align*}
    \begin{bmatrix}
   0 & \textbf{I}_n \\ \textbf{I}_n & 0  \end{bmatrix} . 
\end{align*}

\noindent The above projection operator plays a significant role in computations performed for CSS probability distributions that have been previously identified in {\color{blue}[40]}. Albeit the fact that one would expect for computations with similarly defined operations to be introduced for determining the occurrence of bit flip, and phase flip, errors for CSS probability distributions identified in this work, differences arise from the fact that we do not seek to approximate $g_1$ and $g_2$ (the previously described functions in \textit{1.1} that are dependent upon the parity check matrices) with suitably chosen functions that are dependent upon \textit{single} elements of $\textbf{F}_2$ at a time. In {\color{blue}[40]}, decompositions for functions of the two parity check matrices are expressed in terms of $\alpha, \beta \in  \textbf{F}_2$.

Instead, we seek an approximation for $g_1$ and $g_2$ with other functions of the parity check matrices which can depend upon $\alpha, \beta$ simultaneously. With such a decomposition over $\alpha, \beta^{\prime}$ and over $\alpha^{\prime} \beta^{\prime\prime}$, we then perform bit flip and phase flip measurements through parity check that are denoted with $\mathcal{P}_1$ and $\mathcal{P}_2$. To obtain such a desired approximation, we propose, and characterize, the probability distributions,

\begin{align*}
      \textbf{P}  \big[         \textit{CSS code can be implemented into a two-universal QKD hashing protocol} \\ \textit{using a random matrix drawn uniformly at random from L}  \sim \textbf{L}     \big]  , \tag{1} \\    \\     \textbf{P} \big[  \textit{CSS code can be implemented into a two-universal QKD hashing protocol} \\  \textit{using a random matrix drawn uniformly at random from } (L^{-1} )^{\mathrm{T}} \sim  (\textbf{L}^{-1} )^{\mathrm{T}} \big]     \tag{2} , 
\end{align*}

\noindent over CSS codes. As a matter of notation, to simplify probabilities related to the two above, from,

\begin{align*}
  \textbf{P}_{\textbf{L}} \equiv \underset{L \in \textbf{L}}{\bigcup } \textbf{P}_L  , 
\end{align*}

\noindent denote,

\begin{align*}
      (1) \equiv \textbf{P}_L \big[ \cdot \big]   , \\ \\ (2) \equiv \textbf{P}_{(L^{-1} )^{\mathrm{T}}} \big[ \cdot \big] . 
\end{align*}

\noindent To obtain the desired parity check matrices one takes marginals of the above CSS probability distribution, specifically either through the second column, or through the first column. The parity check matrices, as the marginal distributions,

\begin{align*}
     \textbf{P}_L \big[ \cdot \big] \bigg|_{\textit{first column}} \equiv \mathcal{P}_1   , \\ \\ \textbf{P}_{(L^{-1} )^{\mathrm{T}}} \big[ \cdot \big] \bigg|_{\textit{second column}} \equiv \mathcal{P}_2  , 
\end{align*}

\noindent which respectively correspond to the two parity check matrices of Alice and Bob, respectively. The above CSS probability measure, equal, over the support $L \sim \mathrm{Mat}_{\textbf{F}^n_2} \big[ \textit{2} \times \textit{2} \big]$ - the space of \textit{two by two} matrices over the two element base field over $n$ qubits, introduce the following CSS probability measure:

\bigskip

\noindent \textbf{Definition} \textit{1} (\textit{probability distributions over CSS codes, for which suitable collision bounds exist, and also for which efficient computational routines for functions of the parity check matrices can be formulated, from sets of positive operator valued measurements}). Fix a random matrix $L$, and $\big( L^{-1} \big)^{\mathrm{T}}$ as its inverse transpose. The probability measures over Quantum CSS codes that will be considered in this work,

\begin{align*}
  \textbf{P}_{L} \big[ \cdot \big]  , \end{align*}

  \begin{align*} \textbf{P}_{(L^{-1})^{\mathrm{T}}} \big[ \cdot \big] \equiv   \textbf{P}_{M} \big[ \cdot \big]  , 
\end{align*}

\noindent assigns the mass

\begin{align*}
     \textbf{P}_L \big[  \textit{There exists a CSS code such that the first parity check matrix, from the finite dimensional} \\ \textit{representation of L, which can be used for error correction}       \big]    \equiv  \mathscr{P} \mathscr{O} \mathcal{V}  \mathscr{M}_{\mathscr{U}}      , 
\end{align*}

\noindent where,

\begin{align*}
 \underset{\mathscr{U}}{\sum } \mathscr{P} \mathscr{O} \mathcal{V}  \mathscr{M}_{\mathscr{U}} = \textbf{I}   , 
\end{align*}

\noindent and, for $M = \big( L^{-1} \big)^{\mathrm{T}}$, equal,

\begin{align*}
       \textbf{P}_M \big[  \textit{There exists a CSS code such that the second parity check matrix, from the finite dimensional} \\ \textit{representation of M, which can be used for error correction}          \big]  \equiv          \mathscr{P} \mathscr{O} \mathcal{V}  \mathscr{M}_{\mathscr{V}}         , 
\end{align*}

\noindent where,

\begin{align*}
   \underset{\mathscr{V}}{\sum } \mathscr{P} \mathscr{O} \mathcal{V}  \mathscr{M}_{\mathscr{V}} = \textbf{I}  , 
\end{align*}

\noindent The forthcoming expression provided for decomposing the outer products,

\begin{align*}
     \underset{\alpha \neq \beta^{\prime} \in \textbf{F}^n_2}{\sum}        \ket{\psi_{\alpha \beta^{\prime} }} \bra{\psi_{\alpha \beta^{\prime} }}  , \\  \underset{\alpha^{\prime} \neq \beta^{\prime\prime} \in \textbf{F}^n_2}{\sum}    \ket{\psi_{\alpha^{\prime}  \beta^{\prime\prime} }} \bra{\psi_{\alpha^{\prime} \beta^{\prime\prime} }}   , 
\end{align*}

\noindent The above expressions provided for probability distributions over random matrices is closely related to the following outer product decomposition. Under the same choice of $n$ provided previously, denote the maximally entangled state as,

\begin{align*}
   \ket{\psi} \equiv 2^{-\frac{n}{2}} \underset{z \in \textbf{F}^n_2}{\sum} \ket{zz}     , 
\end{align*}

\noindent from which one can also introduce,

\begin{align*}
  \ket{\psi_{\alpha \beta}} \equiv \big[ \textbf{I} \otimes   \sigma^{\alpha^{\mathrm{T}}}_1 \sigma^{\beta^{\mathrm{T}}}_3       \big] \ket{\psi}       . 
\end{align*}

\noindent The spanning set,

\begin{align*}
    \underset{\alpha,\beta \in \textbf{F}^n_2}{\mathrm{span}} \big\{        \ket{\psi_{\alpha \beta}}   \big\}     , 
\end{align*}

\noindent corresponds to a Bell basis over $\textbf{C}^{2^n} \otimes \textbf{C}^{2^n} $. The following result, which is used to establish a security result on the \textit{two-universal} hashing QKD protocol, provides an outer product decomposition.

\bigskip

\noindent \textbf{Lemma} (\textit{obtaining approximations for outer products of maximally entangled states, through Bell basis elements}). For all $n$, $\alpha, \beta \in \textbf{F}^n_2$,

\begin{align*}
     \underset{\alpha \neq \beta^{\prime} \in \textbf{F}^n_2}{\sum}        \ket{\psi_{\alpha \beta^{\prime} }} \bra{\psi_{\alpha \beta^{\prime} }} = \underset{z_B \in \textbf{F}^n_2}{\sum}     \ket{z_B, z_B + \alpha + \beta^{\prime}}_{B(B+A)}     \bigg[      \underset{z_A \in \textbf{F}^n_2}{\sum}    \ket{z_A , z_A + \alpha} \bra{z_A , z_A + \alpha}    \bigg]_{AA}  \\ \times \bra{z_B , z_B  + \alpha + \beta^{\prime}}_{B(B+A)}    , \\ \\          \underset{\alpha^{\prime} \neq \beta^{\prime\prime} \in \textbf{F}^n_2}{\sum}    \ket{\psi_{\alpha^{\prime}  \beta^{\prime\prime} }} \bra{\psi_{\alpha^{\prime} \beta^{\prime\prime} }}             = \underset{x_B \in \textbf{F}^n_2}{\sum}  H^{\otimes 2 n}  \bigg[  \ket{x_B + \alpha^{\prime} + \beta^{\prime\prime}}_B   \bigg[         \underset{x_A \in \textbf{F}^2_n}{\sum}    \ket{x_A , x_A + \alpha} \bra{x_A , x_A + \alpha }            \bigg]_{AA}   \\ \times  \bra{x_B + \alpha^{\prime} + \beta^{\prime\prime}}_{B} \bigg]   H^{\otimes 2 n}      , 
\end{align*}

\noindent In the following result below, one makes use of a purification state applied to random matrices $L$. In the form of the random matrix state obtained after purification, the expected value of $LL^{\prime}$ is taken with respect to the computational bases $L \neq L^{\prime}$.

\bigskip

\noindent \textbf{Proposition} \textit{1} (\textit{Eve's quantum register, from the input state $\rho_{ABE}$, can be expressed in terms of the tensor product of an inner product of a randomly chosen matrix supported over the field with two elements, and with bit flip and phase flip operations}). Under notation that has been provided for the input state to the \textit{two-universal} QKD hashing protocol, the state in Eve's register of the protocol can be expressed through the trace,

\begin{align*}
    \mathrm{Tr}_{AB \textbf{L}^{\prime} S^{\prime} T^{\prime} U^{\prime}_A U^{\prime}_B  V^{\prime}_A V^{\prime}_B  W^{\prime}_A W^{\prime}_B       } \bigg[        \underset{x_A, x_B, z_A, z_B}{\underset{\alpha \neq \beta^{\prime} , a^{\prime} \neq \beta^{\prime\prime} \in \textbf{F}^n_2}{\underset{L \in \textit{Mat}_{\textbf{F}^n_2} [ \textit{2} \times \textit{2}]}{\sum}}}           \mathscr{W} \mathscr{V}_{\textit{Real}} \mathscr{U}_{\textit{Real}}    \big[ \rho \otimes \ket{\mathcal{L}} \bra{\mathcal{L}}  \big]      \mathscr{U}^{\dagger}_{\textit{Real}}      \mathscr{V}^{\dagger}_{\textit{Real}}        \mathscr{W}^{\dagger}      \bigg]      , 
\end{align*}

\noindent where the subscript on the trace corresponds to the computational basis over the purification, for the collection, $P_L \equiv \big\{ \forall L>0  : p_L >0 \big\}$,

\begin{align*}
 \ket{\mathcal{L}}_{P_L} \equiv \underset{l \in L}{\bigcup} \ket{\mathcal{L}}_{P_l} \equiv    \underset{L}{\sum} \sqrt{p_L}  \ket{LL}_{L L^{\prime}}    , 
\end{align*}

\noindent of uniformly distributed random matrices over the field with two elements, and the remaining indices in the subscript denote,

\begin{align*}
U^{\prime}_A U^{\prime}_B  \equiv \textit{System spanned by the Bell basis associated with the outputs } u_A, u_B   , \\   \\ V^{\prime}_A V^{\prime}_B  \equiv \textit{System spanned by the Bell basis associated with the outputs } v_A, v_B  ,  \\ \\    W^{\prime}_A W^{\prime}_B \equiv \textit{System spanned by the Bell basis associated with the outputs } w_A, w_B   , \\ \\ LL^{\prime} \equiv \textit{System spanned by the two random matrices associated with the outputs } LL^{\prime} , 
\end{align*}

\noindent and the isometry,

\begin{align*}
   \mathscr{W} \equiv    \ket{L} \bra{L}_{\textbf{L}} \otimes  \frac{1}{\sqrt{2}}  H^{\otimes 2n} \bigg[  \ket{x_A + x_B} \bra{x_A }  +    \ket{x_A + x_B }    \bra{x_B}             \bigg]_{(A+B)A} H^{\otimes 2n} \\ \otimes       \ket{\mathscr{M} x_A,  \mathscr{M} x_A , \mathscr{M} x_B \mathscr{M} x_B     }_{W_A W^{\prime}_A W_B W^{\prime}_B}    , 
\end{align*}

\noindent for,

\begin{align*}
  \mathscr{M} \equiv \textit{Third column row from the representation of } \big( M^{-1} \big)^{\mathrm{T}}  . 
\end{align*}

\noindent In the next result, to consider suitable isometries, which are ultimately related to computations of the states,

\begin{align*}
  \bra{\mathcal{L}}  \big[ \otimes \textit{Product representations for} (\textit{Simulated})^{\dagger}, (\textit{Ideal})^{\dagger}, \textit{and Real isometries}    \big]             \ket{\mathcal{L}} \\ \equiv    \bra{\mathcal{L}}  \big[  \big(\textit{Ideal Isometry}\big)^{\dagger} (\textit{Simulated Isometry})^{\dagger} \textit{Real Isometry}  \big]             \ket{\mathcal{L}}  , 
\end{align*}

\noindent one performs computations, applying the form of outer product decompositions provided in \textbf{Lemma}. Against the \textit{real} isometries, denoted with $\mathscr{U}_{\textit{Real}}$ and $\mathscr{V}_{\textit{Real}}$ in the above trace, are defined in terms of tensor products of the two parity check matrices,

{\small \begin{align*}
   \mathscr{U}_{\textit{Real}}\equiv    \ket{L} \bra{L}_{\textbf{L}} \otimes \bigg\{ \frac{1}{\sqrt{2}} \bigg[       \ket{z_B + z_A} \bra{z_B}_{(B+A)B} + \ket{z_B + z_A} \bra{z_A }_{(B+A)A}      \bigg] \bigg\}  \end{align*}   \begin{align*}     \otimes   \ket{\mathcal{P}_1  z_A,  \mathcal{P}_1 z_A , \mathcal{P}_1  z_B,  \mathcal{P}_1 z_B ,  g_1 \big( \mathcal{P}_1 , \mathcal{P}_1 \big( z_A + z_B \big)  \big)    ,  g_1 \big( \mathcal{P}_1 , \mathcal{P}_1 \big( z_A + z_B \big) \big)              }_{U_A U^{\prime}_A U_B U^{\prime}_B S S^{\prime} }    , \\ \\ \mathscr{V}_{\textit{Real}} \equiv   \ket{L} \bra{L}_{\textbf{L}} \otimes \bigg\{ \frac{1}{\sqrt{2}}  H^{\otimes 2n} \bigg[       \ket{x_B + x_A} \bra{x_B}_{(B+A)B} + \ket{x_B + x_A} \bra{x_A }_{(B+A)A}      \bigg]  H^{\otimes 2n} \bigg\}  \end{align*}   \begin{align*}   \otimes   \ket{\mathcal{P}_2  x_A,  \mathcal{P}_2 x_A , \mathcal{P}_2  x_B,  \mathcal{P}_2 x_B ,  g_2 \big( \mathcal{P}_1 , \mathcal{P}_2 \big( x_A + x_B \big) \big)  ,  g_2 \big( \mathcal{P}_1 , \mathcal{P}_2 \big( x_A + x_B \big) \big)         }_{U_A U^{\prime}_A U_B U^{\prime}_B T T^{\prime}}      , 
\end{align*} }

\noindent the \textit{idealized} isometries, provided below, are \textit{independent} of the two parity check matrices.

\bigskip

\noindent To define the \textit{ideal} isometries in the \textbf{Propositions} below, recall that $F$ denotes the function of bit flip and phase flip errors for the \textit{two-universal} QKD hashing protocol.

\bigskip

\noindent \textbf{Proposition} \textit{2} (\textit{performing operations for determining the occurrence of bit flip errors given the outer product decomposition provided in Lemma 1}). Denote the ideal isometry with,

{\small \begin{align*}
    \mathscr{U}_{\textit{Ideal}} \equiv     \underset{\alpha \neq \beta^{\prime} \in \textbf{F}^n_2}{\sum}      \ket{\psi_{\alpha \beta^{\prime} }} \bra{\psi_{\alpha \beta^{\prime} }}  \otimes   \bigg\{  \frac{1}{\sqrt{2}} \bigg[  \ket{F \big( \alpha \big) , F \big( \alpha + \beta^{\prime} \big)}_{S(S^{\prime} + T^{\prime})} +    \ket{F \big( \beta  \big) , F \big( \alpha + \beta^{\prime} \big)}_{T(S^{\prime} + T^{\prime})}    \bigg]   \bigg\}   .     
\end{align*} }

\noindent If the real isometry,

{\small \begin{align*}
      \mathscr{U}^{\prime}_{\textit{Real}} \equiv       \ket{L} \bra{L}_{\textbf{L}} \otimes \frac{1}{\sqrt{2}} \bigg\{ \bigg[       \ket{z_B + z_A} \bra{z_B}_{(B+A)B} + \ket{z_B + z_A} \bra{z_A }_{(B+A)A}      \bigg] \\ \otimes    \bigg[  \ket{\mathcal{P}_1  z_A,  \mathcal{P}_1 z_A , \mathcal{P}_1  z_B,  \mathcal{P}_1 z_B ,  g_1 \big( \mathcal{P}_1 , \mathcal{P}_1 \big( z_A + z_B \big)  \big)    , \mathcal{P}_1 z_B     }_{U_A U^{\prime}_A U_B U^{\prime}_B S U^{\prime\prime}_B }       \\ +   \ket{\mathcal{P}_1  z_A,  \mathcal{P}_1 z_A , \mathcal{P}_1  z_B,  \mathcal{P}_1 z_B , \mathcal{P}_1 z_B   ,  g_1 \big( \mathcal{P}_1 , \mathcal{P}_1 \big( z_A + z_B \big) \big)              }_{U_A U^{\prime}_A U_B U^{\prime}_B     U^{\prime\prime}_B   S }       \bigg]  \bigg\}    , 
\end{align*} }

\noindent determines the number of bit flip errors, given the \textit{simulator} isometry,

{\small \begin{align*}
  \mathscr{U}_{\textit{Simulator} } \equiv \ket{L} \bra{L}_{\textbf{L}} \otimes \frac{1}{\sqrt{2}} \bigg\{  \bigg[       \ket{z_B + z_A} \bra{z_B}_{(B+A)B} + \ket{z_B + z_A} \bra{z_A }_{(B+A)A}      \bigg] \\ \otimes      \bigg[  \ket{\mathcal{P}_1  z_A,  \mathcal{P}_1 z_A , \mathcal{P}_1  z_B,  \mathcal{P}_1 z_B ,  \mathcal{P}_1  z_A     , \mathcal{P}_1 z_B     }_{U_A U^{\prime}_A U_B U^{\prime}_B S S^{\prime} }       \\ +   \ket{\mathcal{P}_1  z_A,  \mathcal{P}_1 z_A , \mathcal{P}_1  z_B,  \mathcal{P}_1 z_B , \mathcal{P}_1 z_B   ,  \mathcal{P}_1  z_A }_{U_A U^{\prime}_A U_B U^{\prime}_B S S^{\prime} }       \bigg] \bigg\}    , 
\end{align*} }

\noindent then,

{\small \begin{align*}
   \bra{\mathcal{L}} \bigg[ \underset{x_A, x_B, z_A, z_B}{\underset{\alpha \neq \beta^{\prime} \in \textbf{F}^n_2}{\underset{L \in \textit{Mat}_{\textbf{F}^n_2} [ \textit{2} \times \textit{2}]}{\sum}}}  \mathscr{U}^{\dagger}_{\textit{Ideal}}  \mathscr{U}^{\dagger}_{\textit{Simulator}}  \mathscr{U}^{\prime}_{\textit{Real}}      \bigg] \ket{\mathcal{L}}   \gtrsim          \big[ 2^{5-\frac{3}{2}} - 2^{-k + n h ( \frac{r}{n} ) + 5-\frac{3}{2} }     \big] \textbf{I}_{AB}         . 
\end{align*} }

\noindent As one can expect, the next result below provides a statement of how one might go about constructing ideal isometries corresponding to phase flip errors.

\bigskip

\noindent \textbf{Proposition} \textit{3} (\textit{performing operations for determining the occurrence of phase flip errors given the outer product decomposition provided in Lemma 1}). Denote the ideal isometry with,

{\small \begin{align*}
    \mathscr{V}_{\textit{Ideal}} \equiv        \underset{\alpha^{\prime} \neq \beta^{\prime\prime} \in \textbf{F}^n_2}{\sum}  \ket{\psi_{\alpha^{\prime}  \beta^{\prime\prime} }} \bra{\psi_{\alpha^{\prime} \beta^{\prime\prime} }} \otimes   \bigg\{ \frac{1}{\sqrt{2}} \bigg[   \ket{F ( \alpha^{\prime} + \beta^{\prime\prime} ) , F \big( \alpha^{\prime}  \big) }_{(S+T) S^{\prime}} +   \ket{F ( \alpha^{\prime} + \beta^{\prime\prime} ) , F \big( \beta^{\prime\prime} \big) }_{(S+T) T^{\prime}}                \bigg]  \bigg\}   .    
\end{align*} }

\noindent If the real isometry,

{\small \begin{align*}
      \mathscr{V}^{\prime}_{\textit{Real}} \equiv        \ket{L} \bra{L}_{\textbf{L}} \otimes \frac{1}{\sqrt{2}} \bigg\{ H^{\otimes 2n}  \bigg[       \ket{x_B + x_A} \bra{x_B}_{(B+A)B} + \ket{x_B + x_A} \bra{x_A }_{(B+A)A}      \bigg] H^{\otimes 2n}   \\ \otimes   \bigg[     \ket{\mathcal{P}_2  x_A,  \mathcal{P}_2 x_A , \mathcal{P}_2  x_B,  \mathcal{P}_2 x_B ,  g_2 \big( \mathcal{P}_1 , \mathcal{P}_2 \big( x_A + x_B \big) \big)  ,  \mathcal{P}_2  x_A  }_{U_A U^{\prime}_A U_B U^{\prime}_B T U^{\prime\prime}_A }    \\ +    \ket{\mathcal{P}_2  x_A,  \mathcal{P}_2 x_A , \mathcal{P}_2  x_B,  \mathcal{P}_2 x_B , \mathcal{P}_2 x_B   ,  g_2 \big( \mathcal{P}_1 , \mathcal{P}_2 \big( x_A + x_B \big) \big)         }_{U_A U^{\prime}_A U_B U^{\prime}_B U^{\prime\prime}_B T }        \bigg]    \bigg\}             , 
\end{align*} }

\noindent determines the the number of phase flip errors, given the \textit{simulator} isometry,

{\small \begin{align*}
  \mathscr{V}_{\textit{Simulator} } \equiv \ket{L} \bra{L}_{\textbf{L}} \otimes \frac{1}{\sqrt{2}} \bigg\{  H^{\otimes 2n}   \bigg[       \ket{x_B + x_A} \bra{x_B}_{(B+A)B} + \ket{x_B + x_A} \bra{x_A }_{(B+A)A}      \bigg] H^{\otimes 2n}    \\ \otimes        \bigg[     \ket{\mathcal{P}_2  x_A,  \mathcal{P}_2 x_A , \mathcal{P}_2  x_B,  \mathcal{P}_2 x_B , \mathcal{P}_2  x_A  ,  \mathcal{P}_2  x_B  }_{U_A U^{\prime}_A U_B U^{\prime}_B U^{\prime\prime}_A U^{\prime\prime}_B  }    \\ +    \ket{\mathcal{P}_2  x_A,  \mathcal{P}_2 x_A , \mathcal{P}_2  x_B,  \mathcal{P}_2 x_B , \mathcal{P}_2 x_B   ,  \mathcal{P}_2 x_A }_{U_A U^{\prime}_A U_B U^{\prime}_B U^{\prime\prime}_B U^{\prime\prime}_A }        \bigg]  \bigg\}   , 
\end{align*} }

\noindent then,

{\small \begin{align*}
     \bra{\mathcal{L}} \bigg[ \underset{x_A, x_B, z_A, z_B}{\underset{\alpha^{\prime} \neq \beta^{\prime\prime} \in \textbf{F}^n_2}{\underset{L \in \textit{Mat}_{\textbf{F}^n_2} [ \textit{2} \times \textit{2}]}{\sum}}}     \mathscr{V}^{\dagger}_{\textit{Ideal}}   \mathscr{V}^{\dagger}_{\textit{Simulator}}  \mathscr{V}^{\prime}_{\textit{Real}}      \bigg] \ket{\mathcal{L}}   \gtrsim         \big[ 2^{5-\frac{3}{2}} - 2^{-k + n h ( \frac{r}{n} ) + 5-\frac{3}{2} }     \big]  \textbf{I}_{AB}  . 
\end{align*} }

\noindent Besides properties of the isometries $\mathscr{U}$ and $\mathscr{V}$ provided above, the remaining \textbf{Propositions} below establish how purification implies that the desired security level of the \textit{two-universal} QKD hashing protocol holds.

\bigskip

\noindent \textbf{Proposition} \textit{4} (\textit{commutativity of the standard multiplication operation between the simulator and real isometries}). One has that,

{\small \begin{align*}
 \underset{L \in \textit{Mat}_{\textbf{F}^n_2} [ \textit{2} \times \textit{2} ]}{\sum} \underset{\alpha \neq \beta^{\prime} \in \textbf{F}^n_2}{\sum}     \underset{\alpha^{\prime} \neq \beta^{\prime\prime} \in \textbf{F}^n_2}{\sum}   \big\{  \mathscr{U}_{\textit{Simulator}} \mathscr{V}^{\prime}_{\textit{Real} } \big\}  = \mathscr{F} \bigg\{ \underset{L \in \textit{Mat}_{\textbf{F}^n_2} [ \textit{2} \times \textit{2} ]}{\sum} \underset{\alpha \neq \beta^{\prime} \in \textbf{F}^n_2}{\sum}     \underset{\alpha^{\prime} \neq \beta^{\prime\prime} \in \textbf{F}^n_2}{\sum}  \big\{  \mathscr{V}^{\prime}_{\textit{Real}}\\ \times  \mathscr{U}_{\textit{Simulator}} \big\} \bigg\}  ,
\end{align*} }

\noindent where,

{\small \begin{align*}
  \mathscr{F} =    \frac{  P  \bigg[    P \bigg[ \mathcal{P}_1  \big( \vec{\sigma_1}  \big)     , u^{\prime}  \bigg] ,  u   \bigg]  P  \bigg[       P \bigg[ \mathcal{P}_2  \big( \vec{\sigma_3}  \big)  , v^{\prime} \bigg] ,  v   \bigg]}{  P  \bigg[    P \bigg[ \mathcal{P}_2  \big( \vec{\sigma_3}  \big)     , u^{\prime}  \bigg] ,  u   \bigg]  P  \bigg[       P \bigg[  \mathcal{P}_1  \big( \vec{\sigma_1}  \big)  , v^{\prime} \bigg] ,  v   \bigg]}       . 
\end{align*} }

\bigskip

\noindent \textbf{Corollary} (\textit{closed form representation for the anticommutation prefactor}). For the family of functions,

{\small \[ \left\{\!\begin{array}{ll@{}l} 
   \mathscr{T}_1 \big(                  \vec{\sigma_1} , \vec{\sigma_3} ,               u , u^{\prime} , v  , v^{\prime}        \big)         \equiv \mathscr{T}_1   =   \frac{1}{1 + \underset{2 \leq k \leq 8}{\sum} \big[       \mathcal{T}^{\prime}_k  \big]^{-1}   }  , \\   \mathscr{T}_2 \big(                  \vec{\sigma_1} , \vec{\sigma_3} ,               u , u^{\prime} , v  , v^{\prime}        \big)  \equiv \mathscr{T}_2                =         \frac{1}{1 +  \big[ \mathcal{T}_1 \big]^{-1} + \underset{3 \leq k \leq 8}{\sum} \big[       \mathcal{T}^{\prime}_k  \big]^{-1}   }         ,   \\  \mathscr{T}_3 \big(                  \vec{\sigma_1} , \vec{\sigma_3} ,               u , u^{\prime} , v  , v^{\prime}        \big)  \equiv \mathscr{T}_3 =             \frac{1}{1 + \underset{1 \leq k \leq 2}{\sum} \big[       \mathcal{T}^{\prime}_k  \big]^{-1} + \underset{4 \leq k \leq 8}{\sum} \big[    \mathcal{T}^{\prime}_k    \big]^{-1}   }         , \\       \mathscr{T}_4 \big(                  \vec{\sigma_1} , \vec{\sigma_3} ,               u , u^{\prime} , v  , v^{\prime}        \big)  \equiv \mathscr{T}_4                   =              \frac{1}{1 + \underset{1 \leq k \leq 3}{\sum} \big[       \mathcal{T}^{\prime}_k  \big]^{-1} + \underset{5 \leq k \leq 8}{\sum} \big[    \mathcal{T}^{\prime}_k    \big]^{-1}   }      , \\              \mathscr{T}_5\big(                  \vec{\sigma_1} , \vec{\sigma_3} ,               u , u^{\prime} , v  , v^{\prime}        \big)  \equiv \mathscr{T}_5 =         \frac{1}{1+ \underset{1 \leq k \leq 4}{\sum} \big[           \mathcal{T}^{\prime}_k    \big]^{-1}  +    \underset{6 \leq k \leq 8}{\sum} \big[           \mathcal{T}^{\prime}_k    \big]^{-1}     }     , \\           \mathscr{T}_6\big(                  \vec{\sigma_1} , \vec{\sigma_3} ,               u , u^{\prime} , v  , v^{\prime}        \big)  \equiv \mathscr{T}_6 =  \frac{1}{1 + \underset{1 \leq k \leq 5}{\sum} \big[       \mathcal{T}^{\prime}_k  \big]^{-1} + \underset{7 \leq k \leq 8}{\sum} \big[    \mathcal{T}^{\prime}_k    \big]^{-1}   }   , \\       \mathscr{T}_7 \big(                  \vec{\sigma_1} , \vec{\sigma_3} ,               u , u^{\prime} , v  , v^{\prime}        \big)  \equiv \mathscr{T}_7 = \frac{1}{1 + \underset{1 \leq k \leq 6}{\sum} \big[ \mathcal{T}^{\prime}_k \big]^{-1} + \big[ \mathcal{T}^{\prime}_8  \big]^{-1} }                      ,  \\  \mathscr{T}_8 \big(                  \vec{\sigma_1} , \vec{\sigma_3} ,               u , u^{\prime} , v  , v^{\prime}        \big)  \equiv \mathscr{T}_8 =    \frac{1}{1 + \underset{1 \leq k \leq 7}{\sum}  \big[ \mathcal{T}^{\prime}_k \big]^{-1}  }, 
\end{array}\right. 
\]  }

\noindent one has that,

\begin{align*}
   \mathscr{F} =  \underset{1 \leq k \leq 8}{\sum} \mathscr{T}_k        . 
\end{align*}

\noindent \textbf{Proposition} \textit{5} (\textit{the distance between an ideal transformation, and the real transformation, that one can apply for obtaining the purified state of random matrices, with respect to the diamond norm, is equal to the desired security level of the two-universal QKD hashing protocol}). Denote $\mathscr{E}_{\textit{Ideal}}$ and $\mathscr{E}_{\textit{Real}}$ as the ideal, and real, transformations, respectively, for preparing the purified state $\ket{\mathcal{L}}$, and a strictly positive constant $C$, for which,

{\small \begin{align*}
 \bra{\mathcal{L}} \bigg[ \underset{x_A, x_B, z_A, z_B}{\underset{\alpha \neq \beta^{\prime} \in \textbf{F}^n_2}{\underset{L \in \textit{Mat}_{\textbf{F}^n_2} [ \textit{2} \times \textit{2}]}{\sum}}}  \mathscr{U}^{\dagger}_{\textit{Ideal}}  \mathscr{U}^{\dagger}_{\textit{Simulator}}  \mathscr{U}^{\prime}_{\textit{Real}}      \bigg] \ket{\mathcal{L}}  \bigg/     \big[ 2^{5-\frac{3}{2}} - 2^{-k + n h ( \frac{r}{n} ) + 5-\frac{3}{2} }     \big] \textbf{I}_{AB}       \\ = \bra{\mathcal{L}} \bigg[ \underset{x_A, x_B, z_A, z_B}{\underset{\alpha^{\prime} \neq \beta^{\prime\prime} \in \textbf{F}^n_2}{\underset{L \in \textit{Mat}_{\textbf{F}^n_2} [ \textit{2} \times \textit{2}]}{\sum}}}     \mathscr{V}^{\dagger}_{\textit{Ideal}}   \mathscr{V}^{\dagger}_{\textit{Simulator}}  \mathscr{V}^{\prime}_{\textit{Real}}      \bigg] \ket{\mathcal{L}}  \bigg/    \big[ 2^{5-\frac{3}{2}} - 2^{-k + n h ( \frac{r}{n} ) + 5-\frac{3}{2} }     \big]  \textbf{I}_{AB} \end{align*}

 \begin{align*} \geq C  \equiv C  \big( z_A, z_B, x_A, x_B, L \big)     . 
\end{align*} }

\noindent After applying $\mathscr{E}_{\textit{Ideal}}$, one applies the isometries $\mathscr{U}_{\textit{Ideal}}$, $\mathscr{V}_{\textit{Ideal}}$, $\mathscr{U}_{\textit{Simulator}}$, $\mathscr{W}$, and finally, $  \mathrm{Tr}_{AB \textbf{L}^{\prime} S^{\prime} T^{\prime} U^{\prime}_A U^{\prime}_B  V^{\prime}_A V^{\prime}_B  W^{\prime}_A W^{\prime}_B       } \big[ \cdot \big]$. Then,

\begin{align*}
  d_{\textit{Diamond}}  \big( \mathscr{E}_{\textit{Ideal}} ,  \mathscr{E}_{\textit{Real}} \big)  \lesssim  \sqrt{2^{-k + n h ( \frac{r}{n} ) + 5-\frac{3}{2} }    }  , 
\end{align*}

\noindent implies,

\begin{align*}
  d_{\textit{Diamond}}  \big( \mathscr{E}_{\textit{Ideal}} ,  \mathscr{E}_{\textit{Real}} \big) \leq   \sqrt{ C 2^{-k + n h ( \frac{r}{n} ) + 5-\frac{3}{2} }    }  . 
\end{align*}

\noindent The diamond norm is formally introduced in \textit{2.2}. 
\bigskip

\noindent \textbf{Proposition} \textit{6} (\textit{expansion of purified states in the Bell basis}). For any input state $\rho_{ABE}$ into the \textit{two-universal} QKD hashing protocol, the expansion of the purified state, $\ket{\varphi}_{ABEE^{\prime}}$, with respect to Alice and Bob's Bell basis takes the form,

{\small \begin{align*}
  \ket{\varphi}_{ABEE^{\prime}} =  \underset{\alpha^{\prime} \neq \beta^{\prime\prime} \in \textbf{F}^n_2}{\underset{\alpha \neq \beta^{\prime} \in \textbf{F}^n_2}{\sum}}    \bigg[    \big[ \ket{\psi_{\alpha \beta^{\prime}}}_{AB^{\prime}} \big[ \ket{\psi_{\alpha^{\prime} \beta^{\prime\prime}}}_{A^{\prime} B^{\prime\prime}} \big]  \big] \otimes \big[ \ket{\gamma_{\alpha \beta^{\prime}}}_{E E^{\prime}}  \ket{\gamma_{\alpha^{\prime} \beta^{\prime\prime}}}_{E^{\prime} E^{\prime\prime}} \big] \bigg]   , 
\end{align*} }

\noindent and Eve's vectors,  

{\small \begin{align*}
        \big\{ \ket{\gamma_{\alpha\beta^{\prime}}}_{E E^{\prime}} \big\}_{\alpha \neq \beta^{\prime} \in \textbf{F}^n_2}   , \end{align*}

        \begin{align*} \big\{ \ket{\gamma_{\alpha^{\prime} \beta^{\prime\prime}}}_{E^{\prime} E^{\prime\prime}} \big\}_{\alpha^{\prime} \neq \beta^{\prime\prime} \in \textbf{F}^n_2}  , 
\end{align*}    }

\noindent which satisfy,

\begin{align*}
\underset{\alpha^{\prime} \neq \beta^{\prime\prime} \in \textbf{F}^n_2}{\underset{\alpha \neq \beta^{\prime} \in \textbf{F}^n_2}{\sum}}  \bigg[ \big[ \ket{\gamma_{\alpha^{\prime} \beta^{\prime\prime}}} \big]_{EE^{\prime}} \big[   \ket{\gamma_{\alpha \beta^{\prime}}} \bra{\gamma_{\alpha \beta^{\prime}}} \big]_{AB} \big[ \bra{\gamma_{\alpha^{\prime} \beta^{\prime\prime}}} \big]_{EE^{\prime}} \bigg]   = \textbf{I}  ,
\end{align*}

\noindent then there exists a \textit{rejection} scalar,

\begin{align*}
 \sigma^{\textit{Reject}}_{LEE^{\prime} E^{\prime} E^{\prime\prime}S T U_A V_A W_A U_B V_B W_B   },
\end{align*}

\noindent over the register,

\begin{align*}
  LEE^{\prime} E^{\prime} E^{\prime\prime}S T U_A V_A W_A U_B V_B W_B    ,
\end{align*}

\noindent for which, given the Hamming ball of radius $r>0$, supported over $n$ qubits and centered at the origin, $B_n \big( 0 , r \big)$,

{\small \begin{align*}
       \mathscr{E}_{\textit{Ideal}} \big[        \ket{\varphi} \bra{\varphi}    \big]  =   \sigma^{\textit{Reject}}_{LEE^{\prime} E^{\prime} E^{\prime\prime}S T U_A V_A W_A U_B V_B W_B   } + \underset{u_A, v_A, w_A}{\underset{\alpha \neq \beta^{\prime} , \alpha^{\prime} \neq \beta^{\prime\prime} \in B_n ( 0 , r ) }{\underset{L \in \textit{Mat}_{\textbf{F}^n_2} [ \textit{2} \times \textit{2}]}{\sum}}}   \bigg\{        p_L \ket{L} \bra{L}_{\textbf{L}}         \\ \otimes       \ket{\gamma_{\alpha^{\prime} \beta^{\prime\prime}}} \big[   \ket{\gamma_{\alpha \beta^{\prime}}} \\ \times \bra{\gamma_{\alpha \beta^{\prime}}} \big] \bra{\gamma_{\alpha^{\prime} \beta^{\prime\prime}}}      \otimes \ket{\alpha^{\prime} \beta^{\prime\prime}} \big[  \ket{\alpha \beta^{\prime}} \bra{\alpha \beta^{\prime}} \big] \bra{\alpha^{\prime} \beta^{\prime\prime}}   \\ \otimes 2^{-n}      \ket{u_A, v_A, w_A} \bra{u_A, v_A, w_A}_{U_A V_A W_B}   \\   \otimes \ket{u_A +  \mathcal{P}_1 \alpha^{\prime}  , v_A + \mathcal{P}_2  \beta^{\prime\prime} , w_A + \mathscr{M} \beta^{\prime\prime}}   \\ \times \big[  \ket{u_A +  \mathcal{P}_1 \alpha  , v_A + \mathcal{P}_2  \beta^{\prime} , w_A + \mathscr{M} \beta^{\prime}} \\ \times     \bra{u_A +  \mathcal{P}_1  \alpha , v_A + \mathcal{P}_2 \beta^{\prime} , w_A + \mathscr{M} \beta^{\prime}}  \big]  \\ \times   \bra{u_A +  \mathcal{P}_1 \alpha^{\prime} , v_A + \mathcal{P}_2  \beta^{\prime\prime} , w_A + \mathscr{M} \beta^{\prime\prime}}               \bigg\}         .
\end{align*} }

\noindent Equipped with the two ideal isometries $\mathscr{U}$ and $\mathscr{V}$ above, one can conclude the following two main results:

\bigskip

\noindent \textbf{Theorem} \textit{1} (\textit{probabilistic union bound of the function of bit flip and phase flip errors}). For functions $F$ of bit flip and phase flip errors of the \textit{two-universal} QKD hashing protocol,  and probabilistic collision bound $\Theta$, over the set of all possible errors $\mathscr{S}$,

\begin{align*}
  \textbf{P} \big[              \big\{ f_{\mathscr{S}} \neq g_{1,\mathscr{S}} \big\}  , \big\{ f_{\mathscr{S}}  \neq g_{2, \mathscr{S}} \big\} \big] =    \textbf{P} \big[              \big\{ f_{\mathscr{S}} \neq g_{1,\mathscr{S}} , g_{2, \mathscr{S}} \big\} \big]   \\ \leq \Theta \big| \mathscr{S} \big|   . 
\end{align*}

\bigskip

\noindent Often times, it is informative to take the set of all possible errors $\mathscr{S}$ to be  supported over the Hamming ball $B_n \big( 0 , r \big)$, that is, the Hamming ball supported over $n$ qubits that is centered at the origin with radius $r>0$.

\bigskip

\noindent \textbf{Theorem} \textit{2} (\textit{security of the novel two-universal QKD hashing protocol}). Denote the hashing protocol with $\pi^{\prime} \big( n,k,r\big)$, along with the same constant $C$ provided in \textbf{Proposition} \textit{5}. For the number of bit flip, and phase flip, errors $2r$, and total number of qubits $n$ that are encoded in states that Alice and Bob initially input to the QKD hashing protocol, the protocol is,

\begin{align*}
  2^{-\frac{k}{2} + n \frac{h}{2} ( \frac{r}{n} ) + \frac{5}{2} ( 5 - \frac{3}{2} ) + \mathrm{log}_2 \sqrt{C}}  ,
\end{align*}

\noindent secure.

\section{Quantum information-theoretic objects}

\subsection{Two-universal QKD hashing protocols where the computation of functions of parity check matrices is not efficient}

\noindent We introduce additional objects discussed in {\color{blue}[40]}, beyond the projection operator and the $n$-qubit Pauli group in the previous section. One assumes that the final state that is produced by the QKD protocol, after implementing \textit{two-universal} hashing, takes the form,

\begin{align*}
     \widetilde{\rho}_{W_A W_B C E} \equiv \ket{\bot\bot}    \bra{\bot\bot}_{W_A W_B} \otimes \widetilde{\rho}_{CE} \big( \bot \big) + \underset{w_A, w_B}{\sum} \ket{w_A w_B} \bra{w_A w_B}_{W_A W_B}  \otimes \widetilde{\rho}_{CE} \big( w_A , w_B \big)    , 
\end{align*}

\noindent where the subscript of the above quantum state $\rho$ denotes the registers that Alice and Bob would use for outputting the desired secret key, which has a corresponding transcript of classical communication, $C$, and $\bot$ denotes the abort message, after which the QKD hashing protocol would terminate.

\bigskip

\noindent \textbf{Definition} \textit{1}, {\color{blue}[40]} ($\epsilon$ \textit{-security of the QKD protocol}). A QKD protocol is said to be $\epsilon$-secure if the trace distance between  $\widetilde{\rho}_{W_A W_B C E}$, and,

\begin{align*}
    \ket{\bot\bot} \bra{\bot \bot}_{W_A W_B} \otimes \widetilde{\rho}_{CE} \big( \bot \big)       +   \underset{w}{\sum} \frac{1}{\big| W \big|} \ket{ww} \bra{ww}_{W_A W_B} \otimes \big( \widetilde{\rho}_{CE} - \widetilde{\rho}_{CE} \big( \bot \big) \big)         ,
\end{align*}

\noindent is equal to $\epsilon$, where $\big| W \big|$ denotes the size of the secret key space.

\bigskip

\noindent \textbf{Definition} \textit{2}, {\color{blue}[40]} ($\epsilon$-\textit{correctness of the QKD protocol}). A QKD protocol is said to be $\epsilon$-correct if for all input states $\rho_{ABE}$ the probability,

\begin{align*}
    \textbf{P} \big[ W_A \neq W_B \big] =     \underset{w_A \neq w_B}{\sum}     \mathrm{Tr} \big( \widetilde{\rho}_{CE}   \big( w_A , w_B \big) \big)     , 
\end{align*}

\noindent that the keys outputted by Alice and Bob are not equal, is bounded by $\epsilon$.

\bigskip

\noindent \textbf{Definition} \textit{3}, {\color{blue}[40]} ($\epsilon$-\textit{closeness with respect to the trace distance}). Alice's key is $\epsilon$ \textit{secret} if for all input states $\rho_{ABE}$, the reduced output state $\widetilde{\rho}_{W_A C E}$ is $\epsilon$ close to the corresponding ideal state,

\begin{align*}
   \ket{\bot} \bra{\bot}_{W_A} \otimes \widetilde{\rho}_{CE} \big( \bot \big) + \underset{w}{\sum} \frac{1}{\big| W \big| }   \ket{w} \bra{w}_{W_A} \otimes \big( \widetilde{\rho}_{CE} - \widetilde{\rho}_{CE} \big( \bot \big) \big)   .
\end{align*}

\noindent In comparison to the outer product decomposition provided in the previous section, specifically with \textbf{Lemma}, maximally entangled states have previously been expressed through the following decompositions. Instead of obtaining an expansion for the maximally entangled state with elements of the Bell basis with \textit{both} elements of the base field, expansions with one element from the base field are given by:

\bigskip

\noindent \textbf{Lemma} \textit{7}, {\color{blue}[40]} (\textit{outer products of maximally entangled states}). For all $n$, $\alpha, \beta \in \textbf{F}^n_2$,

{\small \begin{align*}
 \underset{\beta^{\prime} \in \textbf{F}^n_2}{\sum} \ket{\psi_{\alpha \beta^{\prime}}} \bra{\psi_{\alpha \beta^{\prime}}} = \underset{z_A \in \textbf{F}^n_2}{\sum} \ket{z_A, z_A + \alpha} \bra{z_A , z_A + \alpha }   , \end{align*}

 \begin{align*}     \underset{\alpha^{\prime} \in \textbf{F}^n_2}{\sum} \ket{\psi_{\alpha^{\prime} \beta}} \bra{\psi_{\alpha^{\prime} \beta}} = \underset{x_A \in \textbf{F}^n_2}{\sum} \big\{ H^{\otimes 2n} \ket{x_A , x_A + \beta} \bra{x_A , x_A + \beta}  H^{\otimes 2n}  \big\}  .
\end{align*} }

\noindent As alluded to in previous remarks throughout \textit{1.1}, one must determine whether Quantum states manipulated in the hashing protocol exhibit "typical" errors. As a function of a subset $S$ of the field with two elements supported over $n$ qubits, denote the image of,

\begin{align*}
   f_S : \textbf{F}^n_2 \longrightarrow S \cup \big\{ \bot \big\}  ,
\end{align*}

\noindent with, 

\[ \left\{\!\begin{array}{ll@{}l} 
\alpha \Longleftrightarrow \alpha \in  S , \\ \bot \text{ otherwise} , 
\end{array}\right. 
\]

\noindent corresponding to the set of errors over $S$. Simply put, $S$ denotes the set of possible acceptable errors, namely those corresponding to bit flip and phase flip errors. Besides quantifying the occurence of such errors, the \textit{collision bound} is defined with the following probabilistic quantity:

\bigskip

\noindent \textbf{Definition} \textit{4}, {\color{blue}[40]} (\textit{collision bounds from hashing functions}). Denote a family of functions $\textbf{H}$, which can be expressed as a mapping with codomain $\textbf{X}$ and image $\textbf{Y}$. The \textit{two-universal} collision probability from the family of hashing functions $\textbf{H}$, $\epsilon$, satisfies,

\begin{align*}
   \underset{h \in \textbf{H}}{\textbf{P}} \big[ h \big( x \big) = h \big( x^{\prime} \big) \big]  \leq \epsilon . 
\end{align*}

\noindent In the above inequality, the mass that the probability measure over hashing functions assigns is distributed uniformly for $h \in \textbf{H}$. Otherwise, the probability measure. over hashing functions assigns an equal mass of $\big| \textbf{Y} \big|^{-1}$ for each such $h$.

\bigskip

\noindent Below, we introduce the collision bound from POVMs for $\mathscr{U}$ and $\mathscr{V}$ isometries defined in \textbf{Definition} \textit{1}.

\bigskip

\noindent \textbf{Definition} \textit{2} (\textit{suitable collision probability bounds from POVMs over isometries}). Denote a family of hashing functions, $\mathscr{H}$, where $\mathscr{H} : \mathscr{X} \longrightarrow \mathscr{Y}$, for two finite sets $\mathscr{X}$ and $\mathscr{Y}$. The \textit{two-universal} collision probability bound, $\Theta > 0 $, satisfies,

\begin{align*}
  \mathscr{P}\mathscr{O}\mathcal{V}\mathscr{M}_{\mathscr{U}} \big[ H \in \mathscr{H} : H \big( x \big) = H \big( x^{\prime} \big)  \big] = \mathscr{P}\mathscr{O}\mathcal{V}\mathscr{M}_{\mathscr{V}} \big[ H \in \mathscr{H} : H \big( x \big) = H \big( x^{\prime} \big)  \big] \\ \textbf{P}_L \big[ H \in \mathscr{H} : H \big( x \big) = H \big( x^{\prime} \big)  \big]  \leq \Theta . 
\end{align*}

\noindent Given the \textit{collision bound} probability above, we conclude our overview of the results from {\color{blue}[40]} with the following statement of the security result. Below, denote the \textit{two-universal} QKD hashing protocol considered in {\color{blue}[40]} with $\pi \big( n , k , r \big)$, where $n$ denotes the total number of qubits that Alice or Bob initially transmit given a suitable encoding, $k$ denotes the size of the syndrome measurements, and $r$ denotes the number of bit flip, and phase flip, errors.

\bigskip

\noindent \textbf{Theorem} \textit{2}, {\color{blue}[40]} (\textit{security of the two-universal QKD hashing protocol}). Fix $n,k,r \in \textbf{N}$ such that $2n h \big( \frac{r}{n} \big) < 2k < n $. Then, the \textit{two-universal} QKD hashing protocol $\pi \big( n , k , r \big)$ is,

\begin{align*}
  2^{-\frac{k}{2} + n h ( \frac{r}{n} ) + \frac{5}{2}}  , 
\end{align*}

\noindent secure.

\bigskip

\noindent Determining how the security of the QKD hashing protocol depends upon the decomposition of outer products of the maximally entangled states is of interest to explore with the following objects. Intuitively, the security level of the hashing protocol will be shown to depend upon braket states,

\begin{align*}
  \bra{\mathcal{L}}  \big[ \otimes \textit{Product representations for} (\textit{Simulated})^{\dagger}, (\textit{Ideal})^{\dagger}, \textit{and Real isometries}    \big]             \ket{\mathcal{L}} \\ \equiv    \bra{\mathcal{L}}  \big[ (\textit{Simulated Isometry})^{\dagger} \big(\textit{Ideal Isometry}\big)^{\dagger} \textit{Real Isometry}  \big]             \ket{\mathcal{L}}  , 
\end{align*}

\noindent of the purified states of random matrices, which were previously described in \textit{1.4}.

\subsection{Two-universal QKD hashing protocols, $\pi^{\prime} \big( n ,k,r \big)$ where the computation of functions of parity check matrices does not make use of idealized states, and is efficient}

\noindent To make the computation of functions of parity check matrices more efficient in the \textit{two-universal} QKD hashing protocol $\pi^{\prime}\big( n , k ,r \big)$, introduce the states,

\begin{align*}
 \ket{\mathcal{P}_1  z_A,  \mathcal{P}_1 z_A , \mathcal{P}_1  z_B,  \mathcal{P}_1 z_B ,  g_1 \big( \mathcal{P}_1 , \mathcal{P}_1 \big( z_A + z_B \big)  \big)    , \mathcal{P}_1 z_B     }_{U_A U^{\prime}_A U_B U^{\prime}_B S U^{\prime\prime}_B}      , \\ \\    \ket{\mathcal{P}_1  z_A,  \mathcal{P}_1 z_A , \mathcal{P}_1  z_B,  \mathcal{P}_1 z_B , \mathcal{P}_1 z_B   ,  g_1 \big( \mathcal{P}_1 , \mathcal{P}_1 \big( z_A + z_B \big) \big)              }_{U_A U^{\prime}_A U_B U^{\prime}_B  U_B U^{\prime\prime}_B  S }       , \\ \\    \ket{\mathcal{P}_1  z_A,  \mathcal{P}_1 z_A , \mathcal{P}_1  z_B,  \mathcal{P}_1 z_B ,  \mathcal{P}_1  z_A     , \mathcal{P}_1 z_B     }_{U_A U^{\prime}_A U_B U^{\prime}_B U^{\prime\prime}_A U^{\prime\prime}_B}          ,  \\ \\   \ket{\mathcal{P}_1  z_A,  \mathcal{P}_1 z_A , \mathcal{P}_1  z_B,  \mathcal{P}_1 z_B , \mathcal{P}_1 z_B   ,  \mathcal{P}_1  z_A }_{U_A U^{\prime}_A U_B U^{\prime}_B U^{\prime\prime}_B U^{\prime\prime}_A  }        , \\ \\ \ket{\mathcal{P}_2  x_A,  \mathcal{P}_2 x_A , \mathcal{P}_2  x_B,  \mathcal{P}_2 x_B ,  g_2 \big( \mathcal{P}_1 , \mathcal{P}_2 \big( x_A + x_B \big) \big)  ,  \mathcal{P}_2  x_A  }_{U_A U^{\prime}_A U_B U^{\prime}_B T U^{\prime\prime}_A }       , \\ \\   \ket{\mathcal{P}_2  x_A,  \mathcal{P}_2 x_A , \mathcal{P}_2  x_B,  \mathcal{P}_2 x_B , \mathcal{P}_2 x_B   ,  g_2 \big( \mathcal{P}_1 , \mathcal{P}_2 \big( x_A + x_B \big) \big)         }_{U_A U^{\prime}_A U_B     U^{\prime}_B U^{\prime\prime}_A T     }    , \\ \\  \ket{\mathcal{P}_2  x_A,  \mathcal{P}_2 x_A , \mathcal{P}_2  x_B,  \mathcal{P}_2 x_B , \mathcal{P}_2  x_A  ,  \mathcal{P}_2  x_B  }_{U_A U^{\prime}_A U_B U^{\prime}_B U^{\prime\prime}_A U^{\prime\prime}_B }  , \\ \\   \ket{\mathcal{P}_2  x_A,  \mathcal{P}_2 x_A , \mathcal{P}_2  x_B,  \mathcal{P}_2 x_B , \mathcal{P}_2 x_B   ,  \mathcal{P}_2 x_A }_{U_A U^{\prime}_A U_B U^{\prime}_B U^{\prime\prime}_B U^{\prime\prime\prime}_B U^{\prime\prime}_A }        . 
\end{align*}

\noindent The states above, as functions of $\mathcal{P}_1$ and $\mathcal{P}_2$, are used to define real, ideal, and simulator isometries. Through expectation values of the form,

\begin{align*}
       \bra{\mathcal{L}}  \big[  \big(\textit{Ideal Isometry}\big)^{\dagger} (\textit{Simulated Isometry})^{\dagger}  \textit{Real Isometry}  \big]             \ket{\mathcal{L}} , 
\end{align*}

\noindent in the purified state of random matrices, one is able to readily conclude that the desired security level for the \textit{two-univrsal} QKD hashing protocol holds. However, in comparison to other security levels that have been previously identified for \textit{two-universal} hashing protocols, as first obtained in {\color{blue}[40]}, being able to efficiently compute $g_1$ and $g_2$ raises the following implications:

\begin{itemize}
    \item[$\bullet$] \textit{Composition of real, ideal, and simulator, isomteries}. In comparison to the isometries introduced in {\color{blue}[40]}, those introduced in this work for obtained the security level oif the \textit{two-universal} QKD hashing protocol are dependent upon:
        \begin{itemize}
            \item[$\bullet$] $\ket{z_B + z_A} \bra{z_B} + \ket{z_B + z_A} \bra{z_A }$, with a normalization of $\sqrt{2}$, 
            \item[$\bullet$] $   \ket{x_B + x_A} \bra{x_B} + \ket{x_B + x_A} \bra{x_A }$, with a normalization of $\sqrt{2}$,
            \item[$\bullet$] Functions $g_1$ of the first parity check matrix, $\mathcal{P}_1$, 
            \item[$\bullet$] Functions $g_2$ of the first parity check matrix, $\mathcal{P}_2$, 
            \item[$\bullet$] $\ket{\mathcal{P}_1  z_A,  \mathcal{P}_1 z_A , \mathcal{P}_1  z_B,  \mathcal{P}_1 z_B ,  \mathcal{P}_1  z_A     , \mathcal{P}_1 z_B     }_{U_A U^{\prime}_A U_B U^{\prime}_B U^{\prime\prime}_A U^{\prime\prime}_B}   $,

           \item[$\bullet$]  $   \ket{\mathcal{P}_1  z_A,  \mathcal{P}_1 z_A , \mathcal{P}_1  z_B,  \mathcal{P}_1 z_B , \mathcal{P}_1 z_B   ,  \mathcal{P}_1  z_A }_{U_A U^{\prime}_A U_B U^{\prime}_B S S^{\prime} }  $, 
            \item[$\bullet$] $\ket{\mathcal{P}_2  x_A,  \mathcal{P}_2 x_A , \mathcal{P}_2  x_B,  \mathcal{P}_2 x_B ,  g_2 \big( \mathcal{P}_1 , \mathcal{P}_2 \big( x_A + x_B \big) \big)  ,  \mathcal{P}_2  x_A  }_{U_A U^{\prime}_A U_B U^{\prime}_B U^{\prime\prime}_B T U^{\prime\prime}_A }     +    \ket{\mathcal{P}_2  x_A,  \mathcal{P}_2 x_A , \mathcal{P}_2  x_B,  \mathcal{P}_2 x_B , \mathcal{P}_2 x_B   ,  g_2 \big( \mathcal{P}_1 , \mathcal{P}_2 \big( x_A + x_B \big) \big)         }_{U_A U^{\prime}_A U_B U^{\prime}_B U^{\prime\prime}_B T }   $, with a normalization of $\sqrt{2}$,
            \item[$\bullet$] functions $F$ of the set of possible errors, which depend upon the logical $X$ and $Z$ operations. 
        \end{itemize}

     \item[$\bullet$] \textit{Expectation values of purified states with ideal, simulated, and real, isometries}. Expectation values of the form,

     \begin{align*}   \bra{\mathcal{L}}  \big[  \big(\textit{Ideal Isometry}\big)^{\dagger} (\textit{Simulated Isometry})^{\dagger} \textit{Real Isometry}  \big]             \ket{\mathcal{L}}  , 
\end{align*}

\noindent can be readily computed by distributed the above terms over operators of tensor products, which will be provided in the next section.

          \item[$\bullet$] \textit{Outer product decompositions}. To demonstrate the lower bounds, up to constants, of the form,

    \begin{align*}
            \big[ 2^{5-\frac{3}{2}} - 2^{-k + n h ( \frac{r}{n} ) + 5-\frac{3}{2} }     \big]  \textbf{I}_{AB}   , 
    \end{align*}

          \noindent respectively hold for,

{\small \begin{align*}
  \bra{\mathcal{L}} \bigg[ \underset{x_A, x_B, z_A, z_B}{\underset{\alpha \neq \beta^{\prime} \in \textbf{F}^n_2}{\underset{L \in \textit{Mat}_{\textbf{F}^n_2} [ \textit{2} \times \textit{2}]}{\sum}}}  \mathscr{U}^{\dagger}_{\textit{Ideal}}  \mathscr{U}^{\dagger}_{\textit{Simulator}}  \mathscr{U}^{\prime}_{\textit{Real}}      \bigg] \ket{\mathcal{L}}    , \\ \\   \bra{\mathcal{L}} \bigg[ \underset{x_A, x_B, z_A, z_B}{\underset{\alpha^{\prime} \neq \beta^{\prime\prime} \in \textbf{F}^n_2}{\underset{L \in \textit{Mat}_{\textbf{F}^n_2} [ \textit{2} \times \textit{2}]}{\sum}}}     \mathscr{V}^{\dagger}_{\textit{Ideal}}   \mathscr{V}^{\dagger}_{\textit{Simulator}}  \mathscr{V}^{\prime}_{\textit{Real}}      \bigg] \ket{\mathcal{L}}   , 
\end{align*} }

          \noindent we make use of the following expansions of outer products, with respect to the Bell basis,

{\small \begin{align*}
     \underset{\alpha \neq \beta^{\prime} \in \textbf{F}^n_2}{\sum}        \ket{\psi_{\alpha \beta^{\prime} }} \bra{\psi_{\alpha \beta^{\prime} }} = \underset{z_B \in \textbf{F}^n_2}{\sum}     \ket{z_B, z_B + \alpha + \beta^{\prime}}_{B(B+A)}     \bigg[      \underset{z_A \in \textbf{F}^n_2}{\sum}    \ket{z_A , z_A + \alpha}  \\ \times  \bra{z_A , z_A + \alpha}    \bigg]_{AA}  \bra{z_B , z_B  + \alpha + \beta^{\prime}}_{B(B+A)}    , \\ \\          \underset{\alpha^{\prime} \neq \beta^{\prime\prime} \in \textbf{F}^n_2}{\sum}    \ket{\psi_{\alpha^{\prime}  \beta^{\prime\prime} }} \bra{\psi_{\alpha^{\prime} \beta^{\prime\prime} }}             = \underset{x_B \in \textbf{F}^n_2}{\sum} \bigg\{  H^{\otimes 2 n}  \bigg[  \ket{x_B + \alpha^{\prime} + \beta^{\prime\prime}}_B   \bigg[         \underset{x_A \in \textbf{F}^2_n}{\sum}    \ket{x_A , x_A + \alpha}  \\ \times  \bra{x_A , x_A + \alpha }            \bigg]_{AA}   \bra{x_B + \alpha^{\prime} + \beta^{\prime\prime}}_{B} \bigg]   H^{\otimes 2 n}   \bigg\}    . 
\end{align*} }

      \item[$\bullet$] \textit{Computation of the diamond distance between real and ideal resources applied throughout the purification routine}. The desired security threshold for the \textit{two-universal} QKD hashing protocol is computed with respect to the distance,

        \begin{align*}
           d_{\textit{Diamond}} \big( \Phi  , X \big)  \equiv  \big| \big|  \big( \Phi \otimes \textbf{I}_n \big) X \big| \big|_1      , 
        \end{align*}

\noindent for the mapping $\Phi : \textbf{M}_n \big( \textbf{C} \big) \longrightarrow \textbf{M}_m \big( \textbf{C}\big)$, namely the collection of $m\times n$ matrices with elements over the base field $\textbf{C}$, $X \in \textbf{M}_{n^2} \big( \textbf{C} \big)$ and the identity map $\textbf{I}_n : \textbf{M}_n \big( \textbf{C} \big) \longrightarrow \textbf{M}_n \big( \textbf{C} \big)$.
      
\end{itemize}

\section{Arguments of main results}

\subsection{$\textbf{Lemma}$}

\subsubsection{Description of computations of entangled outer products with respect to the Bell basis}

\noindent To demonstrate that the desired expansions of entangled outer products with respect to the Bell basis hold, given a choice of $\alpha \neq \beta^{\prime}$ from the base field we incorporate terms from the measurements of,

{\small

\begin{align*}
 \underset{z_B \in \textbf{F}^n_2}{\sum}     \ket{z_B, z_B + \alpha + \beta^{\prime}}_{B(B+A)}    , \end{align*}

 \begin{align*} \underset{z_A \in \textbf{F}^n_2}{\sum}    \ket{z_A , z_A + \alpha}_{AA}   ,  \\   \\  \underset{z_A \in \textbf{F}^n_2}{\sum}    \bra{z_A , z_A + \alpha}_{AA}    , \end{align*}

 \begin{align*}    \underset{z_B \in \textbf{F}^n_2}{\sum}     \bra{z_B , z_B  + \alpha + \beta^{\prime}}_{B(B+A)}   ,  \\ 
\end{align*}

}

\noindent corresponding to the errors of the above bra and ket states from bit-flip errors, respectively, and from the measurements of,

{\small

\begin{align*}
       \underset{x_B \in \textbf{F}^n_2}{\sum}  H^{\otimes 2 n}    \ket{x_B + \alpha^{\prime} + \beta^{\prime\prime}}_B       H^{\otimes 2 n}        , \\ \\               \underset{x_B \in \textbf{F}^n_2}{\sum}      H^{\otimes 2 n}             \bra{x_B + \alpha^{\prime} + \beta^{\prime\prime}}_{B}  H^{\otimes 2 n}             , \\ \\   \underset{x_B \in \textbf{F}^n_2}{\sum} H^{\otimes 2 n}     \bra{x_A , x_A + \alpha }_{AA}             H^{\otimes 2 n}  , \\ \\   \underset{x_B \in \textbf{F}^n_2}{\sum} H^{\otimes 2 n}             \ket{x_A , x_A + \alpha }_{AA}                   H^{\otimes 2 n}  , \\ 
\end{align*}

}

\noindent corresponding to the errors of the above bra and ket states from phase-flip errors, respectively. As a computation involving $\mathscr{P}$ the above entangled outer products are spanned by products of bra and ket states from elements distributed over the base field. As a result we compute the composition of a projection operator with itself, which relates the outer products,

{\small

\begin{align*}
   \ket{\alpha \beta^{\prime}} \bra{\alpha \beta^{\prime}}_{AB}   , \end{align*}

   \begin{align*} \ket{\alpha^{\prime} \beta^{\prime\prime}} \bra{\alpha^{\prime} \beta^{\prime\prime}}_{AB}  , \\ 
\end{align*}

}

\noindent over the base field as the projection operator with respect to $\sigma_3$ of $A$ and of $B$, $\sigma^3_A$ and $\sigma^3_B$, respectively, as well as with respect to $\sigma_3$ of $AB$, $\sigma^3_{AB}$. 

\subsubsection{Proof}

\noindent \textit{Proof of Lemma}. Denote an $n$ element basis of $\textbf{F}^n_2$ with $e_1, \cdots, e_n$, $AB$ Alice and Bob's system at some register $R$, where $R = AB$. The desired expansions, in the Bell states of the outer products,

{\small\begin{align*}
     \underset{\alpha \neq \beta^{\prime} \in \textbf{F}^n_2}{\sum}        \ket{\psi_{\alpha \beta^{\prime} }} \bra{\psi_{\alpha \beta^{\prime} }}   , \\ \\          \underset{\alpha^{\prime} \neq \beta^{\prime\prime} \in \textbf{F}^n_2}{\sum}    \ket{\psi_{\alpha^{\prime}  \beta^{\prime\prime} }} \bra{\psi_{\alpha^{\prime} \beta^{\prime\prime} }}              , 
\end{align*} }

\noindent follows from the fact that the first maximally entangled outer product,

{\small \begin{align*}
   \underset{\alpha \neq \beta^{\prime} \in \textbf{F}^n_2}{\sum}        \ket{\psi_{\alpha \beta^{\prime} }} \bra{\psi_{\alpha \beta^{\prime} }}    , 
\end{align*} }

\noindent can be expressed with the projection,

{\small \begin{align*}
  \ket{\alpha \beta^{\prime}} \bra{\alpha \beta^{\prime}}_{AB} = \mathscr{P} \bigg[  \big[  \vec{\sigma^A_3} , \vec{\sigma^B_3}   \big]^{\mathrm{T}}  , \big[  \alpha , \beta^{\prime} \big]^{\mathrm{T}}      \bigg]   ,  \\ \\   \ket{\psi_{\alpha \beta^{\prime}}} \bra{\psi_{\alpha \beta^{\prime}}}_{AB} = \mathscr{P} \bigg[  \big[  \vec{\sigma^{AB}_3} , \vec{\sigma^{AB}_3}   \big]^{\mathrm{T}}  , \big[  \alpha , \beta^{\prime} \big]^{\mathrm{T}}      \bigg]   , 
\end{align*} }

\noindent while the second maximally entangled outer product,

{\small \begin{align*}
\underset{\alpha^{\prime} \neq \beta^{\prime\prime} \in \textbf{F}^n_2}{\sum}    \ket{\psi_{\alpha^{\prime}  \beta^{\prime\prime} }} \bra{\psi_{\alpha^{\prime} \beta^{\prime\prime} }}              , 
\end{align*} }

\noindent can be expressed with the projection,

{\small \begin{align*}
     \ket{\alpha^{\prime} \beta^{\prime\prime}} \bra{\alpha^{\prime} \beta^{\prime\prime}}_{AB} = \mathscr{P} \bigg[  \big[  \vec{\sigma^A_3} , \vec{\sigma^B_3}   \big]^{\mathrm{T}}  , \big[  \alpha^{\prime} , \beta^{\prime\prime} \big]^{\mathrm{T}}      \bigg]  ,  \\ \\   \ket{\psi_{\alpha^{\prime} \beta^{\prime\prime}}} \bra{\psi_{\alpha^{\prime} \beta^{\prime\prime}}}_{AB} = \mathscr{P} \bigg[   \big[  \vec{\sigma^{AB}_3} , \vec{\sigma^{AB}_3}   \big]^{\mathrm{T}}  , \big[  \alpha^{\prime} , \beta^{\prime\prime} \big]^{\mathrm{T}}      \bigg]     . 
\end{align*} }

\noindent The projection used for expressing each maximally entangled outer product above is given by,

{\small \begin{align*}
  \mathscr{P} \big( \vec{g} \circ  \vec{g^{\prime}}, x \big)     =  \mathscr{P} \big( \vec{g} , \vec{g^{\prime}}, x \big)      =    2^{-m} \underset{1 \leq j \leq m}{\prod}  \big[  \textbf{I}  + \big( - 1 \big)^{x_j} \big( g_j \circ g^{\prime}_j \big)   \big]     \\  =  2^{-m} \underset{1 \leq j \leq m}{\prod}  \bigg[  \textbf{I}  + \big( - 1 \big)^{x_j} \bigg[ g_j  \bigg[   \underset{1 \leq j^{\prime} \leq m}{\prod}  \bigg[  \textbf{I}  + \big( - 1 \big)^{x_j^{\prime}} g_{j^{\prime}}             \bigg]          \bigg]    \bigg]    \\    \equiv   P \bigg[ P \bigg[        \big[  \vec{\sigma^{AB}_3} , \vec{\sigma^{AB}_3}   \big]^{\mathrm{T}}    ,  \big[ \alpha , \beta^{\prime} \big]^{\mathrm{T}}      \bigg] ,   \big[ \alpha^{\prime} , \beta^{\prime\prime} \big]^{\mathrm{T}}          \bigg]        ,        
\end{align*} }

\noindent from which we conclude the argument. \boxed{}

\subsection{$\textbf{Proposition}$ \textit{1}}

\subsubsection{Description of computations against the trace operation}

\noindent To obtain the desired closed form representation for the outer product $\ket{\mathcal{L}} \bra{\mathcal{L}}$ of a purified state $\mathcal{L}$ of a random-matrix, one computes,

{\small

\begin{align*}
 \underset{x_A, x_B, z_A, z_B}{\underset{\alpha \neq \beta^{\prime} , a^{\prime} \neq \beta^{\prime\prime} \in \textbf{F}^n_2}{\underset{L \in \textit{Mat}_{\textbf{F}^n_2} [ \textit{2} \times \textit{2}]}{\sum}}} \big\{  \mathscr{W}  \big( L , \alpha , \beta^{\prime} \big) \mathscr{V}_{\textit{Real}}  \big( L , \alpha , \beta^{\prime} \big)  \mathscr{U}_{\textit{Real}}  \big( L , \alpha , \beta^{\prime} \big) \big\}  \equiv \underset{x_A, x_B, z_A, z_B}{\underset{\alpha \neq \beta^{\prime} , a^{\prime} \neq \beta^{\prime\prime} \in \textbf{F}^n_2}{\underset{L \in \textit{Mat}_{\textbf{F}^n_2} [ \textit{2} \times \textit{2}]}{\sum}}} \big\{  \mathscr{W}  \mathscr{V}_{\textit{Real}}  \mathscr{U}_{\textit{Real}}     \big\}      , \\ \\  \underset{x_A, x_B, z_A, z_B}{\underset{\alpha \neq \beta^{\prime} , a^{\prime} \neq \beta^{\prime\prime} \in \textbf{F}^n_2}{\underset{L \in \textit{Mat}_{\textbf{F}^n_2} [ \textit{2} \times \textit{2}]}{\sum}}}     \big\{ \mathscr{U}^{\dagger}_{\textit{Real}}  \big( L , \alpha , \beta^{\prime} \big)    \mathscr{V}^{\dagger}_{\textit{Real}}   \big( L , \alpha , \beta^{\prime} \big)       \mathscr{W}^{\dagger} \big( L , \alpha , \beta^{\prime} \big) \big\} \equiv  \underset{x_A, x_B, z_A, z_B}{\underset{\alpha \neq \beta^{\prime} , a^{\prime} \neq \beta^{\prime\prime} \in \textbf{F}^n_2}{\underset{L \in \textit{Mat}_{\textbf{F}^n_2} [ \textit{2} \times \textit{2}]}{\sum}}}  \big\{ \mathscr{U}^{\dagger}_{\textit{Real}}     \mathscr{V}^{\dagger}_{\textit{Real}}     \mathscr{W}^{\dagger}  \big\}      , \\ 
\end{align*}

}

\noindent which can then determine how the tensor product of the density $\rho$ with the outer product $\ket{\mathcal{L}} \bra{\mathcal{L}}$ behaves with respect to the trace operation obtained from the correspondence,

{\small

\begin{align*}
  \bigg\{  \underset{\textit{QKD Protocol}}{\prod} \mathrm{Tr} \big[ \rho \big]  \bigg\}  \longleftrightarrow \bigg\{            \underset{n \geq 0}{\underset{\textit{QKD Subprotocols}}{\prod}} \mathrm{Tr} \big[ \rho_1 \otimes \cdots \otimes \rho_n \big]         \bigg\}  \longleftrightarrow \bigg\{                     \underset{n \geq 0}{\underset{\textit{QKD Subprotocols}}{\prod}} \big\{  \mathrm{Tr} \big[ \rho_1 \big] \times  \cdots  \\ \times  \mathrm{Tr} \big[ \rho_n \big]  \big\}               \bigg\}     \longleftrightarrow \bigg\{  \bigg\{   \underset{n \geq 0}{\underset{\textit{First QKD Subprotocol}}{\prod}}  \mathrm{Tr} \big[ \rho_1 \big] \bigg\} \times  \cdots \times  \bigg\{  \underset{n \geq 0}{\underset{\textit{$n$th QKD Subprotocol}}{\prod}} \mathrm{Tr} \big[ \rho_n \big]     \bigg\}          \bigg\}   ,
\end{align*}

}

\noindent from the input density state $\rho_{\textit{init}}$ prepared with a suitably chosen encoding protocol, namely the multiplicativity of the trace over all subprotocols of a QKD protocol.

\subsubsection{Proof}

\noindent \textit{Proof of Proposition 1}. We adapt the argument of \textbf{Proposition} \textit{1} from {\color{blue}[40]}, in place of the real, simulator, and ideal, isometries $\mathscr{U}$ and $\mathscr{V}$ introduced in \textit{1.4}. That is, to argue that,

\begin{align*}
           \underset{x_A, x_B, z_A, z_B}{\underset{\alpha \neq \beta^{\prime} , a^{\prime} \neq \beta^{\prime\prime} \in \textbf{F}^n_2}{\underset{L \in \textit{Mat}_{\textbf{F}^n_2} [ \textit{2} \times \textit{2}]}{\sum}}}         \bigg\{    \mathscr{W} \mathscr{V}_{\textit{Real}} \mathscr{U}_{\textit{Real}}    \big[ \rho \otimes \ket{\mathcal{L}} \bra{\mathcal{L}}  \big]      \mathscr{U}^{\dagger}_{\textit{Real}}      \mathscr{V}^{\dagger}_{\textit{Real}}        \mathscr{W}^{\dagger}  \bigg\}         , 
\end{align*}

\noindent against the trace operation,

\begin{align*}
    \mathrm{Tr}_{AB \textbf{L}^{\prime} S^{\prime} T^{\prime} U^{\prime}_A U^{\prime}_B  V^{\prime}_A V^{\prime}_B  W^{\prime}_A W^{\prime}_B       } \big[     \cdot  \big]      , 
\end{align*}

\noindent takes the desired form, observe that it suffices to compute, sequentially:



{\small \[ \left\{\!\begin{array}{ll@{}l} 
\textbf{Initialization}, \textit{(0)}: \textit{Alice and Bob initialize a system, } AB, \textit{ that they share between themselves,} \\ \textit{and  which they can compute the trace of with the operator } \mathrm{Tr}_{AB} \big[ \cdot \big], \textit{from n qubit states that} \\   \textit{ they receive from Eve}.  \\ \\  \textbf{Random matrix state purification}, \textit{(1)}: \textit{Alice and Bob pick a random matrix}, L, \textit{which they purify} \\ \textit{to } \ket{\mathcal{L}}_{\textbf{L} \textbf{L}^{\prime}} \textit{ and which they can compute the trace of with the operator }  \mathrm{Tr}_{\textbf{L}^{\prime}} \big[ \cdot \big].  \\ \\ \textbf{Isometry}, \textit{(2)}: \textit{Alice and Bob apply the isometry } L_1 \textit{ to } z_A, \textit{ and to } z_B, \textit{respectively, which they can } \\ \textit{compute the trace of with the operator } \mathrm{Tr}_{U^{\prime}_A U^{\prime}_B} \big[ \cdot \big]  .  \\ \\ \textbf{Transpose of random matrix inverse}, \textit{(3)}:     \textit{Alice and Bob apply the second, and third, columns of } 
\\  \big( L^{-1} \big)^{\mathrm{T}} \textit{ to the results obtained from the previous step above, which they can compute the trace of} 
\\  \textit{with the operator } \mathrm{Tr}_{V^{\prime}_A V^{\prime}_B} \big[ \cdot \big]     . \\ \\ \textbf{Parity check matrices computations}, \textit{(4)}:        \textit{Alice and Bob compute functions } g_1 \textit{ and } g_2 \textit{ of } \\  \textit{ the parity check matrices, which they can compute the trace of with the operator } \mathrm{Tr}_{S^{\prime} T^{\prime}} \big[ \cdot \big] , \\ \\ \textbf{Secret key output}, \textit{(5)}: \textit{If the two-universal hashing protocol is not aborted, Alice and Bob output} \\   \textit{ the secret key, which she first outputs as } w_B, \textit{and after which Bob accepts } w_B \\ + \big[ \textit{third column of } \big( L^{-1} \big)^{\mathrm{T}} \big] t \textit{ as a copy of, which they can compute the trace of} \\ \textit{ with the operator } \mathrm{Tr}_{W^{\prime}_A W^{\prime}_B} \big[ \cdot \big] .     
\end{array}\right. 
\]  }

\noindent Putting together the trace operations for each step listed above yields the desired trace operation over,

\begin{align*}
     \mathscr{W} \mathscr{V}_{\textit{Real}} \mathscr{U}_{\textit{Real}}    \big[ \rho \otimes \ket{\mathcal{L}} \bra{\mathcal{L}}  \big]      \mathscr{U}^{\dagger}_{\textit{Real}}      \mathscr{V}^{\dagger}_{\textit{Real}}        \mathscr{W}^{\dagger} ,
\end{align*}

\noindent for each possible choice of $L, \alpha, \beta^{\prime}, \alpha^{\prime}, \beta^{\prime\prime}, x_A, x_B, z_A$, and $z_B$, particularly through,

{\small \begin{align*}
  \underset{\textit{Computational bases provided in steps 0-5 above}}{\prod}    \mathrm{Tr} \big[ \cdot \big]  = \mathrm{Tr}_{AB} \big[ \cdot \big] \mathrm{Tr}_{\textbf{L}^{\prime}} \big[ \cdot \big] \mathrm{Tr}_{U^{\prime}_A U^{\prime}_B} \big[ \cdot \big] \mathrm{Tr}_{V^{\prime}_A V^{\prime}_B} \big[ \cdot \big] \mathrm{Tr}_{S^{\prime} T^{\prime}} \big[ \cdot \big]  \\ \times   \mathrm{Tr}_{W^{\prime}_A W^{\prime}_B} \big[ \cdot \big]     \\ \\ =  \mathrm{Tr}_{AB \textbf{L}^{\prime} S^{\prime} T^{\prime} U^{\prime}_A U^{\prime}_B  V^{\prime}_A V^{\prime}_B  W^{\prime}_A W^{\prime}_B       } \big[     \cdot  \big]             , \\ 
\end{align*} }

\noindent yields the desired trace operation provided in the statement of \textbf{Proposition} \textit{1}, from which we conclude the argument. \boxed{}

\subsection{$\textbf{Proposition}$ \textit{2}}

\subsubsection{Description of the purified random matrix state from the Ideal, Simulator and Real $\mathscr{U}$ Isometries}

\noindent Recall,

{\small 
               
               \begin{align*}   \mathscr{U}_{\textit{Ideal}} \equiv     \underset{\alpha \neq \beta^{\prime} \in \textbf{F}^n_2}{\sum}      \ket{\psi_{\alpha \beta^{\prime} }} \bra{\psi_{\alpha \beta^{\prime} }}  \otimes   \bigg\{  \frac{1}{\sqrt{2}} \bigg[  \ket{F \big( \alpha \big) , F \big( \alpha + \beta^{\prime} \big)}_{S(S^{\prime} + T^{\prime})}  \\ +    \ket{F \big( \beta  \big) , F \big( \alpha + \beta^{\prime} \big)}_{T(S^{\prime} + T^{\prime})}    \bigg]   \bigg\}  ,  \end{align*}
               
               \begin{align*}
       \mathscr{U}_{\textit{Simulator} } \equiv \ket{L} \bra{L}_{\textbf{L}} \otimes \frac{1}{\sqrt{2}} \bigg\{  \bigg[       \ket{z_B + z_A} \bra{z_B}_{(B+A)B} + \ket{z_B + z_A} \bra{z_A }_{(B+A)A}      \bigg] \\ \otimes      \bigg[  \ket{\mathcal{P}_1  z_A,  \mathcal{P}_1 z_A , \mathcal{P}_1  z_B,  \mathcal{P}_1 z_B ,  \mathcal{P}_1  z_A     , \mathcal{P}_1 z_B     }_{U_A U^{\prime}_A U_B U^{\prime}_B S S^{\prime} }       \\ +   \ket{\mathcal{P}_1  z_A,  \mathcal{P}_1 z_A , \mathcal{P}_1  z_B,  \mathcal{P}_1 z_B , \mathcal{P}_1 z_B   ,  \mathcal{P}_1  z_A }_{U_A U^{\prime}_A U_B U^{\prime}_B S S^{\prime} }       \bigg] \bigg\}       ,     \end{align*}
                
               \begin{align*} \mathscr{U}_{\textit{Real}}\equiv    \ket{L} \bra{L}_{\textbf{L}} \otimes \bigg\{ \frac{1}{\sqrt{2}} \bigg[       \ket{z_B + z_A} \bra{z_B}_{(B+A)B} + \ket{z_B + z_A} \bra{z_A }_{(B+A)A}      \bigg] \bigg\} \end{align*}

               \begin{align*} \otimes   \ket{\mathcal{P}_1  z_A,  \mathcal{P}_1 z_A , \mathcal{P}_1  z_B,  \mathcal{P}_1 z_B ,  g_1 \big( \mathcal{P}_1 , \mathcal{P}_1 \big( z_A + z_B \big)  \big)    ,  g_1 \big( \mathcal{P}_1 , \mathcal{P}_1 \big( z_A + z_B \big) \big)              }_{U_A U^{\prime}_A U_B U^{\prime}_B S S^{\prime} }   . 
\end{align*}
}

\noindent To obtained the desired up to constants lower bound for,

{\small \begin{align*}
   \bra{\mathcal{L}} \bigg[ \underset{z_A, z_B}{\underset{\alpha \neq \beta^{\prime} \in \textbf{F}^n_2}{\underset{L \in \textit{Mat}_{\textbf{F}^n_2} [ \textit{2} \times \textit{2}]}{\sum}}}  \mathscr{U}^{\dagger}_{\textit{Ideal}}  \mathscr{U}^{\dagger}_{\textit{Simulator}}  \mathscr{U}^{\prime}_{\textit{Real}}      \bigg] \ket{\mathcal{L}}      , 
\end{align*} }

\noindent observe that it suffices to, by straightforward expansion of the mixed purified state above and linearity of the expectation, compute a superposition from the following terms,

{\small

\begin{align*}
   \underset{z_A, z_B}{\underset{\alpha \neq \beta^{\prime} \in \textbf{F}^n_2}{\underset{L \in \textit{Mat}_{\textbf{F}^n_2} [ \textit{2} \times \textit{2}]}{\sum}}}   p_L   \bigg\{     \bigg[      \ket{z_B, z_B + \alpha + \beta^{\prime}}_{B(B+A)}     \bigg[       \ket{z_A , z_A + \alpha} \bra{z_A , z_A + \alpha}    \bigg]_{AA}  \bra{z_B , z_B  + \alpha + \beta^{\prime}}_{B(B+A)}    \\   \times     \frac{1}{\sqrt{2}}  \ket{F \big( \alpha \big) , F \big( \alpha + \beta^{\prime} \big)}^{\dagger}_{S(S^{\prime} + T^{\prime})} \bigg\}_{(B(B+A))(AA)(B(B+A))(S(S^{\prime} + T^{\prime}))}     , \\ 
\end{align*}

}

 \noindent corresponding to the outer product of,

{\small

\begin{align*}
   \underset{z_A, z_B}{\underset{\alpha \neq \beta^{\prime} \in \textbf{F}^n_2}{\underset{L \in \textit{Mat}_{\textbf{F}^n_2} [ \textit{2} \times \textit{2}]}{\sum}}}   p_L      \bigg\{       \ket{z_B, z_B + \alpha + \beta^{\prime}}_{B(B+A)}     \bigg[       \ket{z_A , z_A + \alpha} \bra{z_A , z_A + \alpha}    \bigg]_{AA} \bigg\}_{(B(B+A))(AA)} , \\ \\     \underset{z_A, z_B}{\underset{\alpha \neq \beta^{\prime} \in \textbf{F}^n_2}{\underset{L \in \textit{Mat}_{\textbf{F}^n_2} [ \textit{2} \times \textit{2}]}{\sum}}}   \bra{z_B , z_B  + \alpha + \beta^{\prime}}_{B(B+A)}      \frac{1}{\sqrt{2}}  \ket{F \big( \alpha \big) , F \big( \alpha + \beta^{\prime} \big)}^{\dagger}_{S(S^{\prime} + T^{\prime})}     , \\ 
\end{align*}

}

 \noindent for the first term of the superposition,

{\small

\begin{align*}
   \underset{z_A, z_B}{\underset{\alpha \neq \beta^{\prime} \in \textbf{F}^n_2}{\underset{L \in \textit{Mat}_{\textbf{F}^n_2} [ \textit{2} \times \textit{2}]}{\sum}}}   p_L     \bigg\{        \bigg[       \ket{z_B, z_B + \alpha + \beta^{\prime}}_{B(B+A)}         \bigg[       \ket{z_A , z_A + \alpha} \bra{z_A , z_A + \alpha}    \bigg]_{AA}  \bra{z_B , z_B  + \alpha + \beta^{\prime}}_{B(B+A)}          \\ \times \frac{1}{\sqrt{2}} \ket{F \big( \beta  \big) , F \big( \alpha + \beta^{\prime} \big)}^{\dagger}_{T(S^{\prime} + T^{\prime})}       \bigg\}_{(B(B+A))(AA)(B(B+A))(T(S^{\prime} + T^{\prime}))}     , \\ 
\end{align*}

}

 \noindent corresponding to the outer product of, 

 {\small

\begin{align*}
   \underset{z_A, z_B}{\underset{\alpha \neq \beta^{\prime} \in \textbf{F}^n_2}{\underset{L \in \textit{Mat}_{\textbf{F}^n_2} [ \textit{2} \times \textit{2}]}{\sum}}}   p_L         \bigg\{       \ket{z_B, z_B + \alpha + \beta^{\prime}}_{B(B+A)}         \bigg[       \ket{z_A , z_A + \alpha} \bra{z_A , z_A + \alpha}    \bigg]_{AA} \bigg\}_{( B ( B + A ))(AA) }  ,  \\ \\   \underset{z_A, z_B}{\underset{\alpha \neq \beta^{\prime} \in \textbf{F}^n_2}{\underset{L \in \textit{Mat}_{\textbf{F}^n_2} [ \textit{2} \times \textit{2}]}{\sum}}}  \bra{z_B , z_B  + \alpha + \beta^{\prime}}_{B(B+A)}         \frac{1}{\sqrt{2}} \ket{F \big( \beta  \big) , F \big( \alpha + \beta^{\prime} \big)}^{\dagger}_{T(S^{\prime} + T^{\prime})}        , \\ 
\end{align*}

}

\noindent for the second term of the superposition,

{\small

\begin{align*}
  \bigg\{  \bigg[     \ket{z_B + z_A} \bra{z_B}_{(B+A)B} + \ket{z_B + z_A} \bra{z_A }_{(B+A)A}     \bigg]  \ket{\mathcal{P}_1  z_A,  \mathcal{P}_1 z_A , \mathcal{P}_1  z_B,  \mathcal{P}_1 z_B ,  \mathcal{P}_1  z_A     , \mathcal{P}_1 z_B     }^{\dagger}_{U_A U^{\prime}_A U_B U^{\prime}_B S S^{\prime} } \\ \bigg\}_{((B+A)A U_A U^{\prime}_A U^{\prime}_B S S^{\prime})}    , \\ 
\end{align*}

}

 \noindent corresponding to third term of the superposition, 

{\small

\begin{align*}
  \bigg\{  \bigg[     \ket{z_B + z_A} \bra{z_B}_{(B+A)B} + \ket{z_B + z_A} \bra{z_A }_{(B+A)A}     \bigg]    \ket{\mathcal{P}_1  z_A,  \mathcal{P}_1 z_A , \mathcal{P}_1  z_B,  \mathcal{P}_1 z_B , \mathcal{P}_1 z_B   ,  \mathcal{P}_1  z_A }^{\dagger}_{U_A U^{\prime}_A U_B U^{\prime}_B S S^{\prime} }  \\ \bigg\}_{((B+A)A U_A U^{\prime}_A U_B U^{\prime}_B S S^{\prime})}   , \\ 
\end{align*}

}

 \noindent corresponding to fourth term of the superposition, and, lastly,

{\small

\begin{align*}
      \bigg\{   \bigg[       \ket{z_B + z_A} \bra{z_B}_{(B+A)B} + \ket{z_B + z_A} \bra{z_A }_{(B+A)A}      \bigg]    \\ \times  \ket{\mathcal{P}_1  z_A,  \mathcal{P}_1 z_A , \mathcal{P}_1  z_B,  \mathcal{P}_1 z_B ,  g_1 \big( \mathcal{P}_1 , \mathcal{P}_1 \big( z_A + z_B \big)  \big)    , \mathcal{P}_1 z_B     }_{U_A U^{\prime}_A U_B U^{\prime}_B S S^{\prime} } \bigg\}_{((B+A)B + (B+A)A)(U_A U^{\prime}_A U_B U^{\prime}_B S S^{\prime} )}                       , \\ 
\end{align*}

}

 \noindent corresponding to fifth term of the superposition. Denoting the above terms as $\mathscr{P}_1$, $\mathscr{P}_2$, $\mathscr{P}^{\prime}_1$,$ \mathscr{P}^{\prime}_2 $, $\mathscr{P}^{\prime\prime}_1$,$ \mathscr{P}^{\prime\prime}_2 $, respectively, one computes an expansion of the form,

 {\small \begin{align*}   \big\{ \mathscr{P}_1 + \mathscr{P}_2  \big\} \big\{  \mathscr{P}^{\prime}_1 + \mathscr{P}^{\prime}_2 \big\}  \big\{ \mathscr{P}^{\prime\prime}_1 + \mathscr{P}^{\prime\prime}_2 \big\}    ,    \\ 
\end{align*}}

\noindent where,

{\small

\begin{align*}
       \mathscr{P}_1 \mathscr{P}^{\prime}_1 \mathscr{P}^{\prime\prime}_1  = \mathcal{O}_1 + \mathcal{O}_2  + \mathcal{O}_3   + \mathcal{O}_4           , \\  \\  \mathscr{P}_1 \mathscr{P}^{\prime}_1  \mathscr{P}^{\prime\prime}_2  = \mathcal{O}_5 + \mathcal{O}_6  + \mathcal{O}_7   + \mathcal{O}_8       , \\ \\  \mathscr{P}_1 \mathscr{P}^{\prime}_2   \mathscr{P}^{\prime\prime}_1 = \mathcal{O}_9 + \mathcal{O}_{10} + \mathcal{O}_{11}   + \mathcal{O}_{12}   , \\ \\   \mathscr{P}_1  \mathscr{P}^{\prime}_2    \mathscr{P}^{\prime\prime}_2  = \mathcal{O}_{13} + \mathcal{O}_{14}  + \mathcal{O}_{15}   + \mathcal{O}_{16}     , \\ \\        \mathscr{P}_2 \mathscr{P}^{\prime}_1 \mathscr{P}^{\prime\prime}_2 = \mathcal{O}_{17} + \mathcal{O}_{18}  + \mathcal{O}_{19}   + \mathcal{O}_{20}        , \\ \\ \mathscr{P}_2 \mathscr{P}^{\prime}_1 \mathscr{P}^{\prime\prime}_2 = \mathcal{O}_{21} + \mathcal{O}_{22}  + \mathcal{O}_{23}   + \mathcal{O}_{24}   , \\ \\  \mathscr{P}_2 \mathscr{P}^{\prime\prime}_2 \mathscr{P}^{\prime\prime}_1 = \mathcal{O}_{25} + \mathcal{O}_{26}  + \mathcal{O}_{27}   + \mathcal{O}_{28}  , \\ \\  \mathscr{P}_2 \mathscr{P}^{\prime\prime}_2 \mathscr{P}^{\prime\prime}_2 = \mathcal{O}_{29} + \mathcal{O}_{30}  + \mathcal{O}_{31}   + \mathcal{O}_{32}  . 
\end{align*}

}

\subsubsection{Proof}

\noindent \textit{Proof of Proposition 2}. By direct computation, to write,

{\small \begin{align*}
   \bra{\mathcal{L}} \bigg[ \underset{z_A, z_B}{\underset{\alpha \neq \beta^{\prime} \in \textbf{F}^n_2}{\underset{L \in \textit{Mat}_{\textbf{F}^n_2} [ \textit{2} \times \textit{2}]}{\sum}}}  \mathscr{U}^{\dagger}_{\textit{Ideal}}  \mathscr{U}^{\dagger}_{\textit{Simulator}}  \mathscr{U}^{\prime}_{\textit{Real}}      \bigg] \ket{\mathcal{L}}      , 
\end{align*} }

\noindent observe,

{\small   \begin{align*}
 \bigg[  \underset{L}{\sum} \sqrt{p_L}  \bra{LL}_{L L^{\prime}}        \bigg]    \bigg\{     \bigg[    \underset{x_A, x_B, z_A, z_B}{\underset{\alpha \neq \beta^{\prime} \in \textbf{F}^n_2}{\sum}}      \ket{\psi_{\alpha \beta^{\prime} }} \bra{\psi_{\alpha \beta^{\prime} }}   \otimes    \frac{1}{\sqrt{2}} \bigg[  \ket{F \big( \alpha \big) , F \big( \alpha + \beta^{\prime} \big)}_{S(S^{\prime} + T^{\prime})} \\   +    \ket{F \big( \beta^{\prime}  \big) , F \big( \alpha + \beta^{\prime} \big)}_{T(S^{\prime} + T^{\prime})}    \bigg]            \bigg]^{\dagger}   \bigg[  \ket{L} \bra{L}_{\textbf{L}} \otimes \frac{1}{\sqrt{2}} \bigg\{  \bigg[       \ket{z_B + z_A} \bra{z_B}_{(B+A)B} + \ket{z_B + z_A} \bra{z_A }_{(B+A)A}      \bigg] \\ \otimes      \bigg[  \ket{\mathcal{P}_1  z_A,  \mathcal{P}_1 z_A , \mathcal{P}_1  z_B,  \mathcal{P}_1 z_B ,  \mathcal{P}_1  z_A     , \mathcal{P}_1 z_B     }_{U_A U^{\prime}_A U_B U^{\prime}_B S S^{\prime} }       \\ +   \ket{\mathcal{P}_1  z_A,  \mathcal{P}_1 z_A , \mathcal{P}_1  z_B,  \mathcal{P}_1 z_B , \mathcal{P}_1 z_B   ,  \mathcal{P}_1  z_A }_{U_A U^{\prime}_A U_B U^{\prime}_B S S^{\prime} }       \bigg] \bigg\}        \bigg]^{\dagger} \\ \times      \ket{L} \bra{L}_{\textbf{L}} \otimes \frac{1}{\sqrt{2}} \bigg\{ \bigg[       \ket{z_B + z_A} \bra{z_B}_{(B+A)B} + \ket{z_B + z_A} \bra{z_A }_{(B+A)A}      \bigg] \\ \otimes    \bigg[  \ket{\mathcal{P}_1  z_A,  \mathcal{P}_1 z_A , \mathcal{P}_1  z_B,  \mathcal{P}_1 z_B ,  g_1 \big( \mathcal{P}_1 , \mathcal{P}_1 \big( z_A + z_B \big)  \big)    , \mathcal{P}_1 z_B     }_{U_A U^{\prime}_A U_B U^{\prime}_B S S^{\prime} } \\  +   \ket{\mathcal{P}_1  z_A,  \mathcal{P}_1 z_A , \mathcal{P}_1  z_B,  \mathcal{P}_1 z_B , \mathcal{P}_1 z_B   ,  g_1 \big( \mathcal{P}_1 , \mathcal{P}_1 \big( z_A + z_B \big) \big)              }_{U_A U^{\prime}_A U_B U^{\prime}_B S S^{\prime} }       \bigg]  \bigg\}      \bigg\}       \end{align*}

 \begin{align*}  \times \bigg[ \underset{L}{\sum} \sqrt{p_L}  \ket{LL}_{L L^{\prime}}  \bigg] \\ \\ =   \underset{z_A, z_B}{\underset{\alpha \neq \beta^{\prime} \in \textbf{F}^n_2}{\underset{L \in \textit{Mat}_{\textbf{F}^n_2} [ \textit{2} \times \textit{2}]}{\sum}}}  \bigg[ p_L   \bigg\{     \bigg[          \ket{\psi_{\alpha \beta^{\prime} }} \bra{\psi_{\alpha \beta^{\prime} }} \times    \frac{1}{\sqrt{2}} \bigg[  \ket{F \big( \alpha \big) , F \big( \alpha + \beta^{\prime} \big)}_{S(S^{\prime} + T^{\prime})}    +    \ket{F \big( \beta  \big) , F \big( \alpha + \beta^{\prime} \big)}_{T(S^{\prime} + T^{\prime})}    \bigg]            \bigg]^{\dagger} \\ \times   \bigg[  \frac{1}{\sqrt{2}} \bigg\{  \bigg[       \ket{z_B + z_A} \bra{z_B}_{(B+A)B} + \ket{z_B + z_A} \bra{z_A }_{(B+A)A}      \bigg] \\ \times      \bigg[  \ket{\mathcal{P}_1  z_A,  \mathcal{P}_1 z_A , \mathcal{P}_1  z_B,  \mathcal{P}_1 z_B ,  \mathcal{P}_1  z_A     , \mathcal{P}_1 z_B     }_{U_A U^{\prime}_A U_B U^{\prime}_B S S^{\prime} }       \\ +   \ket{\mathcal{P}_1  z_A,  \mathcal{P}_1 z_A , \mathcal{P}_1  z_B,  \mathcal{P}_1 z_B , \mathcal{P}_1 z_B   ,  \mathcal{P}_1  z_A }_{U_A U^{\prime}_A U_B U^{\prime}_B S S^{\prime} }       \bigg] \bigg\}        \bigg]^{\dagger} \\ \times    \frac{1}{\sqrt{2}} \bigg\{ \bigg[       \ket{z_B + z_A} \bra{z_B}_{(B+A)B} + \ket{z_B + z_A} \bra{z_A }_{(B+A)A}      \bigg]       \end{align*}

 \begin{align*}   \times    \bigg[  \ket{\mathcal{P}_1  z_A,  \mathcal{P}_1 z_A , \mathcal{P}_1  z_B,  \mathcal{P}_1 z_B ,  g_1 \big( \mathcal{P}_1 , \mathcal{P}_1 \big( z_A + z_B \big)  \big)    , \mathcal{P}_1 z_B     }_{U_A U^{\prime}_A U_B U^{\prime}_B S S^{\prime} }       \\ +   \ket{\mathcal{P}_1  z_A,  \mathcal{P}_1 z_A , \mathcal{P}_1  z_B,  \mathcal{P}_1 z_B , \mathcal{P}_1 z_B   ,  g_1 \big( \mathcal{P}_1 , \mathcal{P}_1 \big( z_A + z_B \big) \big)              }_{U_A U^{\prime}_A U_B U^{\prime}_B S S^{\prime} }       \bigg]  \bigg\}      \bigg\}   \bigg]    \end{align*}

 \begin{align*}  \overset{(\mathrm{\textbf{Lemma}})}{=}  {\underset{\alpha \neq \beta^{\prime} \in \textbf{F}^n_2}{\underset{L \in \textit{Mat}_{\textbf{F}^n_2} [ \textit{2} \times \textit{2}]}{\sum}}}  \bigg[ p_L   \bigg\{     \bigg[     \underbrace{\underset{z_B \in \textbf{F}^n_2}{\sum}     \ket{z_B, z_B + \alpha + \beta^{\prime}}_{B(B+A)}     \bigg[      \underset{z_A \in \textbf{F}^n_2}{\sum}    \ket{z_A , z_A + \alpha} \bra{z_A , z_A + \alpha}    \bigg]_{AA}}  \\ \times \underbrace{\ket{z_B , z_B  + \alpha + \beta^{\prime}}_{B(B+A)}}               \times     \frac{1}{\sqrt{2}} \bigg[  \ket{F \big( \alpha \big) , F \big( \alpha + \beta^{\prime} \big)}_{S(S^{\prime} + T^{\prime})} \\   +    \ket{F \big( \beta  \big) , F \big( \alpha + \beta^{\prime} \big)}_{T(S^{\prime} + T^{\prime})}    \bigg]            \bigg]^{\dagger} \times   \bigg[  \frac{1}{\sqrt{2}} \bigg\{  \bigg[       \ket{z_B + z_A} \bra{z_B}_{(B+A)B} + \ket{z_B + z_A} \bra{z_A }_{(B+A)A}      \bigg]  \\  \times    \bigg[  \ket{\mathcal{P}_1  z_A,  \mathcal{P}_1 z_A , \mathcal{P}_1  z_B,  \mathcal{P}_1 z_B ,  \mathcal{P}_1  z_A     , \mathcal{P}_1 z_B     }_{U_A U^{\prime}_A U_B U^{\prime}_B S S^{\prime} }            \\  +   \ket{\mathcal{P}_1  z_A,  \mathcal{P}_1 z_A , \mathcal{P}_1  z_B,  \mathcal{P}_1 z_B , \mathcal{P}_1 z_B   ,  \mathcal{P}_1  z_A }_{U_A U^{\prime}_A U_B U^{\prime}_B S S^{\prime} }       \bigg] \bigg\}        \bigg]^{\dagger} \\ \times   \frac{1}{\sqrt{2}} \bigg\{ \bigg[       \ket{z_B + z_A} \bra{z_B}_{(B+A)B} + \ket{z_B + z_A} \bra{z_A }_{(B+A)A}      \bigg] \\ \times    \bigg[  \ket{\mathcal{P}_1  z_A,  \mathcal{P}_1 z_A , \mathcal{P}_1  z_B,  \mathcal{P}_1 z_B ,  g_1 \big( \mathcal{P}_1 , \mathcal{P}_1 \big( z_A + z_B \big)  \big)    , \mathcal{P}_1 z_B     }_{U_A U^{\prime}_A U_B U^{\prime}_B S S^{\prime} }     \\    +   \ket{\mathcal{P}_1  z_A,  \mathcal{P}_1 z_A , \mathcal{P}_1  z_B,  \mathcal{P}_1 z_B , \mathcal{P}_1 z_B   ,  g_1 \big( \mathcal{P}_1 , \mathcal{P}_1 \big( z_A + z_B \big) \big)              }_{U_A U^{\prime}_A U_B U^{\prime}_B S S^{\prime} }       \bigg]  \bigg\}      \bigg\}    \bigg]   \\ \\    =   \bigg(   \underset{z_A, z_B}{\underset{\alpha \neq \beta^{\prime} \in \textbf{F}^n_2}{\underset{L \in \textit{Mat}_{\textbf{F}^n_2} [ \textit{2} \times \textit{2}]}{\sum}}}   p_L   \bigg\{     \bigg[      \ket{z_B, z_B + \alpha + \beta^{\prime}}_{B(B+A)}     \bigg[       \ket{z_A , z_A + \alpha} \bra{z_A , z_A + \alpha}    \bigg]_{AA}  \bra{z_B , z_B  + \alpha + \beta^{\prime}}_{B(B+A)}    \\   \times     \frac{1}{\sqrt{2}}  \ket{F \big( \alpha \big) , F \big( \alpha + \beta^{\prime} \big)}^{\dagger}_{S(S^{\prime} + T^{\prime})} \bigg\}       +               \underset{z_A, z_B}{\underset{\alpha \neq \beta^{\prime} \in \textbf{F}^n_2}{\underset{L \in \textit{Mat}_{\textbf{F}^n_2} [ \textit{2} \times \textit{2}]}{\sum}}}   p_L     \bigg\{        \bigg[       \ket{z_B, z_B + \alpha + \beta^{\prime}}_{B(B+A)}    \\   \times      \bigg[       \ket{z_A , z_A + \alpha} \bra{z_A , z_A + \alpha}    \bigg]_{AA}  \bra{z_B , z_B  + \alpha + \beta^{\prime}}_{B(B+A)}          \\ \times \frac{1}{\sqrt{2}} \ket{F \big( \beta  \big) , F \big( \alpha + \beta^{\prime} \big)}^{\dagger}_{T(S^{\prime} + T^{\prime})}       \bigg\} \bigg)   \\  \times         \bigg(    \frac{1}{\sqrt{2}}  \bigg\{     \bigg[     \ket{z_B + z_A} \bra{z_B}_{(B+A)B} + \ket{z_B + z_A} \bra{z_A }_{(B+A)A}     \bigg] \\ \times  \ket{\mathcal{P}_1  z_A,  \mathcal{P}_1 z_A , \mathcal{P}_1  z_B,  \mathcal{P}_1 z_B ,  \mathcal{P}_1  z_A     , \mathcal{P}_1 z_B     }^{\dagger}_{U_A U^{\prime}_A U_B U^{\prime}_B S S^{\prime} }  \end{align*}

 \begin{align*}    +    \bigg[     \ket{z_B + z_A} \bra{z_B}_{(B+A)B} + \ket{z_B + z_A} \bra{z_A }_{(B+A)A}     \bigg] \\ \times     \ket{\mathcal{P}_1  z_A,  \mathcal{P}_1 z_A , \mathcal{P}_1  z_B,  \mathcal{P}_1 z_B , \mathcal{P}_1 z_B   ,  \mathcal{P}_1  z_A }^{\dagger}_{U_A U^{\prime}_A U_B U^{\prime}_B S S^{\prime} }           \bigg\}   \\  \times \frac{1}{\sqrt{2}}  \bigg\{           \bigg[       \ket{z_B + z_A} \bra{z_B}_{(B+A)B} + \ket{z_B + z_A} \bra{z_A }_{(B+A)A}      \bigg] \\ \times    \ket{\mathcal{P}_1  z_A,  \mathcal{P}_1 z_A , \mathcal{P}_1  z_B,  \mathcal{P}_1 z_B ,  g_1 \big( \mathcal{P}_1 , \mathcal{P}_1 \big( z_A + z_B \big)  \big)    , \mathcal{P}_1 z_B     }_{U_A U^{\prime}_A U_B U^{\prime}_B S S^{\prime} }       \\ +  \bigg[       \ket{z_B + z_A} \bra{z_B}_{(B+A)B} + \ket{z_B + z_A} \bra{z_A }_{(B+A)A}      \bigg] \\ \times   \ket{\mathcal{P}_1  z_A,  \mathcal{P}_1 z_A , \mathcal{P}_1  z_B,  \mathcal{P}_1 z_B , \mathcal{P}_1 z_B   ,  g_1 \big( \mathcal{P}_1 , \mathcal{P}_1 \big( z_A + z_B \big) \big)              }_{U_A U^{\prime}_A U_B U^{\prime}_B S S^{\prime} }  \bigg\}  \bigg)        \end{align*}

 \begin{align*} =   \big\{ \mathscr{P}_1 + \mathscr{P}_2  \big\} \big\{  \mathscr{P}^{\prime}_1 + \mathscr{P}^{\prime}_2 \big\}  \big\{ \mathscr{P}^{\prime\prime}_1 + \mathscr{P}^{\prime\prime}_2 \big\}    .    \\ 
\end{align*} }

\noindent In the following, we perform computations from the terms identified in the above product expansion,

{\small

\begin{align*}
  \mathscr{P}_1 =      \underset{z_A, z_B}{\underset{\alpha \neq \beta^{\prime} \in \textbf{F}^n_2}{\underset{L \in \textit{Mat}_{\textbf{F}^n_2} [ \textit{2} \times \textit{2}]}{\sum}}}   p_L    \bigg[      \ket{z_B, z_B + \alpha + \beta^{\prime}}_{B(B+A)}     \bigg[       \ket{z_A , z_A + \alpha} \bra{z_A , z_A + \alpha}    \bigg]_{AA}  \bra{z_B , z_B  + \alpha + \beta^{\prime}}_{B(B+A)}    \\   \times     \frac{1}{\sqrt{2}}  \ket{F \big( \alpha \big) , F \big( \alpha + \beta^{\prime} \big)}^{\dagger}_{S(S^{\prime} + T^{\prime})} \bigg\}                  , \\ \\ \mathscr{P}_2 =   \underset{z_A, z_B}{\underset{\alpha \neq \beta^{\prime} \in \textbf{F}^n_2}{\underset{L \in \textit{Mat}_{\textbf{F}^n_2} [ \textit{2} \times \textit{2}]}{\sum}}}   p_L     \bigg\{        \bigg[       \ket{z_B, z_B + \alpha + \beta^{\prime}}_{B(B+A)}     \bigg[       \ket{z_A , z_A + \alpha} \bra{z_A , z_A + \alpha}    \bigg]_{AA}  \bra{z_B , z_B  + \alpha + \beta^{\prime}}_{B(B+A)}          \\ \times \frac{1}{\sqrt{2}} \ket{F \big( \beta  \big) , F \big( \alpha + \beta^{\prime} \big)}^{\dagger}_{T(S^{\prime} + T^{\prime})}                  , \end{align*}
  
  \begin{align*} \mathscr{P}^{\prime}_1 =     \bigg[     \ket{z_B + z_A} \bra{z_B}_{(B+A)B} + \ket{z_B + z_A} \bra{z_A }_{(B+A)A}     \bigg] \\ \times  \ket{\mathcal{P}_1  z_A,  \mathcal{P}_1 z_A , \mathcal{P}_1  z_B,  \mathcal{P}_1 z_B ,  \mathcal{P}_1  z_A     , \mathcal{P}_1 z_B     }^{\dagger}_{U_A U^{\prime}_A U_B U^{\prime}_B S S^{\prime} }                  , \\ \\ \mathscr{P}^{\prime}_2  =                           \bigg[     \ket{z_B + z_A} \bra{z_B}_{(B+A)B} + \ket{z_B + z_A} \bra{z_A }_{(B+A)A}     \bigg] \\ \times     \ket{\mathcal{P}_1  z_A,  \mathcal{P}_1 z_A , \mathcal{P}_1  z_B,  \mathcal{P}_1 z_B , \mathcal{P}_1 z_B   ,  \mathcal{P}_1  z_A }^{\dagger}_{U_A U^{\prime}_A U_B U^{\prime}_B S S^{\prime} }        , \end{align*}
  
  \begin{align*}  \mathscr{P}^{\prime\prime}_1 =                  \bigg[       \ket{z_B + z_A} \bra{z_B}_{(B+A)B} + \ket{z_B + z_A} \bra{z_A }_{(B+A)A}      \bigg] \\ \times    \ket{\mathcal{P}_1  z_A,  \mathcal{P}_1 z_A , \mathcal{P}_1  z_B,  \mathcal{P}_1 z_B ,  g_1 \big( \mathcal{P}_1 , \mathcal{P}_1 \big( z_A + z_B \big)  \big)    , \mathcal{P}_1 z_B     }_{U_A U^{\prime}_A U_B U^{\prime}_B S S^{\prime} }         , \\ \\ \mathscr{P}^{\prime\prime}_2 =  \bigg[       \ket{z_B + z_A} \bra{z_B}_{(B+A)B} + \ket{z_B + z_A} \bra{z_A }_{(B+A)A}      \bigg] \\ \times  \ket{\mathcal{P}_1  z_A,  \mathcal{P}_1 z_A , \mathcal{P}_1  z_B,  \mathcal{P}_1 z_B , \mathcal{P}_1 z_B   ,  g_1 \big( \mathcal{P}_1 , \mathcal{P}_1 \big( z_A + z_B \big) \big)              }_{U_A U^{\prime}_A U_B U^{\prime}_B S S^{\prime} }   , \\ 
\end{align*}

}

\noindent Further rearrangements of the above superposition, term by term, imply,

{\small \begin{align*}
\mathscr{P}_1 \mathscr{P}^{\prime}_1 \mathscr{P}^{\prime\prime}_1 \end{align*}

\begin{align*} =  \underset{z_A, z_B}{\underset{\alpha \neq \beta^{\prime} \in \textbf{F}^n_2}{\underset{L \in \textit{Mat}_{\textbf{F}^n_2} [ \textit{2} \times \textit{2}]}{\sum}}}   p_L   \bigg\{     \bigg[         \ket{z_B, z_B + \alpha + \beta^{\prime}}_{B(B+A)}     \bigg[       \ket{z_A , z_A + \alpha} \bra{z_A , z_A + \alpha}    \bigg]_{AA}  \\ \times \bra{z_B , z_B  + \alpha + \beta^{\prime}}_{B(B+A)}               \times     \frac{1}{\sqrt{2}}  \ket{F \big( \alpha \big) , F \big( \alpha + \beta^{\prime} \big)}^{\dagger}_{S(S^{\prime} + T^{\prime})} \bigg\}  \\ \times \frac{1}{\sqrt{2}}  \bigg\{     \bigg[     \ket{z_B + z_A} \bra{z_B}_{(B+A)B} + \ket{z_B + z_A} \bra{z_A }_{(B+A)A}     \bigg] \\ \times  \ket{\mathcal{P}_1  z_A,  \mathcal{P}_1 z_A , \mathcal{P}_1  z_B,  \mathcal{P}_1 z_B ,  \mathcal{P}_1  z_A     , \mathcal{P}_1 z_B     }^{\dagger}_{U_A U^{\prime}_A U_B U^{\prime}_B S S^{\prime} }         \bigg\}  \\ \times \frac{1}{\sqrt{2}}  \bigg\{           \bigg[       \ket{z_B + z_A} \bra{z_B}_{(B+A)B} + \ket{z_B + z_A} \bra{z_A }_{(B+A)A}      \bigg] \\ \times    \ket{\mathcal{P}_1  z_A,  \mathcal{P}_1 z_A , \mathcal{P}_1  z_B,  \mathcal{P}_1 z_B ,  g_1 \big( \mathcal{P}_1 , \mathcal{P}_1 \big( z_A + z_B \big)  \big)    , \mathcal{P}_1 z_B     }_{U_A U^{\prime}_A U_B U^{\prime}_B S S^{\prime} }     \bigg\}       , \\  \tag{1}
\end{align*} }

\noindent corresponding to the first term,

{\small \begin{align*}
 \mathscr{P}_1 \mathscr{P}^{\prime}_1  \mathscr{P}^{\prime\prime}_2       \end{align*}

 \begin{align*} = \underset{z_A, z_B}{\underset{\alpha \neq \beta^{\prime} \in \textbf{F}^n_2}{\underset{L \in \textit{Mat}_{\textbf{F}^n_2} [ \textit{2} \times \textit{2}]}{\sum}}}   p_L   \bigg\{     \bigg[        \ket{z_B, z_B + \alpha + \beta^{\prime}}_{B(B+A)}     \bigg[        \ket{z_A , z_A + \alpha} \bra{z_A , z_A + \alpha}    \bigg]_{AA}  \\ \times \bra{z_B , z_B  + \alpha + \beta^{\prime}}_{B(B+A)}               \times     \frac{1}{\sqrt{2}}  \ket{F \big( \alpha \big) , F \big( \alpha + \beta^{\prime} \big)}^{\dagger}_{S(S^{\prime} + T^{\prime})} \bigg\}  \\ \times \frac{1}{\sqrt{2}}  \bigg\{     \bigg[     \ket{z_B + z_A} \bra{z_B}_{(B+A)B} + \ket{z_B + z_A} \bra{z_A }_{(B+A)A}     \bigg] \end{align*}

 \begin{align*} \times  \ket{\mathcal{P}_1  z_A,  \mathcal{P}_1 z_A , \mathcal{P}_1  z_B,  \mathcal{P}_1 z_B ,  \mathcal{P}_1  z_A     , \mathcal{P}_1 z_B     }^{\dagger}_{U_A U^{\prime}_A U_B U^{\prime}_B S S^{\prime} }         \bigg\}  \\ \times \frac{1}{\sqrt{2}}  \bigg\{        \bigg[       \ket{z_B + z_A} \bra{z_B}_{(B+A)B} + \ket{z_B + z_A} \bra{z_A }_{(B+A)A}      \bigg] \\ \times   \ket{\mathcal{P}_1  z_A,  \mathcal{P}_1 z_A , \mathcal{P}_1  z_B,  \mathcal{P}_1 z_B , \mathcal{P}_1 z_B   ,  g_1 \big( \mathcal{P}_1 , \mathcal{P}_1 \big( z_A + z_B \big) \big)              }_{U_A U^{\prime}_A U_B U^{\prime}_B S S^{\prime} }        \bigg\}    \bigg\}       , \\  \tag{2}
\end{align*} }

\noindent corresponding to the second term,

{\small \begin{align*}
    \mathscr{P}_1 \mathscr{P}^{\prime}_2   \mathscr{P}^{\prime\prime}_1   \end{align*}
    
    \begin{align*} =  \underset{z_A, z_B}{\underset{\alpha \neq \beta^{\prime} \in \textbf{F}^n_2}{\underset{L \in \textit{Mat}_{\textbf{F}^n_2} [ \textit{2} \times \textit{2}]}{\sum}}}   p_L   \bigg\{     \bigg[     \ket{z_B, z_B + \alpha + \beta^{\prime}}_{B(B+A)}     \bigg[    \ket{z_A , z_A + \alpha} \bra{z_A , z_A + \alpha}    \bigg]_{AA}  \\ \times \bra{z_B , z_B  + \alpha + \beta^{\prime}}_{B(B+A)}               \times     \frac{1}{\sqrt{2}}  \ket{F \big( \alpha \big) , F \big( \alpha + \beta^{\prime} \big)}^{\dagger}_{S(S^{\prime} + T^{\prime})} \bigg\}  \\ \times \frac{1}{\sqrt{2}}  \bigg\{  \bigg[     \ket{z_B + z_A} \bra{z_B}_{(B+A)B} + \ket{z_B + z_A} \bra{z_A }_{(B+A)A}     \bigg] \\ \times     \ket{\mathcal{P}_1  z_A,  \mathcal{P}_1 z_A , \mathcal{P}_1  z_B,  \mathcal{P}_1 z_B , \mathcal{P}_1 z_B   ,  \mathcal{P}_1  z_A }^{\dagger}_{U_A U^{\prime}_A U_B U^{\prime}_B S S^{\prime} }        \bigg\}  \\ \times  \frac{1}{\sqrt{2}} \bigg\{           \bigg[       \ket{z_B + z_A} \bra{z_B}_{(B+A)B} + \ket{z_B + z_A} \bra{z_A }_{(B+A)A}      \bigg] \\ \times    \ket{\mathcal{P}_1  z_A,  \mathcal{P}_1 z_A , \mathcal{P}_1  z_B,  \mathcal{P}_1 z_B ,  g_1 \big( \mathcal{P}_1 , \mathcal{P}_1 \big( z_A + z_B \big)  \big)    , \mathcal{P}_1 z_B     }_{U_A U^{\prime}_A U_B U^{\prime}_B S S^{\prime} }        \bigg\}     , \\  \tag{3}
\end{align*} } 

\noindent corresponding to the third term,

{\small \begin{align*}
   \mathscr{P}_1  \mathscr{P}^{\prime}_2    \mathscr{P}^{\prime\prime}_2   \end{align*}

   \begin{align*} = \underset{z_A, z_B}{\underset{\alpha \neq \beta^{\prime} \in \textbf{F}^n_2}{\underset{L \in \textit{Mat}_{\textbf{F}^n_2} [ \textit{2} \times \textit{2}]}{\sum}}}   p_L   \bigg\{     \bigg[      \ket{z_B, z_B + \alpha + \beta^{\prime}}_{B(B+A)}     \bigg[     \ket{z_A , z_A + \alpha} \bra{z_A , z_A + \alpha}    \bigg]_{AA}  \\ \times \bra{z_B , z_B  + \alpha + \beta^{\prime}}_{B(B+A)}               \times     \frac{1}{\sqrt{2}}  \ket{F \big( \alpha \big) , F \big( \alpha + \beta^{\prime} \big)}^{\dagger}_{S(S^{\prime} + T^{\prime})} \bigg\}  \end{align*}

   \begin{align*}  \times \frac{1}{\sqrt{2}}  \bigg\{  \bigg[     \ket{z_B + z_A} \bra{z_B}_{(B+A)B} + \ket{z_B + z_A} \bra{z_A }_{(B+A)A}     \bigg] \\ \times     \ket{\mathcal{P}_1  z_A,  \mathcal{P}_1 z_A , \mathcal{P}_1  z_B,  \mathcal{P}_1 z_B , \mathcal{P}_1 z_B   ,  \mathcal{P}_1  z_A }^{\dagger}_{U_A U^{\prime}_A U_B U^{\prime}_B S S^{\prime} }        \bigg\} \\  \times \frac{1}{\sqrt{2}} \bigg\{    \bigg[       \ket{z_B + z_A} \bra{z_B}_{(B+A)B} + \ket{z_B + z_A} \bra{z_A }_{(B+A)A}      \bigg] \\ \times   \ket{\mathcal{P}_1  z_A,  \mathcal{P}_1 z_A , \mathcal{P}_1  z_B,  \mathcal{P}_1 z_B , \mathcal{P}_1 z_B   ,  g_1 \big( \mathcal{P}_1 , \mathcal{P}_1 \big( z_A + z_B \big) \big)              }_{U_A U^{\prime}_A U_B U^{\prime}_B S S^{\prime} }        \bigg\}     , \\  \tag{4}
\end{align*} }

\noindent corresponding to the fourth term,

{\small \begin{align*}
 \mathscr{P}_2 \mathscr{P}^{\prime}_1 \mathscr{P}^{\prime\prime}_1 \end{align*}

 \begin{align*} =  \underset{z_A, z_B}{\underset{\alpha \neq \beta^{\prime} \in \textbf{F}^n_2}{\underset{L \in \textit{Mat}_{\textbf{F}^n_2} [ \textit{2} \times \textit{2}]}{\sum}}}   p_L     \bigg\{        \bigg[      \ket{z_B, z_B + \alpha + \beta^{\prime}}_{B(B+A)}     \bigg[      \ket{z_A , z_A + \alpha} \bra{z_A , z_A + \alpha}    \bigg]_{AA}  \\ \times \bra{z_B , z_B  + \alpha + \beta^{\prime}}_{B(B+A)}          \times \frac{1}{\sqrt{2}} \ket{F \big( \beta  \big) , F \big( \alpha + \beta^{\prime} \big)}^{\dagger}_{T(S^{\prime} + T^{\prime})}        \bigg\}   \\  \times \frac{1}{\sqrt{2}}  \bigg\{    \bigg[     \ket{z_B + z_A} \bra{z_B}_{(B+A)B} + \ket{z_B + z_A} \bra{z_A }_{(B+A)A}     \bigg] \\ \times  \ket{\mathcal{P}_1  z_A,  \mathcal{P}_1 z_A , \mathcal{P}_1  z_B,  \mathcal{P}_1 z_B ,  \mathcal{P}_1  z_A     , \mathcal{P}_1 z_B     }^{\dagger}_{U_A U^{\prime}_A U_B U^{\prime}_B S S^{\prime} }         \bigg\}  \\ \times \frac{1}{\sqrt{2}}  \bigg\{           \bigg[       \ket{z_B + z_A} \bra{z_B}_{(B+A)B} + \ket{z_B + z_A} \bra{z_A }_{(B+A)A}      \bigg] \\ \times    \ket{\mathcal{P}_1  z_A,  \mathcal{P}_1 z_A , \mathcal{P}_1  z_B,  \mathcal{P}_1 z_B ,  g_1 \big( \mathcal{P}_1 , \mathcal{P}_1 \big( z_A + z_B \big)  \big)    , \mathcal{P}_1 z_B     }_{U_A U^{\prime}_A U_B U^{\prime}_B S S^{\prime} }    \bigg\}   , \\  \tag{5} 
\end{align*} }

\noindent corresponding to the fifth term,

{\small \begin{align*}
  \mathscr{P}_2 \mathscr{P}^{\prime}_1 \mathscr{P}^{\prime\prime}_2 \end{align*}

  \begin{align*} =  \underset{z_A, z_B}{\underset{\alpha \neq \beta^{\prime} \in \textbf{F}^n_2}{\underset{L \in \textit{Mat}_{\textbf{F}^n_2} [ \textit{2} \times \textit{2}]}{\sum}}}   p_L     \bigg\{        \bigg[         \ket{z_B, z_B + \alpha + \beta^{\prime}}_{B(B+A)}     \bigg[        \ket{z_A , z_A + \alpha} \bra{z_A , z_A + \alpha}    \bigg]_{AA}  \\ \times \bra{z_B , z_B  + \alpha + \beta^{\prime}}_{B(B+A)}          \times \frac{1}{\sqrt{2}} \ket{F \big( \beta  \big) , F \big( \alpha + \beta^{\prime} \big)}^{\dagger}_{T(S^{\prime} + T^{\prime})}        \bigg\}   \\ \times   \frac{1}{\sqrt{2}}  \bigg\{     \bigg[     \ket{z_B + z_A} \bra{z_B}_{(B+A)B} + \ket{z_B + z_A} \bra{z_A }_{(B+A)A}     \bigg] \end{align*}

   \begin{align*} \times  \ket{\mathcal{P}_1  z_A,  \mathcal{P}_1 z_A , \mathcal{P}_1  z_B,  \mathcal{P}_1 z_B ,  \mathcal{P}_1  z_A     , \mathcal{P}_1 z_B     }^{\dagger}_{U_A U^{\prime}_A U_B U^{\prime}_B S S^{\prime} }   \bigg\} \\ \times     \frac{1}{\sqrt{2}}  \bigg\{      \bigg[       \ket{z_B + z_A} \bra{z_B}_{(B+A)B} + \ket{z_B + z_A} \bra{z_A }_{(B+A)A}      \bigg] \\ \times   \ket{\mathcal{P}_1  z_A,  \mathcal{P}_1 z_A , \mathcal{P}_1  z_B,  \mathcal{P}_1 z_B , \mathcal{P}_1 z_B   ,  g_1 \big( \mathcal{P}_1 , \mathcal{P}_1 \big( z_A + z_B \big) \big)              }_{U_A U^{\prime}_A U_B U^{\prime}_B S S^{\prime} }            \bigg\}   ,  \\  \tag{6}
\end{align*} }

\noindent corresponding to the sixth term,

{\small \begin{align*}
   \mathscr{P}_2 \mathscr{P}^{\prime\prime}_2 \mathscr{P}^{\prime\prime}_1 \end{align*}

   \begin{align*} =  \underset{z_A, z_B}{\underset{\alpha \neq \beta^{\prime} \in \textbf{F}^n_2}{\underset{L \in \textit{Mat}_{\textbf{F}^n_2} [ \textit{2} \times \textit{2}]}{\sum}}}   p_L     \bigg\{        \bigg[       \ket{z_B, z_B + \alpha + \beta^{\prime}}_{B(B+A)}     \bigg[       \ket{z_A , z_A + \alpha} \bra{z_A , z_A + \alpha}    \bigg]_{AA}  \\ \times \bra{z_B , z_B  + \alpha + \beta^{\prime}}_{B(B+A)}          \times \frac{1}{\sqrt{2}} \ket{F \big( \beta  \big) , F \big( \alpha + \beta^{\prime} \big)}^{\dagger}_{T(S^{\prime} + T^{\prime})}        \bigg\}  \\ \times   \frac{1}{\sqrt{2}} \bigg\{     \bigg[     \ket{z_B + z_A} \bra{z_B}_{(B+A)B} + \ket{z_B + z_A} \bra{z_A }_{(B+A)A}     \bigg] \\ \times   \ket{\mathcal{P}_1  z_A,  \mathcal{P}_1 z_A , \mathcal{P}_1  z_B,  \mathcal{P}_1 z_B , \mathcal{P}_1 z_B   ,  \mathcal{P}_1  z_A }^{\dagger}_{U_A U^{\prime}_A U_B U^{\prime}_B S S^{\prime} }     \bigg\}      \\ \times  \frac{1}{\sqrt{2}}  \bigg\{                       \bigg[       \ket{z_B + z_A} \bra{z_B}_{(B+A)B} + \ket{z_B + z_A} \bra{z_A }_{(B+A)A}      \bigg] \\ \times    \ket{\mathcal{P}_1  z_A,  \mathcal{P}_1 z_A , \mathcal{P}_1  z_B,  \mathcal{P}_1 z_B ,  g_1 \big( \mathcal{P}_1 , \mathcal{P}_1 \big( z_A + z_B \big)  \big)    , \mathcal{P}_1 z_B     }_{U_A U^{\prime}_A U_B U^{\prime}_B S S^{\prime} }               \bigg\}   , \\  \tag{7}
\end{align*} }

\noindent corresponding to the seventh term,

{\small \begin{align*}
   \mathscr{P}_2 \mathscr{P}^{\prime\prime}_2 \mathscr{P}^{\prime\prime}_2 \end{align*}

   \begin{align*} =  \underset{z_A, z_B}{\underset{\alpha \neq \beta^{\prime} \in \textbf{F}^n_2}{\underset{L \in \textit{Mat}_{\textbf{F}^n_2} [ \textit{2} \times \textit{2}]}{\sum}}}   p_L     \bigg\{        \bigg[       \ket{z_B, z_B + \alpha + \beta^{\prime}}_{B(B+A)}     \bigg[         \ket{z_A , z_A + \alpha} \bra{z_A , z_A + \alpha}    \bigg]_{AA}  \\ \times \bra{z_B , z_B  + \alpha + \beta^{\prime}}_{B(B+A)}          \times \frac{1}{\sqrt{2}} \ket{F \big( \beta  \big) , F \big( \alpha + \beta^{\prime} \big)}^{\dagger}_{T(S^{\prime} + T^{\prime})}        \bigg\}   \\ \times       \frac{1}{\sqrt{2}} \bigg\{     \bigg[     \ket{z_B + z_A} \bra{z_B}_{(B+A)B} + \ket{z_B + z_A} \bra{z_A }_{(B+A)A}     \bigg] \end{align*}

   \begin{align*}  \times   \ket{\mathcal{P}_1  z_A,  \mathcal{P}_1 z_A , \mathcal{P}_1  z_B,  \mathcal{P}_1 z_B , \mathcal{P}_1 z_B   ,  \mathcal{P}_1  z_A }^{\dagger}_{U_A U^{\prime}_A U_B U^{\prime}_B S S^{\prime} }     \bigg\}    \\ \times  \frac{1}{\sqrt{2}}  \bigg\{ \bigg[       \ket{z_B + z_A} \bra{z_B}_{(B+A)B} + \ket{z_B + z_A} \bra{z_A }_{(B+A)A}      \bigg] \\ \times   \ket{\mathcal{P}_1  z_A,  \mathcal{P}_1 z_A , \mathcal{P}_1  z_B,  \mathcal{P}_1 z_B , \mathcal{P}_1 z_B   ,  g_1 \big( \mathcal{P}_1 , \mathcal{P}_1 \big( z_A + z_B \big) \big)              }_{U_A U^{\prime}_A U_B U^{\prime}_B S S^{\prime} }   \bigg\}  , \\  \tag{8}
\end{align*} }

\noindent corresponding to the eighth term. Proceeding further, the desired superposition,

{\small \begin{align*}
   1 + 2  + 3 + 4 + 5  + 6 + 7 + 8   =  1^{\prime} + 2^{\prime}  + 3^{\prime} + 4^{\prime} + 5^{\prime}  + 6^{\prime}  + 7^{\prime}  + 8^{\prime}  , 
\end{align*} }

\noindent is readily obtained upon consolidating the following terms,

{\small \begin{align*}
\mathscr{P}_1 \mathscr{P}^{\prime}_1 \mathscr{P}^{\prime\prime}_1 \end{align*}

\begin{align*} = \mathcal{O}_1 + \mathcal{O}_2  + \mathcal{O}_3   + \mathcal{O}_4   \end{align*}

\begin{align*}  =  \bigg[ \frac{1}{\sqrt{2}} \bigg]^3   \underset{z_A, z_B}{\underset{\alpha \neq \beta^{\prime} \in \textbf{F}^n_2}{\underset{L \in \textit{Mat}_{\textbf{F}^n_2} [ \textit{2} \times \textit{2}]}{\sum}}}  \bigg\{ \bigg\{  p_L  \bra{\mathcal{P}_1  z_A,  \mathcal{P}_1 z_A , \mathcal{P}_1  z_B,  \mathcal{P}_1 z_B ,  \mathcal{P}_1  z_A     , \mathcal{P}_1 z_B     }_{U_A U^{\prime}_A U_B U^{\prime}_B S S^{\prime} }   \\ \times  \bigg\{     \bigg[      \ket{z_B, z_B + \alpha + \beta^{\prime}}_{B(B+A)}     \bigg[        \ket{z_A , z_A + \alpha} \bra{z_A , z_A + \alpha}    \bigg]_{AA}  \\ \times \bra{z_B , z_B  + \alpha + \beta^{\prime}}_{B(B+A)}               \times       \bra{F \big( \alpha \big) , F \big( \alpha + \beta^{\prime} \big)}_{S(S^{\prime} + T^{\prime})} \bigg\} \\ \times   \ket{z_B + z_A} \bra{z_B}_{(B+A)B} \bigg\}         \ket{z_B + z_A} \bra{z_B}_{(B+A)B}  \bigg\}                \\   \times \ket{\mathcal{P}_1  z_A,  \mathcal{P}_1 z_A , \mathcal{P}_1  z_B,  \mathcal{P}_1 z_B ,  g_1 \big( \mathcal{P}_1 , \mathcal{P}_1 \big( z_A + z_B \big)  \big)    , \mathcal{P}_1 z_B     }_{U_A U^{\prime}_A U_B U^{\prime}_B S S^{\prime} }  \\ \\ + \bigg[ \frac{1}{\sqrt{2}} \bigg]^3   \underset{z_A, z_B}{\underset{\alpha \neq \beta^{\prime} \in \textbf{F}^n_2}{\underset{L \in \textit{Mat}_{\textbf{F}^n_2} [ \textit{2} \times \textit{2}]}{\sum}}}  \bigg\{ \bigg\{  p_L  \bra{\mathcal{P}_1  z_A,  \mathcal{P}_1 z_A , \mathcal{P}_1  z_B,  \mathcal{P}_1 z_B ,  \mathcal{P}_1  z_A     , \mathcal{P}_1 z_B     }_{U_A U^{\prime}_A U_B U^{\prime}_B S S^{\prime} }   \\ \times  \bigg\{     \bigg[      \ket{z_B, z_B + \alpha + \beta^{\prime}}_{B(B+A)}     \bigg[        \ket{z_A , z_A + \alpha} \bra{z_A , z_A + \alpha}    \bigg]_{AA}  \\    \times \bra{z_B , z_B  + \alpha + \beta^{\prime}}_{B(B+A)}               \times       \bra{F \big( \alpha \big) , F \big( \alpha + \beta^{\prime} \big)}_{S(S^{\prime} + T^{\prime})} \bigg\}  \end{align*}

 \begin{align*}    \times   \ket{z_B + z_A} \bra{z_B}_{(B+A)B} \bigg\}         \ket{z_B + z_A} \bra{z_A}_{(B+A)B}  \bigg\}               \\   \times \ket{\mathcal{P}_1  z_A,  \mathcal{P}_1 z_A , \mathcal{P}_1  z_B,  \mathcal{P}_1 z_B ,  g_1 \big( \mathcal{P}_1 , \mathcal{P}_1 \big( z_A + z_B \big)  \big)    , \mathcal{P}_1 z_B     }_{U_A U^{\prime}_A U_B U^{\prime}_B S S^{\prime} }  \\ \\ + \bigg[ \frac{1}{\sqrt{2}} \bigg]^3   \underset{z_A, z_B}{\underset{\alpha \neq \beta^{\prime} \in \textbf{F}^n_2}{\underset{L \in \textit{Mat}_{\textbf{F}^n_2} [ \textit{2} \times \textit{2}]}{\sum}}}  \bigg\{ \bigg\{  p_L  \bra{\mathcal{P}_1  z_A,  \mathcal{P}_1 z_A , \mathcal{P}_1  z_B,  \mathcal{P}_1 z_B ,  \mathcal{P}_1  z_A     , \mathcal{P}_1 z_B     }_{U_A U^{\prime}_A U_B U^{\prime}_B S S^{\prime} }   \\ \times  \bigg\{     \bigg[        \ket{z_B, z_B + \alpha + \beta^{\prime}}_{B(B+A)}     \bigg[        \ket{z_A , z_A + \alpha} \bra{z_A , z_A + \alpha}    \bigg]_{AA}  \\ \times \bra{z_B , z_B  + \alpha + \beta^{\prime}}_{B(B+A)}               \times       \bra{F \big( \alpha \big) , F \big( \alpha + \beta^{\prime} \big)}_{S(S^{\prime} + T^{\prime})} \bigg\} \\ \times   \ket{z_B + z_A} \bra{z_A}_{(B+A)B} \bigg\}         \ket{z_B + z_A} \bra{z_B}_{(B+A)B}  \bigg\}            \end{align*}

 \begin{align*}    \times \ket{\mathcal{P}_1  z_A,  \mathcal{P}_1 z_A , \mathcal{P}_1  z_B,  \mathcal{P}_1 z_B ,  g_1 \big( \mathcal{P}_1 , \mathcal{P}_1 \big( z_A + z_B \big)  \big)    , \mathcal{P}_1 z_B     }_{U_A U^{\prime}_A U_B U^{\prime}_B S S^{\prime} }   \\ \\ + \bigg[ \frac{1}{\sqrt{2}} \bigg]^3   \underset{z_A, z_B}{\underset{\alpha \neq \beta^{\prime} \in \textbf{F}^n_2}{\underset{L \in \textit{Mat}_{\textbf{F}^n_2} [ \textit{2} \times \textit{2}]}{\sum}}}  \bigg\{ \bigg\{  p_L  \bra{\mathcal{P}_1  z_A,  \mathcal{P}_1 z_A , \mathcal{P}_1  z_B,  \mathcal{P}_1 z_B ,  \mathcal{P}_1  z_A     , \mathcal{P}_1 z_B     }_{U_A U^{\prime}_A U_B U^{\prime}_B S S^{\prime} }   \\ \times  \bigg\{     \bigg[       \ket{z_B, z_B + \alpha + \beta^{\prime}}_{B(B+A)}     \bigg[       \ket{z_A , z_A + \alpha} \bra{z_A , z_A + \alpha}    \bigg]_{AA}  \\ \times \bra{z_B , z_B  + \alpha + \beta^{\prime}}_{B(B+A)}               \times       \bra{F \big( \alpha \big) , F \big( \alpha + \beta^{\prime} \big)}_{S(S^{\prime} + T^{\prime})} \bigg\} \\ \times   \ket{z_B + z_A} \bra{z_A}_{(B+A)B} \bigg\}         \ket{z_B + z_A} \bra{z_A}_{(B+A)B}  \bigg\}               \\   \times \ket{\mathcal{P}_1  z_A,  \mathcal{P}_1 z_A , \mathcal{P}_1  z_B,  \mathcal{P}_1 z_B ,  g_1 \big( \mathcal{P}_1 , \mathcal{P}_1 \big( z_A + z_B \big)  \big)    , \mathcal{P}_1 z_B     }_{U_A U^{\prime}_A U_B U^{\prime}_B S S^{\prime} } ,  \\ \tag{$1^{\prime}$}
\end{align*} }

\noindent corresponding to the first term,

{\small \begin{align*}
 \mathscr{P}_1 \mathscr{P}^{\prime}_1  \mathscr{P}^{\prime\prime}_2   \end{align*}

\begin{align*} = \mathcal{O}_5 + \mathcal{O}_6  + \mathcal{O}_7   + \mathcal{O}_8   \end{align*}

\begin{align*}   = \bigg[ \frac{1}{\sqrt{2}} \bigg]^3     \underset{z_A, z_B}{\underset{\alpha \neq \beta^{\prime} \in \textbf{F}^n_2}{\underset{L \in \textit{Mat}_{\textbf{F}^n_2} [ \textit{2} \times \textit{2}]}{\sum}}}  \bigg\{ \bigg\{  p_L             \bra{\mathcal{P}_1  z_A,  \mathcal{P}_1 z_A , \mathcal{P}_1  z_B,  \mathcal{P}_1 z_B ,  \mathcal{P}_1  z_A     , \mathcal{P}_1 z_B     }_{U_A U^{\prime}_A U_B U^{\prime}_B S S^{\prime} }   \\  \times \bigg\{    \bigg[        \ket{z_B, z_B + \alpha + \beta^{\prime}}_{B(B+A)}     \bigg[        \ket{z_A , z_A + \alpha} \bra{z_A , z_A + \alpha}    \bigg]_{AA}  \\ \times \bra{z_B , z_B  + \alpha + \beta^{\prime}}_{B(B+A)}               \times   \bra{F \big( \alpha \big) , F \big( \alpha + \beta^{\prime} \big)}_{S(S^{\prime} + T^{\prime})} \bigg\}  \\ \times     \ket{z_B + z_A} \bra{z_B}_{(B+A)B} \bigg\}      \ket{z_B + z_A} \bra{z_B}_{(B+A)B}       \bigg\}       \\ \times    \ket{\mathcal{P}_1  z_A,  \mathcal{P}_1 z_A , \mathcal{P}_1  z_B,  \mathcal{P}_1 z_B , \mathcal{P}_1 z_B   ,  g_1 \big( \mathcal{P}_1 , \mathcal{P}_1 \big( z_A + z_B \big) \big)              }_{U_A U^{\prime}_A U_B U^{\prime}_B S S^{\prime} } \\ \\  + \bigg[ \frac{1}{\sqrt{2}} \bigg]^3     \underset{z_A, z_B}{\underset{\alpha \neq \beta^{\prime} \in \textbf{F}^n_2}{\underset{L \in \textit{Mat}_{\textbf{F}^n_2} [ \textit{2} \times \textit{2}]}{\sum}}}  \bigg\{ \bigg\{  p_L             \bra{\mathcal{P}_1  z_A,  \mathcal{P}_1 z_A , \mathcal{P}_1  z_B,  \mathcal{P}_1 z_B ,  \mathcal{P}_1  z_A     , \mathcal{P}_1 z_B     }_{U_A U^{\prime}_A U_B U^{\prime}_B S S^{\prime} }   \end{align*}

\begin{align*}    \times \bigg\{    \bigg[        \ket{z_B, z_B + \alpha + \beta^{\prime}}_{B(B+A)}     \bigg[        \ket{z_A , z_A + \alpha} \bra{z_A , z_A + \alpha}    \bigg]_{AA}  \\ \times \bra{z_B , z_B  + \alpha + \beta^{\prime}}_{B(B+A)}               \times   \bra{F \big( \alpha \big) , F \big( \alpha + \beta^{\prime} \big)}_{S(S^{\prime} + T^{\prime})} \bigg\}  \\   \times     \ket{z_B + z_A} \bra{z_A}_{(B+A)A} \bigg\} \ket{z_B+ z_A} \bra{z_B}_{(B+A)B}  \bigg\}  \\ \times  \ket{\mathcal{P}_1  z_A,  \mathcal{P}_1 z_A , \mathcal{P}_1  z_B,  \mathcal{P}_1 z_B , \mathcal{P}_1 z_B   ,  g_1 \big( \mathcal{P}_1 , \mathcal{P}_1 \big( z_A + z_B \big) \big)              }_{U_A U^{\prime}_A U_B U^{\prime}_B S S^{\prime} }    \\ \\ + \bigg[ \frac{1}{\sqrt{2}} \bigg]^3     \underset{z_A, z_B}{\underset{\alpha \neq \beta^{\prime} \in \textbf{F}^n_2}{\underset{L \in \textit{Mat}_{\textbf{F}^n_2} [ \textit{2} \times \textit{2}]}{\sum}}}  \bigg\{ \bigg\{  p_L             \bra{\mathcal{P}_1  z_A,  \mathcal{P}_1 z_A , \mathcal{P}_1  z_B,  \mathcal{P}_1 z_B ,  \mathcal{P}_1  z_A     , \mathcal{P}_1 z_B     }_{U_A U^{\prime}_A U_B U^{\prime}_B S S^{\prime} }  \\  \times \bigg\{    \bigg[        \ket{z_B, z_B + \alpha + \beta^{\prime}}_{B(B+A)}     \bigg[        \ket{z_A , z_A + \alpha} \bra{z_A , z_A + \alpha}    \bigg]_{AA}  \\ \times \bra{z_B , z_B  + \alpha + \beta^{\prime}}_{B(B+A)}               \times   \bra{F \big( \alpha \big) , F \big( \alpha + \beta^{\prime} \big)}_{S(S^{\prime} + T^{\prime})} \bigg\}    \end{align*}

 \begin{align*}   \times     \ket{z_B + z_A} \bra{z_A}_{(B+A)A} \bigg\} \ket{z_B+ z_A} \bra{z_A}_{(B+A)B}  \bigg\}  \\ \times  \ket{\mathcal{P}_1  z_A,  \mathcal{P}_1 z_A , \mathcal{P}_1  z_B,  \mathcal{P}_1 z_B , \mathcal{P}_1 z_B   ,  g_1 \big( \mathcal{P}_1 , \mathcal{P}_1 \big( z_A + z_B \big) \big)              }_{U_A U^{\prime}_A U_B U^{\prime}_B S S^{\prime} }   \\ \\    \bigg[ \frac{1}{\sqrt{2}} \bigg]^3     \underset{z_A, z_B}{\underset{\alpha \neq \beta^{\prime} \in \textbf{F}^n_2}{\underset{L \in \textit{Mat}_{\textbf{F}^n_2} [ \textit{2} \times \textit{2}]}{\sum}}}  \bigg\{ \bigg\{  p_L             \bra{\mathcal{P}_1  z_A,  \mathcal{P}_1 z_A , \mathcal{P}_1  z_B,  \mathcal{P}_1 z_B ,  \mathcal{P}_1  z_A     , \mathcal{P}_1 z_B     }_{U_A U^{\prime}_A U_B U^{\prime}_B S S^{\prime} }   \\  \times \bigg\{    \bigg[        \ket{z_B, z_B + \alpha + \beta^{\prime}}_{B(B+A)}     \bigg[        \ket{z_A , z_A + \alpha} \bra{z_A , z_A + \alpha}    \bigg]_{AA}  \\ \times \bra{z_B , z_B  + \alpha + \beta^{\prime}}_{B(B+A)}               \times   \bra{F \big( \alpha \big) , F \big( \alpha + \beta^{\prime} \big)}_{S(S^{\prime} + T^{\prime})} \bigg\}  \\ \times     \ket{z_B + z_A} \bra{z_B}_{(B+A)B} \bigg\}      \ket{z_B + z_A} \bra{z_A}_{(B+A)B}       \bigg\}       \\ \times    \ket{\mathcal{P}_1  z_A,  \mathcal{P}_1 z_A , \mathcal{P}_1  z_B,  \mathcal{P}_1 z_B , \mathcal{P}_1 z_B   ,  g_1 \big( \mathcal{P}_1 , \mathcal{P}_1 \big( z_A + z_B \big) \big)              }_{U_A U^{\prime}_A U_B U^{\prime}_B S S^{\prime} }       , \\   \tag{$2^{\prime}$}
 \end{align*}}

\noindent corresponding to the second term,

{\small \begin{align*}
 \mathscr{P}_1 \mathscr{P}^{\prime}_2   \mathscr{P}^{\prime\prime}_1 \end{align*}

\begin{align*}  = \mathcal{O}_9 + \mathcal{O}_{10} + \mathcal{O}_{11}   + \mathcal{O}_{12}    \end{align*}

\begin{align*}  = \bigg[ \frac{1}{\sqrt{2}} \bigg]^3  \underset{z_A, z_B}{\underset{\alpha \neq \beta^{\prime} \in \textbf{F}^n_2}{\underset{L \in \textit{Mat}_{\textbf{F}^n_2} [ \textit{2} \times \textit{2}]}{\sum}}}  \bigg\{ \bigg\{  p_L     \bra{\mathcal{P}_1  z_A,  \mathcal{P}_1 z_A , \mathcal{P}_1  z_B,  \mathcal{P}_1 z_B , \mathcal{P}_1 z_B   ,  \mathcal{P}_1  z_A }_{U_A U^{\prime}_A U_B U^{\prime}_B S S^{\prime} }   \\ \times \bigg\{         \bigg[     \ket{z_B, z_B + \alpha + \beta^{\prime}}_{B(B+A)}     \bigg[    \ket{z_A , z_A + \alpha} \bra{z_A , z_A + \alpha}    \bigg]_{AA}  \\ \times \bra{z_B , z_B  + \alpha + \beta^{\prime}}_{B(B+A)}               \times     \ket{F \big( \alpha \big) , F \big( \alpha + \beta^{\prime} \big)}^{\dagger}_{S(S^{\prime} + T^{\prime})} \bigg\}   \\ \times              \ket{z_B + z_A} \bra{z_B}_{(B+A)B}        \bigg\}    \ket{z_B + z_A} \bra{z_B}_{(B+A)B} \bigg\} \\ \times            \ket{\mathcal{P}_1  z_A,  \mathcal{P}_1 z_A , \mathcal{P}_1  z_B,  \mathcal{P}_1 z_B ,  g_1 \big( \mathcal{P}_1 , \mathcal{P}_1 \big( z_A + z_B \big)  \big)    , \mathcal{P}_1 z_B     }_{U_A U^{\prime}_A U_B U^{\prime}_B S S^{\prime} }                  \\     \\  + \bigg[ \frac{1}{\sqrt{2}} \bigg]^3  \underset{z_A, z_B}{\underset{\alpha \neq \beta^{\prime} \in \textbf{F}^n_2}{\underset{L \in \textit{Mat}_{\textbf{F}^n_2} [ \textit{2} \times \textit{2}]}{\sum}}}  \bigg\{ \bigg\{  p_L     \bra{\mathcal{P}_1  z_A,  \mathcal{P}_1 z_A , \mathcal{P}_1  z_B,  \mathcal{P}_1 z_B , \mathcal{P}_1 z_B   ,  \mathcal{P}_1  z_A }_{U_A U^{\prime}_A U_B U^{\prime}_B S S^{\prime} }   \end{align*}

\begin{align*}     \times \bigg\{         \bigg[     \ket{z_B, z_B + \alpha + \beta^{\prime}}_{B(B+A)}     \bigg[    \ket{z_A , z_A + \alpha} \bra{z_A , z_A + \alpha}    \bigg]_{AA}  \\ \times \bra{z_B , z_B  + \alpha + \beta^{\prime}}_{B(B+A)}               \times     \ket{F \big( \alpha \big) , F \big( \alpha + \beta^{\prime} \big)}^{\dagger}_{S(S^{\prime} + T^{\prime})} \bigg\}   \\ \times   \ket{z_B + z_A} \bra{z_B}_{(B+A)B}        \bigg\}    \ket{z_B + z_A} \bra{z_A}_{(B+A)B} \bigg\} \\ \times            \ket{\mathcal{P}_1  z_A,  \mathcal{P}_1 z_A , \mathcal{P}_1  z_B,  \mathcal{P}_1 z_B ,  g_1 \big( \mathcal{P}_1 , \mathcal{P}_1 \big( z_A + z_B \big)  \big)    , \mathcal{P}_1 z_B     }_{U_A U^{\prime}_A U_B U^{\prime}_B S S^{\prime} }                       \\   + \bigg[ \frac{1}{\sqrt{2}} \bigg]^3  \underset{z_A, z_B}{\underset{\alpha \neq \beta^{\prime} \in \textbf{F}^n_2}{\underset{L \in \textit{Mat}_{\textbf{F}^n_2} [ \textit{2} \times \textit{2}]}{\sum}}}  \bigg\{ \bigg\{  p_L     \bra{\mathcal{P}_1  z_A,  \mathcal{P}_1 z_A , \mathcal{P}_1  z_B,  \mathcal{P}_1 z_B , \mathcal{P}_1 z_B   ,  \mathcal{P}_1  z_A }_{U_A U^{\prime}_A U_B U^{\prime}_B S S^{\prime} }  \\   \times \bigg\{         \bigg[     \ket{z_B, z_B + \alpha + \beta^{\prime}}_{B(B+A)}     \bigg[    \ket{z_A , z_A + \alpha} \bra{z_A , z_A + \alpha}    \bigg]_{AA}  \\ \times \bra{z_B , z_B  + \alpha + \beta^{\prime}}_{B(B+A)}               \times     \ket{F \big( \alpha \big) , F \big( \alpha + \beta^{\prime} \big)}^{\dagger}_{S(S^{\prime} + T^{\prime})} \bigg\}   \\ \times  \ket{z_B + z_A} \bra{z_A}_{(B+A)B}        \bigg\}    \ket{z_B + z_A} \bra{z_A}_{(B+A)B} \bigg\} \\ \times            \ket{\mathcal{P}_1  z_A,  \mathcal{P}_1 z_A , \mathcal{P}_1  z_B,  \mathcal{P}_1 z_B ,  g_1 \big( \mathcal{P}_1 , \mathcal{P}_1 \big( z_A + z_B \big)  \big)    , \mathcal{P}_1 z_B     }_{U_A U^{\prime}_A U_B U^{\prime}_B S S^{\prime} }                      \\ \\ + \bigg[ \frac{1}{\sqrt{2}} \bigg]^3  \underset{z_A, z_B}{\underset{\alpha \neq \beta^{\prime} \in \textbf{F}^n_2}{\underset{L \in \textit{Mat}_{\textbf{F}^n_2} [ \textit{2} \times \textit{2}]}{\sum}}}  \bigg\{ \bigg\{  p_L     \bra{\mathcal{P}_1  z_A,  \mathcal{P}_1 z_A , \mathcal{P}_1  z_B,  \mathcal{P}_1 z_B , \mathcal{P}_1 z_B   ,  \mathcal{P}_1  z_A }_{U_A U^{\prime}_A U_B U^{\prime}_B S S^{\prime} }  \end{align*}

 \begin{align*}  \times \bigg\{         \bigg[     \ket{z_B, z_B + \alpha + \beta^{\prime}}_{B(B+A)}     \bigg[    \ket{z_A , z_A + \alpha} \bra{z_A , z_A + \alpha}    \bigg]_{AA}  \\ \times \bra{z_B , z_B  + \alpha + \beta^{\prime}}_{B(B+A)}               \times     \ket{F \big( \alpha \big) , F \big( \alpha + \beta^{\prime} \big)}^{\dagger}_{S(S^{\prime} + T^{\prime})} \bigg\}   \\ \times         \ket{z_B + z_A} \bra{z_B}_{(B+A)B}        \bigg\}    \ket{z_B + z_A} \bra{z_A}_{(B+A)B} \bigg\} \\ \times            \ket{\mathcal{P}_1  z_A,  \mathcal{P}_1 z_A , \mathcal{P}_1  z_B,  \mathcal{P}_1 z_B ,  g_1 \big( \mathcal{P}_1 , \mathcal{P}_1 \big( z_A + z_B \big)  \big)    , \mathcal{P}_1 z_B     }_{U_A U^{\prime}_A U_B U^{\prime}_B S S^{\prime} }                     , \\   \tag{$3^{\prime}$}
 \end{align*} }

\noindent corresponding to the third term,

\begin{align*}
      \mathscr{P}_1  \mathscr{P}^{\prime}_2    \mathscr{P}^{\prime\prime}_2   \end{align*}

\begin{align*} = \mathcal{O}_{13} + \mathcal{O}_{14}  + \mathcal{O}_{15}   + \mathcal{O}_{16}    \end{align*}

\begin{align*}  = \bigg[ \frac{1}{\sqrt{2}} \bigg]^3    \underset{z_A, z_B}{\underset{\alpha \neq \beta^{\prime} \in \textbf{F}^n_2}{\underset{L \in \textit{Mat}_{\textbf{F}^n_2} [ \textit{2} \times \textit{2}]}{\sum}}}    \bigg\{ \bigg\{ p_L        \bra{\mathcal{P}_1  z_A,  \mathcal{P}_1 z_A , \mathcal{P}_1  z_B,  \mathcal{P}_1 z_B , \mathcal{P}_1 z_B   ,  \mathcal{P}_1  z_A }_{U_A U^{\prime}_A U_B U^{\prime}_B S S^{\prime} }   \\ \times \bigg\{ \bigg[      \ket{z_B, z_B + \alpha + \beta^{\prime}}_{B(B+A)}     \bigg[     \ket{z_A , z_A + \alpha} \bra{z_A , z_A + \alpha}    \bigg]_{AA}  \\ \times \bra{z_B , z_B  + \alpha + \beta^{\prime}}_{B(B+A)}               \times    \bra{F \big( \alpha \big) , F \big( \alpha + \beta^{\prime} \big)}_{S(S^{\prime} + T^{\prime})} \bigg\}  \\ \times    \ket{z_B + z_A} \bra{z_B}_{(B+A)B}  \bigg\}  \ket{z_B + z_A} \bra{z_B}_{(B+A) B}   \bigg\}  \\ \times \ket{\mathcal{P}_1  z_A,  \mathcal{P}_1 z_A , \mathcal{P}_1  z_B,  \mathcal{P}_1 z_B , \mathcal{P}_1 z_B   ,  g_1 \big( \mathcal{P}_1 , \mathcal{P}_1 \big( z_A + z_B \big) \big)              }_{U_A U^{\prime}_A U_B U^{\prime}_B S S^{\prime} }       \\      \\ + \bigg[ \frac{1}{\sqrt{2}} \bigg]^3    \underset{z_A, z_B}{\underset{\alpha \neq \beta^{\prime} \in \textbf{F}^n_2}{\underset{L \in \textit{Mat}_{\textbf{F}^n_2} [ \textit{2} \times \textit{2}]}{\sum}}}    \bigg\{ \bigg\{ p_L        \bra{\mathcal{P}_1  z_A,  \mathcal{P}_1 z_A , \mathcal{P}_1  z_B,  \mathcal{P}_1 z_B , \mathcal{P}_1 z_B   ,  \mathcal{P}_1  z_A }_{U_A U^{\prime}_A U_B U^{\prime}_B S S^{\prime} }   \\ \times \bigg\{ \bigg[      \ket{z_B, z_B + \alpha + \beta^{\prime}}_{B(B+A)}     \bigg[     \ket{z_A , z_A + \alpha} \bra{z_A , z_A + \alpha}    \bigg]_{AA}  \\ \times \bra{z_B , z_B  + \alpha + \beta^{\prime}}_{B(B+A)}               \times    \bra{F \big( \alpha \big) , F \big( \alpha + \beta^{\prime} \big)}_{S(S^{\prime} + T^{\prime})} \bigg\}  \\ \times    \ket{z_B + z_A} \bra{z_A}_{(B+A)B}  \bigg\}  \ket{z_B + z_A} \bra{z_B}_{(B+A) B}   \bigg\}  \\ \times \ket{\mathcal{P}_1  z_A,  \mathcal{P}_1 z_A , \mathcal{P}_1  z_B,  \mathcal{P}_1 z_B , \mathcal{P}_1 z_B   ,  g_1 \big( \mathcal{P}_1 , \mathcal{P}_1 \big( z_A + z_B \big) \big)              }_{U_A U^{\prime}_A U_B U^{\prime}_B S S^{\prime} }  \end{align*}

 \begin{align*}  + \bigg[ \frac{1}{\sqrt{2}} \bigg]^3    \underset{z_A, z_B}{\underset{\alpha \neq \beta^{\prime} \in \textbf{F}^n_2}{\underset{L \in \textit{Mat}_{\textbf{F}^n_2} [ \textit{2} \times \textit{2}]}{\sum}}}    \bigg\{ \bigg\{ p_L        \bra{\mathcal{P}_1  z_A,  \mathcal{P}_1 z_A , \mathcal{P}_1  z_B,  \mathcal{P}_1 z_B , \mathcal{P}_1 z_B   ,  \mathcal{P}_1  z_A }_{U_A U^{\prime}_A U_B U^{\prime}_B S S^{\prime} }   \\ \times \bigg\{ \bigg[      \ket{z_B, z_B + \alpha + \beta^{\prime}}_{B(B+A)}     \bigg[     \ket{z_A , z_A + \alpha} \bra{z_A , z_A + \alpha}    \bigg]_{AA}  \\ \times \bra{z_B , z_B  + \alpha + \beta^{\prime}}_{B(B+A)}               \times    \bra{F \big( \alpha \big) , F \big( \alpha + \beta^{\prime} \big)}_{S(S^{\prime} + T^{\prime})} \bigg\}  \\ \times    \ket{z_B + z_A} \bra{z_B}_{(B+A)B}  \bigg\}  \ket{z_B + z_A} \bra{z_A}_{(B+A) B}   \bigg\} \\   \times \ket{\mathcal{P}_1  z_A,  \mathcal{P}_1 z_A , \mathcal{P}_1  z_B,  \mathcal{P}_1 z_B , \mathcal{P}_1 z_B   ,  g_1 \big( \mathcal{P}_1 , \mathcal{P}_1 \big( z_A + z_B \big) \big)              }_{U_A U^{\prime}_A U_B U^{\prime}_B S S^{\prime} }   \end{align*}

 \begin{align*}   + \bigg[ \frac{1}{\sqrt{2}} \bigg]^3    \underset{z_A, z_B}{\underset{\alpha \neq \beta^{\prime} \in \textbf{F}^n_2}{\underset{L \in \textit{Mat}_{\textbf{F}^n_2} [ \textit{2} \times \textit{2}]}{\sum}}}    \bigg\{ \bigg\{ p_L        \bra{\mathcal{P}_1  z_A,  \mathcal{P}_1 z_A , \mathcal{P}_1  z_B,  \mathcal{P}_1 z_B , \mathcal{P}_1 z_B   ,  \mathcal{P}_1  z_A }_{U_A U^{\prime}_A U_B U^{\prime}_B S S^{\prime} }   \\ \times \bigg\{ \bigg[      \ket{z_B, z_B + \alpha + \beta^{\prime}}_{B(B+A)}     \bigg[     \ket{z_A , z_A + \alpha} \bra{z_A , z_A + \alpha}    \bigg]_{AA}  \\ \times \bra{z_B , z_B  + \alpha + \beta^{\prime}}_{B(B+A)}               \times    \bra{F \big( \alpha \big) , F \big( \alpha + \beta^{\prime} \big)}_{S(S^{\prime} + T^{\prime})} \bigg\}  \\ \times    \ket{z_B + z_A} \bra{z_A}_{(B+A)B}  \bigg\}  \ket{z_B + z_A} \bra{z_A}_{(B+A) B}   \bigg\}  \\ \times \ket{\mathcal{P}_1  z_A,  \mathcal{P}_1 z_A , \mathcal{P}_1  z_B,  \mathcal{P}_1 z_B , \mathcal{P}_1 z_B   ,  g_1 \big( \mathcal{P}_1 , \mathcal{P}_1 \big( z_A + z_B \big) \big)              }_{U_A U^{\prime}_A U_B U^{\prime}_B S S^{\prime} }    , \\   \tag{$4^{\prime}$}
 \end{align*}

\noindent corresponding to the fourth term,

{\small \begin{align*}
 \mathscr{P}_2 \mathscr{P}^{\prime}_1 \mathscr{P}^{\prime\prime}_2 \end{align*}

\begin{align*}  = \mathcal{O}_{17} + \mathcal{O}_{18}  + \mathcal{O}_{19}   + \mathcal{O}_{20}   \end{align*}

\begin{align*}  =  \bigg[ \frac{1}{\sqrt{2}} \bigg]^3    \underset{z_A, z_B}{\underset{\alpha \neq \beta^{\prime} \in \textbf{F}^n_2}{\underset{L \in \textit{Mat}_{\textbf{F}^n_2} [ \textit{2} \times \textit{2}]}{\sum}}}    \bigg\{ \bigg\{ p_L   \bra{\mathcal{P}_1  z_A,  \mathcal{P}_1 z_A , \mathcal{P}_1  z_B,  \mathcal{P}_1 z_B ,  \mathcal{P}_1  z_A     , \mathcal{P}_1 z_B     }_{U_A U^{\prime}_A U_B U^{\prime}_B S S^{\prime} }  \\ \times \bigg\{      \bigg[      \ket{z_B, z_B + \alpha + \beta^{\prime}}_{B(B+A)}     \bigg[      \ket{z_A , z_A + \alpha} \bra{z_A , z_A + \alpha}    \bigg]_{AA} \\ \times   \bra{z_B , z_B  + \alpha + \beta^{\prime}}_{B(B+A)}          \times \bra{F \big( \beta  \big) , F \big( \alpha + \beta^{\prime} \big)}_{T(S^{\prime} + T^{\prime})}      \bigg\}  \\ \times  \ket{z_B + z_A} \bra{z_B}_{(B+A)B}  \bigg\}     \ket{z_B + z_A} \bra{z_B}_{(B+A)B} \bra{z_B}_{(B+A)B}                                                                                                                                                                                      \bigg\}  \\   \times   \ket{\mathcal{P}_1  z_A,  \mathcal{P}_1 z_A , \mathcal{P}_1  z_B,  \mathcal{P}_1 z_B ,  g_1 \big( \mathcal{P}_1 , \mathcal{P}_1 \big( z_A + z_B \big)  \big)    , \mathcal{P}_1 z_B     }_{U_A U^{\prime}_A U_B U^{\prime}_B S S^{\prime} }        \\ \\ + \bigg[ \frac{1}{\sqrt{2}} \bigg]^3    \underset{z_A, z_B}{\underset{\alpha \neq \beta^{\prime} \in \textbf{F}^n_2}{\underset{L \in \textit{Mat}_{\textbf{F}^n_2} [ \textit{2} \times \textit{2}]}{\sum}}}    \bigg\{ \bigg\{ p_L  \bra{\mathcal{P}_1  z_A,  \mathcal{P}_1 z_A , \mathcal{P}_1  z_B,  \mathcal{P}_1 z_B ,  \mathcal{P}_1  z_A     , \mathcal{P}_1 z_B     }_{U_A U^{\prime}_A U_B U^{\prime}_B S S^{\prime} }  \\  \times \bigg\{      \bigg[      \ket{z_B, z_B + \alpha + \beta^{\prime}}_{B(B+A)}     \bigg[      \ket{z_A , z_A + \alpha} \bra{z_A , z_A + \alpha}    \bigg]_{AA}  \end{align*}

 \begin{align*}   \times \bra{z_B , z_B  + \alpha + \beta^{\prime}}_{B(B+A)}          \times  \bra{F \big( \beta  \big) , F \big( \alpha + \beta^{\prime} \big)}_{T(S^{\prime} + T^{\prime})}        \bigg\}   \\ \times  \ket{z_B + z_A} \bra{z_B}_{(B+A)B}  \bigg\}       \ket{z_B + z_A} \bra{z_B}_{(B+A)B} \bra{z_A}_{(B+A)B}                                                                                                                                                                                                                                                   \bigg\}  \\    \times    \ket{\mathcal{P}_1  z_A,  \mathcal{P}_1 z_A , \mathcal{P}_1  z_B,  \mathcal{P}_1 z_B ,  g_1 \big( \mathcal{P}_1 , \mathcal{P}_1 \big( z_A + z_B \big)  \big)    , \mathcal{P}_1 z_B     }_{U_A U^{\prime}_A U_B U^{\prime}_B S S^{\prime} }     \\ \\ + \bigg[ \frac{1}{\sqrt{2}} \bigg]^3    \underset{z_A, z_B}{\underset{\alpha \neq \beta^{\prime} \in \textbf{F}^n_2}{\underset{L \in \textit{Mat}_{\textbf{F}^n_2} [ \textit{2} \times \textit{2}]}{\sum}}}    \bigg\{ \bigg\{ p_L  \bra{\mathcal{P}_1  z_A,  \mathcal{P}_1 z_A , \mathcal{P}_1  z_B,  \mathcal{P}_1 z_B ,  \mathcal{P}_1  z_A     , \mathcal{P}_1 z_B     }_{U_A U^{\prime}_A U_B U^{\prime}_B S S^{\prime} }    \\ \times \bigg\{      \bigg[      \ket{z_B, z_B + \alpha + \beta^{\prime}}_{B(B+A)}     \bigg[      \ket{z_A , z_A + \alpha} \bra{z_A , z_A + \alpha}    \bigg]_{AA}  \end{align*}

 \begin{align*}   \times   \bra{z_B , z_B  + \alpha + \beta^{\prime}}_{B(B+A)}          \times \bra{F \big( \beta  \big) , F \big( \alpha + \beta^{\prime} \big)}_{T(S^{\prime} + T^{\prime})}      \bigg\}   \\   \times  \ket{z_B + z_A} \bra{z_A}_{(B+A)B}  \bigg\}       \ket{z_B + z_A} \bra{z_B}_{(B+A)B}                                                                \bigg\}    \\   \times    \ket{\mathcal{P}_1  z_A,  \mathcal{P}_1 z_A , \mathcal{P}_1  z_B,  \mathcal{P}_1 z_B ,  g_1 \big( \mathcal{P}_1 , \mathcal{P}_1 \big( z_A + z_B \big)  \big)    , \mathcal{P}_1 z_B     }_{U_A U^{\prime}_A U_B U^{\prime}_B S S^{\prime} }      \\ \\ + \bigg[ \frac{1}{\sqrt{2}} \bigg]^3    \underset{z_A, z_B}{\underset{\alpha \neq \beta^{\prime} \in \textbf{F}^n_2}{\underset{L \in \textit{Mat}_{\textbf{F}^n_2} [ \textit{2} \times \textit{2}]}{\sum}}}    \bigg\{ \bigg\{ p_L    \bra{\mathcal{P}_1  z_A,  \mathcal{P}_1 z_A , \mathcal{P}_1  z_B,  \mathcal{P}_1 z_B ,  \mathcal{P}_1  z_A     , \mathcal{P}_1 z_B     }_{U_A U^{\prime}_A U_B U^{\prime}_B S S^{\prime} }   \\ \times \bigg\{      \bigg[      \ket{z_B, z_B + \alpha + \beta^{\prime}}_{B(B+A)}     \bigg[      \ket{z_A , z_A + \alpha} \bra{z_A , z_A + \alpha}    \bigg]_{AA} \\ \times   \bra{z_B , z_B  + \alpha + \beta^{\prime}}_{B(B+A)}          \times \bra{F \big( \beta  \big) , F \big( \alpha + \beta^{\prime} \big)}_{T(S^{\prime} + T^{\prime})}      \bigg\} \\ \times        \ket{z_B + z_A} \bra{z_A}_{(B+A)B} \bigg\}        \ket{z_B + z_A} \bra{z_A}_{(B+A)B}                                                                                    \bigg\} \\ \times   \ket{\mathcal{P}_1  z_A,  \mathcal{P}_1 z_A , \mathcal{P}_1  z_B,  \mathcal{P}_1 z_B ,  g_1 \big( \mathcal{P}_1 , \mathcal{P}_1 \big( z_A + z_B \big)  \big)    , \mathcal{P}_1 z_B     }_{U_A U^{\prime}_A U_B U^{\prime}_B S S^{\prime} }      , \\   \tag{$5^{\prime}$}
 \end{align*} }

\noindent corresponding to the fifth term,

{\small \begin{align*}
 \mathscr{P}_2 \mathscr{P}^{\prime}_1 \mathscr{P}^{\prime\prime}_2 \end{align*}

\begin{align*}  = \mathcal{O}_{21} + \mathcal{O}_{22}  + \mathcal{O}_{23}   + \mathcal{O}_{24}    \end{align*}

\begin{align*}  =  \bigg[ \frac{1}{\sqrt{2}} \bigg]^3    \underset{z_A, z_B}{\underset{\alpha \neq \beta^{\prime} \in \textbf{F}^n_2}{\underset{L \in \textit{Mat}_{\textbf{F}^n_2} [ \textit{2} \times \textit{2}]}{\sum}}}    \bigg\{ \bigg\{ p_L                    \bra{\mathcal{P}_1  z_A,  \mathcal{P}_1 z_A , \mathcal{P}_1  z_B,  \mathcal{P}_1 z_B ,  \mathcal{P}_1  z_A     , \mathcal{P}_1 z_B     }_{U_A U^{\prime}_A U_B U^{\prime}_B S S^{\prime} }    \\ \times   \bigg\{    \bigg[         \ket{z_B, z_B + \alpha + \beta^{\prime}}_{B(B+A)}     \bigg[        \ket{z_A , z_A + \alpha} \bra{z_A , z_A + \alpha}    \bigg]_{AA}  \\ \times \bra{z_B , z_B  + \alpha + \beta^{\prime}}_{B(B+A)}          \times \bra{F \big( \beta  \big) , F \big( \alpha + \beta^{\prime} \big)}_{T(S^{\prime} + T^{\prime})}        \bigg\}                                     \\ \times   \ket{z_B + z_A} \bra{z_B}_{(B+A)B}    \bigg\} \ket{z_B + z_A} \bra{z_B}_{(B+A)B}  \bigg\} \\ \times        \ket{\mathcal{P}_1  z_A,  \mathcal{P}_1 z_A , \mathcal{P}_1  z_B,  \mathcal{P}_1 z_B , \mathcal{P}_1 z_B   ,  g_1 \big( \mathcal{P}_1 , \mathcal{P}_1 \big( z_A + z_B \big) \big)              }_{U_A U^{\prime}_A U_B U^{\prime}_B S S^{\prime} }           \end{align*}

\begin{align*}  + \bigg[ \frac{1}{\sqrt{2}} \bigg]^3    \underset{z_A, z_B}{\underset{\alpha \neq \beta^{\prime} \in \textbf{F}^n_2}{\underset{L \in \textit{Mat}_{\textbf{F}^n_2} [ \textit{2} \times \textit{2}]}{\sum}}}    \bigg\{ \bigg\{ p_L    \bra{\mathcal{P}_1  z_A,  \mathcal{P}_1 z_A , \mathcal{P}_1  z_B,  \mathcal{P}_1 z_B ,  \mathcal{P}_1  z_A     , \mathcal{P}_1 z_B     }_{U_A U^{\prime}_A U_B U^{\prime}_B S S^{\prime} }     \\ \times              \bigg\{                 \bigg[         \ket{z_B, z_B + \alpha + \beta^{\prime}}_{B(B+A)}     \bigg[        \ket{z_A , z_A + \alpha} \bra{z_A , z_A + \alpha}    \bigg]_{AA}  \\ \times \bra{z_B , z_B  + \alpha + \beta^{\prime}}_{B(B+A)}          \times  \bra{F \big( \beta  \big) , F \big( \alpha + \beta^{\prime} \big)}_{T(S^{\prime} + T^{\prime})}        \bigg\}                         \\  \times      \ket{z_B + z_A} \bra{z_B}_{(B+A)B}     \bigg\}  \ket{z_B + z_A} \bra{z_A}_{(B+A)A}     \bigg\} \\ \times     \ket{\mathcal{P}_1  z_A,  \mathcal{P}_1 z_A , \mathcal{P}_1  z_B,  \mathcal{P}_1 z_B , \mathcal{P}_1 z_B   ,  g_1 \big( \mathcal{P}_1 , \mathcal{P}_1 \big( z_A + z_B \big) \big)              }_{U_A U^{\prime}_A U_B U^{\prime}_B S S^{\prime} }            \\ \\ + \bigg[ \frac{1}{\sqrt{2}} \bigg]^3    \underset{z_A, z_B}{\underset{\alpha \neq \beta^{\prime} \in \textbf{F}^n_2}{\underset{L \in \textit{Mat}_{\textbf{F}^n_2} [ \textit{2} \times \textit{2}]}{\sum}}}    \bigg\{ \bigg\{ p_L    \bra{\mathcal{P}_1  z_A,  \mathcal{P}_1 z_A , \mathcal{P}_1  z_B,  \mathcal{P}_1 z_B ,  \mathcal{P}_1  z_A     , \mathcal{P}_1 z_B     }_{U_A U^{\prime}_A U_B U^{\prime}_B S S^{\prime} }     \\ \times     \bigg\{  \bigg[         \ket{z_B, z_B + \alpha + \beta^{\prime}}_{B(B+A)}     \bigg[        \ket{z_A , z_A + \alpha} \bra{z_A , z_A + \alpha}    \bigg]_{AA}  \\ \times  \bra{z_B , z_B  + \alpha + \beta^{\prime}}_{B(B+A)}          \times  \bra{F \big( \beta  \big) , F \big( \alpha + \beta^{\prime} \big)}_{T(S^{\prime} + T^{\prime})}        \bigg\}                                  \\ \times      \ket{z_B + z_A} \bra{z_A}_{(B+A)A}    \bigg\}  \ket{z_B + z_A} \bra{z_B}_{(B+A)A}    \bigg\}       \\   \times       \ket{\mathcal{P}_1  z_A,  \mathcal{P}_1 z_A , \mathcal{P}_1  z_B,  \mathcal{P}_1 z_B , \mathcal{P}_1 z_B   ,  g_1 \big( \mathcal{P}_1 , \mathcal{P}_1 \big( z_A + z_B \big) \big)              }_{U_A U^{\prime}_A U_B U^{\prime}_B S S^{\prime} }              \\ \\   + \bigg[ \frac{1}{\sqrt{2}} \bigg]^3    \underset{z_A, z_B}{\underset{\alpha \neq \beta^{\prime} \in \textbf{F}^n_2}{\underset{L \in \textit{Mat}_{\textbf{F}^n_2} [ \textit{2} \times \textit{2}]}{\sum}}}    \bigg\{ \bigg\{ p_L  \bra{\mathcal{P}_1  z_A,  \mathcal{P}_1 z_A , \mathcal{P}_1  z_B,  \mathcal{P}_1 z_B ,  \mathcal{P}_1  z_A     , \mathcal{P}_1 z_B     }_{U_A U^{\prime}_A U_B U^{\prime}_B S S^{\prime} }       \end{align*}

 \begin{align*}   \times     \bigg\{  \bigg[         \ket{z_B, z_B + \alpha + \beta^{\prime}}_{B(B+A)}     \bigg[        \ket{z_A , z_A + \alpha} \bra{z_A , z_A + \alpha}    \bigg]_{AA}  \\ \times \bra{z_B , z_B  + \alpha + \beta^{\prime}}_{B(B+A)}          \times  \bra{F \big( \beta  \big) , F \big( \alpha + \beta^{\prime} \big)}_{T(S^{\prime} + T^{\prime})}        \bigg\}                  \\ \times                  \ket{z_B + z_A} \bra{z_A}_{(B+A)A}    \bigg\}  \ket{z_B + z_A} \bra{z_A}_{(B+A)A}  \bigg\} \\ \times        \ket{\mathcal{P}_1  z_A,  \mathcal{P}_1 z_A , \mathcal{P}_1  z_B,  \mathcal{P}_1 z_B , \mathcal{P}_1 z_B   ,  g_1 \big( \mathcal{P}_1 , \mathcal{P}_1 \big( z_A + z_B \big) \big)              }_{U_A U^{\prime}_A U_B U^{\prime}_B S S^{\prime} }                                     \\  \tag{$6^{\prime}$}
 \end{align*}}

\noindent corresponding to the sixth term,

{\small \begin{align*}
 \mathscr{P}_2 \mathscr{P}^{\prime\prime}_2 \mathscr{P}^{\prime\prime}_1 \end{align*}

\begin{align*}  = \mathcal{O}_{25} + \mathcal{O}_{26}  + \mathcal{O}_{27}   + \mathcal{O}_{28}   \end{align*}

\begin{align*}  =  \bigg[ \frac{1}{\sqrt{2}} \bigg]^3    \underset{z_A, z_B}{\underset{\alpha \neq \beta^{\prime} \in \textbf{F}^n_2}{\underset{L \in \textit{Mat}_{\textbf{F}^n_2} [ \textit{2} \times \textit{2}]}{\sum}}}    \bigg\{ \bigg\{ p_L           \bra{\mathcal{P}_1  z_A,  \mathcal{P}_1 z_A , \mathcal{P}_1  z_B,  \mathcal{P}_1 z_B , \mathcal{P}_1 z_B   ,  \mathcal{P}_1  z_A }_{U_A U^{\prime}_A U_B U^{\prime}_B S S^{\prime} }    \\ \times      \bigg\{  \bigg[       \ket{z_B, z_B + \alpha + \beta^{\prime}}_{B(B+A)}     \bigg[       \ket{z_A , z_A + \alpha} \bra{z_A , z_A + \alpha}    \bigg]_{AA}  \\ \times \bra{z_B , z_B  + \alpha + \beta^{\prime}}_{B(B+A)}          \times  \bra{F \big( \beta  \big) , F \big( \alpha + \beta^{\prime} \big)}_{T(S^{\prime} + T^{\prime})}        \bigg\}   \\   \times   \ket{z_B + z_A} \bra{z_B}_{(B+A)A}  \bigg\} \ket{ z_B + z_A} \bra{z_B}_{(B+A)A}        \\    \times    \ket{\mathcal{P}_1  z_A,  \mathcal{P}_1 z_A , \mathcal{P}_1  z_B,  \mathcal{P}_1 z_B ,  g_1 \big( \mathcal{P}_1 , \mathcal{P}_1 \big( z_A + z_B \big)  \big)    , \mathcal{P}_1 z_B     }_{U_A U^{\prime}_A U_B U^{\prime}_B S S^{\prime} }                                                 \\ \\    + \bigg[ \frac{1}{\sqrt{2}} \bigg]^3    \underset{z_A, z_B}{\underset{\alpha \neq \beta^{\prime} \in \textbf{F}^n_2}{\underset{L \in \textit{Mat}_{\textbf{F}^n_2} [ \textit{2} \times \textit{2}]}{\sum}}}    \bigg\{ \bigg\{ p_L       \bra{\mathcal{P}_1  z_A,  \mathcal{P}_1 z_A , \mathcal{P}_1  z_B,  \mathcal{P}_1 z_B , \mathcal{P}_1 z_B   ,  \mathcal{P}_1  z_A }_{U_A U^{\prime}_A U_B U^{\prime}_B S S^{\prime} }     \\  \times      \bigg\{  \bigg[       \ket{z_B, z_B + \alpha + \beta^{\prime}}_{B(B+A)}     \bigg[       \ket{z_A , z_A + \alpha} \bra{z_A , z_A + \alpha}    \bigg]_{AA}  \\    \times \bra{z_B , z_B  + \alpha + \beta^{\prime}}_{B(B+A)}          \times  \bra{F \big( \beta  \big) , F \big( \alpha + \beta^{\prime} \big)}_{T(S^{\prime} + T^{\prime})}        \bigg\}   \\ \times   \ket{z_B + z_A} \bra{z_B}_{(B+A)A}  \bigg\} \ket{ z_B + z_A} \bra{z_A}_{(B+A)A} \\     \times    \ket{\mathcal{P}_1  z_A,  \mathcal{P}_1 z_A , \mathcal{P}_1  z_B,  \mathcal{P}_1 z_B ,  g_1 \big( \mathcal{P}_1 , \mathcal{P}_1 \big( z_A + z_B \big)  \big)    , \mathcal{P}_1 z_B     }_{U_A U^{\prime}_A U_B U^{\prime}_B S S^{\prime} }       \end{align*}

\begin{align*}    + \bigg[ \frac{1}{\sqrt{2}} \bigg]^3    \underset{z_A, z_B}{\underset{\alpha \neq \beta^{\prime} \in \textbf{F}^n_2}{\underset{L \in \textit{Mat}_{\textbf{F}^n_2} [ \textit{2} \times \textit{2}]}{\sum}}}    \bigg\{ \bigg\{ p_L       \bra{\mathcal{P}_1  z_A,  \mathcal{P}_1 z_A , \mathcal{P}_1  z_B,  \mathcal{P}_1 z_B , \mathcal{P}_1 z_B   ,  \mathcal{P}_1  z_A }_{U_A U^{\prime}_A U_B U^{\prime}_B S S^{\prime} }  \\ \times   \bigg\{    \bigg[       \ket{z_B, z_B + \alpha + \beta^{\prime}}_{B(B+A)}     \bigg[       \ket{z_A , z_A + \alpha} \bra{z_A , z_A + \alpha}    \bigg]_{AA}  \\ \times \bra{z_B , z_B  + \alpha + \beta^{\prime}}_{B(B+A)}          \times  \bra{F \big( \beta  \big) , F \big( \alpha + \beta^{\prime} \big)}_{T(S^{\prime} + T^{\prime})}        \bigg\}   \\ \times   \ket{z_B + z_A} \bra{z_A}_{(B+A)A}  \bigg\} \ket{ z_B + z_A} \bra{z_B}_{(B+A)A}       \end{align*}

 \begin{align*}        \times    \ket{\mathcal{P}_1  z_A,  \mathcal{P}_1 z_A , \mathcal{P}_1  z_B,  \mathcal{P}_1 z_B ,  g_1 \big( \mathcal{P}_1 , \mathcal{P}_1 \big( z_A + z_B \big)  \big)    , \mathcal{P}_1 z_B     }_{U_A U^{\prime}_A U_B U^{\prime}_B S S^{\prime} }       \\ \\ + \bigg[ \frac{1}{\sqrt{2}} \bigg]^3    \underset{z_A, z_B}{\underset{\alpha \neq \beta^{\prime} \in \textbf{F}^n_2}{\underset{L \in \textit{Mat}_{\textbf{F}^n_2} [ \textit{2} \times \textit{2}]}{\sum}}}    \bigg\{ \bigg\{ p_L        \bra{\mathcal{P}_1  z_A,  \mathcal{P}_1 z_A , \mathcal{P}_1  z_B,  \mathcal{P}_1 z_B , \mathcal{P}_1 z_B   ,  \mathcal{P}_1  z_A }_{U_A U^{\prime}_A U_B U^{\prime}_B S S^{\prime} }  \\  \times    \bigg\{    \bigg[       \ket{z_B, z_B + \alpha + \beta^{\prime}}_{B(B+A)}     \bigg[       \ket{z_A , z_A + \alpha} \bra{z_A , z_A + \alpha}    \bigg]_{AA}  \\ \times \bra{z_B , z_B  + \alpha + \beta^{\prime}}_{B(B+A)}          \times  \bra{F \big( \beta  \big) , F \big( \alpha + \beta^{\prime} \big)}_{T(S^{\prime} + T^{\prime})}        \bigg\}  \\ \times   \ket{z_B + z_A} \bra{z_A}_{(B+A)A}  \bigg\} \ket{ z_B + z_A} \bra{z_A}_{(B+A)A}         \\      \times    \ket{\mathcal{P}_1  z_A,  \mathcal{P}_1 z_A , \mathcal{P}_1  z_B,  \mathcal{P}_1 z_B ,  g_1 \big( \mathcal{P}_1 , \mathcal{P}_1 \big( z_A + z_B \big)  \big)    , \mathcal{P}_1 z_B     }_{U_A U^{\prime}_A U_B U^{\prime}_B S S^{\prime} }        , \\   \tag{$7^{\prime}$}
 \end{align*}}

\noindent corresponding to the seventh term, and,

{\small \begin{align*}
 \mathscr{P}_2 \mathscr{P}^{\prime\prime}_2 \mathscr{P}^{\prime\prime}_2 \end{align*}

\begin{align*}  = \mathcal{O}_{29} + \mathcal{O}_{30}  + \mathcal{O}_{31}   + \mathcal{O}_{32}   \end{align*}

\begin{align*}  =  \bigg[ \frac{1}{\sqrt{2}} \bigg]^3    \underset{z_A, z_B}{\underset{\alpha \neq \beta^{\prime} \in \textbf{F}^n_2}{\underset{L \in \textit{Mat}_{\textbf{F}^n_2} [ \textit{2} \times \textit{2}]}{\sum}}}    \bigg\{ \bigg\{ p_L \bra{\mathcal{P}_1  z_A,  \mathcal{P}_1 z_A , \mathcal{P}_1  z_B,  \mathcal{P}_1 z_B , \mathcal{P}_1 z_B   ,  \mathcal{P}_1  z_A }_{U_A U^{\prime}_A U_B U^{\prime}_B S S^{\prime} }   \\ \times   \bigg\{   \bigg[       \ket{z_B, z_B + \alpha + \beta^{\prime}}_{B(B+A)}     \bigg[         \ket{z_A , z_A + \alpha} \bra{z_A , z_A + \alpha}    \bigg]_{AA}     \end{align*}

\begin{align*}         \times \bra{z_B , z_B  + \alpha + \beta^{\prime}}_{B(B+A)}          \times  \bra{F \big( \beta  \big) , F \big( \alpha + \beta^{\prime} \big)}_{T(S^{\prime} + T^{\prime})}        \bigg\}           \\ \times   \ket{z_B + z_A} \bra{z_B}_{(B+A)B}    \bigg\}      \ket{z_B + z_A} \bra{z_B}_{(B+A)A}      \bigg\}  \\    \times   \ket{\mathcal{P}_1  z_A,  \mathcal{P}_1 z_A , \mathcal{P}_1  z_B,  \mathcal{P}_1 z_B , \mathcal{P}_1 z_B   ,  g_1 \big( \mathcal{P}_1 , \mathcal{P}_1 \big( z_A + z_B \big) \big)              }_{U_A U^{\prime}_A U_B U^{\prime}_B S S^{\prime} }    \end{align*}

 \begin{align*}   + \bigg[ \frac{1}{\sqrt{2}} \bigg]^3    \underset{z_A, z_B}{\underset{\alpha \neq \beta^{\prime} \in \textbf{F}^n_2}{\underset{L \in \textit{Mat}_{\textbf{F}^n_2} [ \textit{2} \times \textit{2}]}{\sum}}}    \bigg\{ \bigg\{ p_L  \bra{\mathcal{P}_1  z_A,  \mathcal{P}_1 z_A , \mathcal{P}_1  z_B,  \mathcal{P}_1 z_B , \mathcal{P}_1 z_B   ,  \mathcal{P}_1  z_A }_{U_A U^{\prime}_A U_B U^{\prime}_B S S^{\prime} }    \\ \times   \bigg\{   \bigg[       \ket{z_B, z_B + \alpha + \beta^{\prime}}_{B(B+A)}     \bigg[         \ket{z_A , z_A + \alpha} \bra{z_A , z_A + \alpha}    \bigg]_{AA}  \\ \times \bra{z_B , z_B  + \alpha + \beta^{\prime}}_{B(B+A)}          \times  \bra{F \big( \beta  \big) , F \big( \alpha + \beta^{\prime} \big)}_{T(S^{\prime} + T^{\prime})}        \bigg\}  \\       \times     \ket{z_B + z_A} \bra{z_B}_{(B+A)B}   \bigg\}     \ket{z_B + z_A} \bra{z_A}_{(B+A)A}       \bigg\}   \\   \times   \ket{\mathcal{P}_1  z_A,  \mathcal{P}_1 z_A , \mathcal{P}_1  z_B,  \mathcal{P}_1 z_B , \mathcal{P}_1 z_B   ,  g_1 \big( \mathcal{P}_1 , \mathcal{P}_1 \big( z_A + z_B \big) \big)              }_{U_A U^{\prime}_A U_B U^{\prime}_B S S^{\prime} }  \\ \\ + \bigg[ \frac{1}{\sqrt{2}} \bigg]^3    \underset{z_A, z_B}{\underset{\alpha \neq \beta^{\prime} \in \textbf{F}^n_2}{\underset{L \in \textit{Mat}_{\textbf{F}^n_2} [ \textit{2} \times \textit{2}]}{\sum}}}    \bigg\{ \bigg\{ p_L  \bra{\mathcal{P}_1  z_A,  \mathcal{P}_1 z_A , \mathcal{P}_1  z_B,  \mathcal{P}_1 z_B , \mathcal{P}_1 z_B   ,  \mathcal{P}_1  z_A }_{U_A U^{\prime}_A U_B U^{\prime}_B S S^{\prime} }   \\ \times   \bigg\{   \bigg[       \ket{z_B, z_B + \alpha + \beta^{\prime}}_{B(B+A)}     \bigg[         \ket{z_A , z_A + \alpha} \bra{z_A , z_A + \alpha}    \bigg]_{AA}    \\  \times \bra{z_B , z_B  + \alpha + \beta^{\prime}}_{B(B+A)}          \times \bra{F \big( \beta  \big) , F \big( \alpha + \beta^{\prime} \big)}_{T(S^{\prime} + T^{\prime})}        \bigg\}         \\  \times     \ket{z_B + z_A} \bra{z_A}_{(B+A)B}   \bigg\}       \ket{z_B + z_A} \bra{z_B}_{(B+A)A}     \bigg\}  \\  \times   \ket{\mathcal{P}_1  z_A,  \mathcal{P}_1 z_A , \mathcal{P}_1  z_B,  \mathcal{P}_1 z_B , \mathcal{P}_1 z_B   ,  g_1 \big( \mathcal{P}_1 , \mathcal{P}_1 \big( z_A + z_B \big) \big)              }_{U_A U^{\prime}_A U_B U^{\prime}_B S S^{\prime} }       \end{align*}

 \begin{align*}  + \bigg[ \frac{1}{\sqrt{2}} \bigg]^3    \underset{z_A, z_B}{\underset{\alpha \neq \beta^{\prime} \in \textbf{F}^n_2}{\underset{L \in \textit{Mat}_{\textbf{F}^n_2} [ \textit{2} \times \textit{2}]}{\sum}}}    \bigg\{ \bigg\{ p_L      \bra{\mathcal{P}_1  z_A,  \mathcal{P}_1 z_A , \mathcal{P}_1  z_B,  \mathcal{P}_1 z_B , \mathcal{P}_1 z_B   ,  \mathcal{P}_1  z_A }_{U_A U^{\prime}_A U_B U^{\prime}_B S S^{\prime} }   \\ \times   \bigg\{   \bigg[       \ket{z_B, z_B + \alpha + \beta^{\prime}}_{B(B+A)}     \bigg[         \ket{z_A , z_A + \alpha} \bra{z_A , z_A + \alpha}    \bigg]_{AA}   \\    \times \bra{z_B , z_B  + \alpha + \beta^{\prime}}_{B(B+A)}          \times  \bra{F \big( \beta  \big) , F \big( \alpha + \beta^{\prime} \big)}_{T(S^{\prime} + T^{\prime})}        \bigg\}          \end{align*}

\begin{align*}    \times    \ket{z_B + z_A} \bra{z_A}_{(B+A)B}    \bigg\}    \ket{z_B + z_A} \bra{z_A}_{(B+A)A}     \bigg\}  \\ \times   \ket{\mathcal{P}_1  z_A,  \mathcal{P}_1 z_A , \mathcal{P}_1  z_B,  \mathcal{P}_1 z_B , \mathcal{P}_1 z_B   ,  g_1 \big( \mathcal{P}_1 , \mathcal{P}_1 \big( z_A + z_B \big) \big)              }_{U_A U^{\prime}_A U_B U^{\prime}_B S S^{\prime} }  . \\   \tag{$8^{\prime}$} \\ 
 \end{align*} }

\noindent corresponding to the eighth term. Altogether, obtaining the desired power, $*$, specifically in the $2^{-*}$ term, appearing in the lower bound,

\begin{align*}
   \bra{\mathcal{L}} \bigg[ \underset{z_A, z_B}{\underset{\alpha \neq \beta^{\prime} \in \textbf{F}^n_2}{\underset{L \in \textit{Mat}_{\textbf{F}^n_2} [ \textit{2} \times \textit{2}]}{\sum}}}  \mathscr{U}^{\dagger}_{\textit{Ideal}}  \mathscr{U}^{\dagger}_{\textit{Simulator}}  \mathscr{U}^{\prime}_{\textit{Real}}      \bigg] \ket{\mathcal{L}}  \gtrsim          \big[ 2^{5-\frac{3}{2}} - 2^{-*} \big] \textbf{I}_{AB}         . 
\end{align*}

\noindent for braket states of the purified state for random matrices $L$,

\begin{align*}
   \bra{\mathcal{L}} \bigg[ \underset{z_A, z_B}{\underset{\alpha \neq \beta^{\prime} \in \textbf{F}^n_2}{\underset{L \in \textit{Mat}_{\textbf{F}^n_2} [ \textit{2} \times \textit{2}]}{\sum}}}  \mathscr{U}^{\dagger}_{\textit{Ideal}}  \mathscr{U}^{\dagger}_{\textit{Simulator}}  \mathscr{U}^{\prime}_{\textit{Real}}      \bigg] \ket{\mathcal{L}}  ,
\end{align*}

\noindent is obtained from the observation that, altogether,

{\small   \begin{align*}
          \underset{1 \leq k \leq 2}{\underset{1 \leq j \leq 2}{\underset{1\leq i \leq 2}{\sum}}} \mathscr{P}_i \mathscr{P}^{\prime}_j \mathscr{P}^{\prime\prime}_k        =  \underset{1 \leq i \leq 32}{\sum}   \mathcal{O}_i <    32 \bigg[     \frac{1}{\sqrt{2}}   \bigg]^3 \big\{ \underset{1 \leq i \leq 32}{\mathrm{sup}} \mathcal{O}_i   \big\}   = 2^{5-\frac{3}{2} }   \big\{   \underset{1 \leq i \leq 32}{\mathrm{sup}} \mathcal{O}_i \big\}   \end{align*}

          \begin{align*} =    2^{5-\frac{3}{2} }  \bigg\{    \underset{1 \leq i \leq 16}{\mathrm{sup}} \mathcal{O}_i   \bigg( \textbf{1}_{\{ \mathcal{O}_i \textit{ has an expectation value with support } \ket{z_B , z_B + \alpha + \beta^{\prime}} \bra{z_B, z_B + \alpha + \beta^{\prime}}_{(B+A) A} \}}  \bigg)        \\  +  \underset{1 \leq i \leq 16}{\mathrm{sup}} \mathcal{O}_i  \bigg(  \textbf{1}_{\{ \mathcal{O}_i \textit{ does not have an expectation value with support } \ket{z_B , z_B + \alpha + \beta^{\prime}} \bra{z_B, z_B + \alpha + \beta^{\prime}}_{(B+A) A}  \}}           \bigg)    \bigg\}  \end{align*}

          \begin{align*}    =       2^{5-\frac{3}{2} }  \bigg\{  2   \underset{\{ X^{\prime} = \mathcal{P}_1 z_A, Y^{\prime} = g_1 ( \mathcal{P}_1 , \mathcal{P}_1 ( z_A + z_B) ) \} }{\underset{\{ X = \mathcal{P}_1 z_A , Y = \mathcal{P}_1 z_B \} }{\underset{ \{ X^{\prime} = \mathcal{P}_1 z_B, Y^{\prime} = g_1 ( \mathcal{P}_1 , \mathcal{P}_1 ( z_A + z_B) ) \}  }{\underset{ \{ X = \mathcal{P}_1 z_B , Y = \mathcal{P}_1 z_A \}}{\underset{1 \leq i \leq 16}{\mathrm{sup}}}}}}   \bigg\{   \underset{z_A, z_B}{\underset{\alpha \neq \beta^{\prime} \in \textbf{F}^n_2}{\underset{L \in \textit{Mat}_{\textbf{F}^n_2} [ \textit{2} \times \textit{2}]}{\sum}}}    \bigg\{ \bigg\{ p_L    \bra{\mathcal{P}_1  z_A,  \mathcal{P}_1 z_A , \mathcal{P}_1  z_B,  \mathcal{P}_1 z_B , X  ,  Y }_{U_A U^{\prime}_A U_B U^{\prime}_B S S^{\prime} }        \\    \times   \bigg\{   \bigg[       \ket{z_B, z_B + \alpha + \beta^{\prime}}_{B(B+A)}     \bigg[         \ket{z_A , z_A + \alpha} \bra{z_A , z_A + \alpha}    \bigg]_{AA}  \\ \times \bra{z_B , z_B  + \alpha + \beta^{\prime}}_{B(B+A)}          \times  \bra{F \big( \beta  \big) , F \big( \alpha + \beta^{\prime} \big)}_{T(S^{\prime} + T^{\prime})}        \bigg\}             \end{align*}

\begin{align*}   \times    \ket{z_B + z_A} \bra{z_A}_{(B+A)B}    \bigg\}    \ket{z_B + z_A} \bra{z_A}_{(B+A)A}     \bigg\}  \\ \times   \ket{\mathcal{P}_1  z_A,  \mathcal{P}_1 z_A , \mathcal{P}_1  z_B,  \mathcal{P}_1 z_B , X^{\prime} , Y^{\prime}          }_{U_A U^{\prime}_A U_B U^{\prime}_B S S^{\prime} }        \bigg\}  \\ \times \big[    \textbf{1}_{\{ \mathcal{O}_i \textit{ has an expectation value with support } \ket{z_B , z_B + \alpha + \beta^{\prime}} \bra{z_B, z_B + \alpha + \beta^{\prime}}_{(B+A) A} \}}           \\  + \textbf{1}_{\{ \mathcal{O}_i \textit{ does not have an expectation value with support } \ket{z_B , z_B + \alpha + \beta^{\prime}} \bra{z_B, z_B + \alpha + \beta^{\prime}}_{(B+A) A}  \}}   \big]   \bigg\} \end{align*}

          \begin{align*}  =        2^{5-\frac{3}{2} }  \bigg\{  2   \underset{\{ X^{\prime} = \mathcal{P}_1 z_A, Y^{\prime} = g_1 ( \mathcal{P}_1 , \mathcal{P}_1 ( z_A + z_B) ) \} }{\underset{\{ X = \mathcal{P}_1 z_A , Y = \mathcal{P}_1 z_B \} }{\underset{ \{ X^{\prime} = \mathcal{P}_1 z_B, Y^{\prime} = g_1 ( \mathcal{P}_1 , \mathcal{P}_1 ( z_A + z_B) ) \}  }{\underset{ \{ X = \mathcal{P}_1 z_B , Y = \mathcal{P}_1 z_A \}}{\underset{1 \leq i \leq 16}{\mathrm{sup}}}}}}   \bigg\{        \bigg\{   {\underset{z_A, z_B}{\underset{\alpha \neq \beta^{\prime} \in \textbf{F}^n_2}{\sum}}}    \bigg\{ \bigg\{    \bra{\mathcal{P}_1  z_A,  \mathcal{P}_1 z_A , \mathcal{P}_1  z_B,  \mathcal{P}_1 z_B , X  ,  Y }_{U_A U^{\prime}_A U_B U^{\prime}_B S S^{\prime} }    \\ \times   \bigg\{   \bigg[       \ket{z_B, z_B + \alpha + \beta^{\prime}}_{B(B+A)}     \bigg[         \ket{z_A , z_A + \alpha} \bra{z_A , z_A + \alpha}    \bigg]_{AA}  \\ \times \bra{z_B , z_B  + \alpha + \beta^{\prime}}_{B(B+A)}             \bigg\}          \times    \ket{z_B + z_A} \bra{z_A}_{(B+A)B}    \bigg\}  \\ \times   \ket{z_B + z_A} \bra{z_A}_{(B+A)A}     \bigg\}  \\ \times  {\underset{L \in \textit{Mat}_{\textbf{F}^n_2} [ \textit{2} \times \textit{2}]}{\sum}}    \bra{F \big( \beta  \big) , F \big( \alpha + \beta^{\prime} \big)}_{T(S^{\prime} + T^{\prime})}        \ket{\mathcal{P}_1  z_A,  \mathcal{P}_1 z_A , \mathcal{P}_1  z_B,  \mathcal{P}_1 z_B , X^{\prime} , Y^{\prime}          }_{U_A U^{\prime}_A U_B U^{\prime}_B S S^{\prime} }                         \bigg\} \\   \times \big[    \textbf{1}_{\{ \mathcal{O}_i \textit{ has an expectation value with support } \ket{z_B , z_B + \alpha + \beta^{\prime}} \bra{z_B, z_B + \alpha + \beta^{\prime}}_{(B+A) A} \}}           \\  + \textbf{1}_{\{ \mathcal{O}_i \textit{ does not have an expectation value with support } \ket{z_B , z_B + \alpha + \beta^{\prime}} \bra{z_B, z_B + \alpha + \beta^{\prime}}_{(B+A) A}  \}}   \big]   \bigg\}       \\ \\ =          2^{5-\frac{3}{2} }  \bigg\{  2   \underset{\{ X^{\prime} = \mathcal{P}_1 z_A, Y^{\prime} = g_1 ( \mathcal{P}_1 , \mathcal{P}_1 ( z_A + z_B) ) \} }{\underset{\{ X = \mathcal{P}_1 z_A , Y = \mathcal{P}_1 z_B \} }{\underset{ \{ X^{\prime} = \mathcal{P}_1 z_B, Y^{\prime} = g_1 ( \mathcal{P}_1 , \mathcal{P}_1 ( z_A + z_B) ) \}  }{\underset{ \{ X = \mathcal{P}_1 z_B , Y = \mathcal{P}_1 z_A \}}{\underset{1 \leq i \leq 16}{\mathrm{sup}}}}}}   \bigg\{        \bigg\{   {\underset{z_A, z_B}{\underset{\alpha \neq \beta^{\prime} \in \textbf{F}^n_2}{\sum}}}    \bigg\{ \bigg\{    \bra{\mathcal{P}_1  z_A,  \mathcal{P}_1 z_A , \mathcal{P}_1  z_B,  \mathcal{P}_1 z_B , X  ,  Y }_{U_A U^{\prime}_A U_B U^{\prime}_B S S^{\prime} }  \\  \times   \bigg\{   \bigg[       \ket{z_B, z_B + \alpha + \beta^{\prime}}_{B(B+A)}     \bigg[         \ket{z_A , z_A + \alpha} \bra{z_A , z_A + \alpha}    \bigg]_{AA}  \\  \times \bra{z_B , z_B  + \alpha + \beta^{\prime}}_{B(B+A)}             \bigg\}          \times    \ket{z_B + z_A} \bra{z_A}_{(B+A)B}    \bigg\}  \\ \times   \ket{z_B + z_A} \bra{z_A}_{(B+A)A}     \bigg\}  \\   \times      \textbf{P}_L         \big[    F \big( \beta \big) = X^{\prime}  , F \big( \alpha + \beta^{\prime} \big) = Y^{\prime}    \big]           \bigg\} \\   \times \big[    \textbf{1}_{\{ \mathcal{O}_i \textit{ has an expectation value with support } \ket{z_B , z_B + \alpha + \beta^{\prime}} \bra{z_B, z_B + \alpha + \beta^{\prime}}_{(B+A) A} \}}           \end{align*}

          \begin{align*}   + \textbf{1}_{\{ \mathcal{O}_i \textit{ does not have an expectation value with support } \ket{z_B , z_B + \alpha + \beta^{\prime}} \bra{z_B, z_B + \alpha + \beta^{\prime}}_{(B+A) A}  \}}   \big]   \bigg\} , \\  \tag{\textit{Product Expectation}   }  \end{align*} }

\noindent which yields the desired, up to constants, lower bound, from the fact that,

{\tiny  \begin{align*}
   {\underset{ \{ X^{\prime} = \mathcal{P}_1 z_A, Y^{\prime} = g_1 ( \mathcal{P}_1 , \mathcal{P}_1 ( z_A + z_B) ) \} }{\underset{ \{ X^{\prime} = \mathcal{P}_1 z_B, Y^{\prime} = g_1 ( \mathcal{P}_1 , \mathcal{P}_1 ( z_A + z_B) ) \} }{\mathrm{sup}}}}   \textbf{P}_L         \big[    F \big( \beta \big) = X^{\prime}  , F \big( \alpha + \beta^{\prime} \big) = Y^{\prime}    \big]  =         \textbf{P}_L         \bigg[    {\underset{ \{ X^{\prime} = \mathcal{P}_1 z_A, Y^{\prime} = g_1 ( \mathcal{P}_1 , \mathcal{P}_1 ( z_A + z_B) ) \} }{\underset{ \{ X^{\prime} = \mathcal{P}_1 z_B, Y^{\prime} = g_1 ( \mathcal{P}_1 , \mathcal{P}_1 ( z_A + z_B) ) \} }{\mathrm{sup}}}}   \big\{   F \big( \beta \big) = X^{\prime} \\  , F \big( \alpha + \beta^{\prime} \big) = Y^{\prime}  \big\}   \bigg]    \\   \\ = 1 - \textbf{P}_L         \bigg[    {\underset{ \{ X^{\prime} = \mathcal{P}_1 z_A, Y^{\prime} = g_1 ( \mathcal{P}_1 , \mathcal{P}_1 ( z_A + z_B) ) \} }{\underset{ \{ X^{\prime} = \mathcal{P}_1 z_B, Y^{\prime} = g_1 ( \mathcal{P}_1 , \mathcal{P}_1 ( z_A + z_B) ) \} }{\mathrm{sup}}}}   \big\{   F \big( \beta \big) \neq  X^{\prime}  , F \big( \alpha + \beta^{\prime} \big) \neq  Y^{\prime}  \big\}   \bigg]   \\ \\ \overset{(\textbf{Corollary} \textit{1}), {\color{blue}[40]}}{>} 1 - 2^{-k + n h ( \frac{r}{n})}  ,
\end{align*}     }

\noindent implies that,

        {\small   \begin{align*}                     (\textit{Product Expectation})  \gtrsim   2^{5-\frac{3}{2}}   \big[ 1 -    2^{-k + n h ( \frac{r}{n} )}  \big] \textbf{I}_{AB}  =   \big[   2^{5-\frac{3}{2}} -     2^{-k + n h ( \frac{r}{n} ) + 5-\frac{3}{2} }  \big] \textbf{I}_{AB}       , 
\end{align*} }

\noindent from which we conclude the argument, as,

\begin{align*}
  * \equiv  2^{-k + n h ( \frac{r}{n} ) + 5-\frac{3}{2} }    .       \boxed{}
\end{align*}

\subsection{$\textbf{Proposition}$ \textit{3}}

\subsubsection{Description of Description of the purified random matrix state from the Ideal, Simulator and Real $\mathscr{V}$ Isometries}

\noindent In the previous result above, in the same way that the expected value was decomposed as,

{\small

\begin{align*}
       \mathscr{P}_1 \mathscr{P}^{\prime}_1 \mathscr{P}^{\prime\prime}_1  = \mathcal{O}_1 + \mathcal{O}_2  + \mathcal{O}_3   + \mathcal{O}_4           , \\  \\  \mathscr{P}_1 \mathscr{P}^{\prime}_1  \mathscr{P}^{\prime\prime}_2  = \mathcal{O}_5 + \mathcal{O}_6  + \mathcal{O}_7   + \mathcal{O}_8       , \\ \\  \mathscr{P}_1 \mathscr{P}^{\prime}_2   \mathscr{P}^{\prime\prime}_1 = \mathcal{O}_9 + \mathcal{O}_{10} + \mathcal{O}_{11}   + \mathcal{O}_{12}   , \\ \\  \mathscr{P}_1  \mathscr{P}^{\prime}_2    \mathscr{P}^{\prime\prime}_2  = \mathcal{O}_{13} + \mathcal{O}_{14}  + \mathcal{O}_{15}   + \mathcal{O}_{16}     , \\ \\        \mathscr{P}_2 \mathscr{P}^{\prime}_1 \mathscr{P}^{\prime\prime}_2 = \mathcal{O}_{17} + \mathcal{O}_{18}  + \mathcal{O}_{19}   + \mathcal{O}_{20}        , \end{align*}
       
       \begin{align*}  \mathscr{P}_2 \mathscr{P}^{\prime}_1 \mathscr{P}^{\prime\prime}_2 = \mathcal{O}_{21} + \mathcal{O}_{22}  + \mathcal{O}_{23}   + \mathcal{O}_{24}   , \\ \\  \mathscr{P}_2 \mathscr{P}^{\prime\prime}_2 \mathscr{P}^{\prime\prime}_1 = \mathcal{O}_{25} + \mathcal{O}_{26}  + \mathcal{O}_{27}   + \mathcal{O}_{28}  , \\ \\  \mathscr{P}_2 \mathscr{P}^{\prime\prime}_2 \mathscr{P}^{\prime\prime}_2 = \mathcal{O}_{29} + \mathcal{O}_{30}  + \mathcal{O}_{31}   + \mathcal{O}_{32}  . 
\end{align*}

}

\noindent through the product,

 {\small \begin{align*}   \big\{ \mathscr{P}_1 + \mathscr{P}_2  \big\} \big\{  \mathscr{P}^{\prime}_1 + \mathscr{P}^{\prime}_2 \big\}  \big\{ \mathscr{P}^{\prime\prime}_1 + \mathscr{P}^{\prime\prime}_2 \big\}    ,    \\ 
\end{align*}}

\noindent one can introduce terms for decomposing the product,

  {\small \begin{align*}   \big\{ \mathscr{T}_1 + \mathscr{T}_2  \big\} \big\{  \mathscr{T}^{\prime}_1 + \mathscr{T}^{\prime}_2 \big\}  \big\{ \mathscr{T}^{\prime\prime}_1 + \mathscr{T}^{\prime\prime}_2 \big\}    ,  
\end{align*} }

\noindent from products of $\mathscr{V}$ isometries.

\subsubsection{Proof}

\noindent \textit{Proof of Proposition 3}. By direct computation, to lower bound,

{\small \begin{align*}
     \bra{\mathcal{L}} \bigg[ \underset{x_A, x_B}{\underset{\alpha^{\prime} \neq \beta^{\prime\prime} \in \textbf{F}^n_2}{\underset{L \in \textit{Mat}_{\textbf{F}^n_2} [ \textit{2} \times \textit{2}]}{\sum}}}     \mathscr{V}^{\dagger}_{\textit{Ideal}}   \mathscr{V}^{\dagger}_{\textit{Simulator}}  \mathscr{V}^{\prime}_{\textit{Real}}      \bigg] \ket{\mathcal{L}}   , 
\end{align*} }

\noindent observe,

{\small \begin{align*}
     \bigg[  \underset{L}{\sum} \sqrt{p_L}  \bra{LL}_{L L^{\prime}}        \bigg] \end{align*}

     \begin{align*} \times  \bigg\{    \bigg[       \underset{x_A, x_B, z_A, z_B}{\underset{\alpha^{\prime} \neq \beta^{\prime\prime}  \in \textbf{F}^n_2}{\sum}}    \ket{\psi_{\alpha^{\prime}  \beta^{\prime\prime} }} \bra{\psi_{\alpha^{\prime} \beta^{\prime\prime} }}    \otimes   \frac{1}{\sqrt{2}} \bigg[   \ket{F ( \alpha^{\prime} + \beta^{\prime\prime} ) , F \big( \alpha^{\prime}  \big) }_{(S+T) S^{\prime}}    +   \ket{F ( \alpha^{\prime} + \beta^{\prime\prime} ) , F \big( \beta^{\prime\prime} \big) }_{(S+T) T^{\prime}}                \bigg]     \bigg]^{\dagger}  \\ \times  \bigg[      \ket{L} \bra{L}_{\textbf{L}} \otimes \frac{1}{\sqrt{2}} \bigg\{  H^{\otimes 2n}   \bigg[       \ket{x_B + x_A} \bra{x_B}_{(B+A)B} + \ket{x_B + x_A}   \bra{x_A }_{(B+A)A}      \bigg] \\   \times H^{\otimes 2n}     \otimes        \bigg[     \ket{\mathcal{P}_2  x_A,  \mathcal{P}_2 x_A , \mathcal{P}_2  x_B,  \mathcal{P}_2 x_B , \mathcal{P}_2  x_A  ,  \mathcal{P}_2  x_B  }_{U_A U^{\prime}_A U_B U^{\prime}_B U^{\prime\prime}_B U^{\prime\prime}_A }  \\  +    \ket{\mathcal{P}_2  x_A,  \mathcal{P}_2 x_A , \mathcal{P}_2  x_B,  \mathcal{P}_2 x_B , \mathcal{P}_2 x_B   ,  \mathcal{P}_2 x_A }_{U_A U^{\prime}_A U_B U^{\prime}_B U^{\prime\prime}_B U^{\prime\prime}_A }        \bigg]  \bigg\}       \bigg]^{\dagger} \\ \times  \bigg[          \ket{L} \bra{L}_{\textbf{L}} \otimes \frac{1}{\sqrt{2}} \bigg\{ H^{\otimes 2n}  \bigg[       \ket{x_B + x_A} \bra{x_B}_{(B+A)B} + \ket{x_B + x_A} \bra{x_A }_{(B+A)A}      \bigg] H^{\otimes 2n}     \end{align*}

          \begin{align*}    \otimes   \bigg[     \ket{\mathcal{P}_2  x_A,  \mathcal{P}_2 x_A , \mathcal{P}_2  x_B,  \mathcal{P}_2 x_B ,  g_2 \big( \mathcal{P}_1 , \mathcal{P}_2 \big( x_A + x_B \big) \big)  ,  \mathcal{P}_2  x_A  }_{U_A U^{\prime}_A U_B U^{\prime}_B U^{\prime\prime}_B U^{\prime\prime}_A }    \\ +    \ket{\mathcal{P}_2  x_A,  \mathcal{P}_2 x_A , \mathcal{P}_2  x_B,  \mathcal{P}_2 x_B , \mathcal{P}_2 x_B   ,  g_2 \big( \mathcal{P}_1 , \mathcal{P}_2 \big( x_A + x_B \big) \big)         }_{U_A U^{\prime}_A U_B U^{\prime}_B U^{\prime\prime}_B U^{\prime\prime}_A }        \bigg]    \bigg\}            \bigg] \bigg\}   \\ \times \bigg[ \underset{L}{\sum} \sqrt{p_L}  \ket{LL}_{L L^{\prime}}  \bigg] \end{align*}

     \begin{align*} =   \underset{x_A, x_B}{\underset{\alpha^{\prime} \neq \beta^{\prime\prime} \in \textbf{F}^n_2}{\underset{L \in \textit{Mat}_{\textbf{F}^n_2} [ \textit{2} \times \textit{2}]}{\sum}}} \bigg[ p_L      \bigg\{    \bigg[         \ket{\psi_{\alpha^{\prime}  \beta^{\prime\prime} }} \bra{\psi_{\alpha^{\prime} \beta^{\prime\prime} }}    \times   \frac{1}{\sqrt{2}} \bigg[   \ket{F ( \alpha^{\prime} + \beta^{\prime\prime} ) , F \big( \alpha^{\prime}  \big) }_{(S+T) S^{\prime}}   \\ +   \ket{F ( \alpha^{\prime}  + \beta^{\prime\prime} ) , F \big( \beta^{\prime\prime} \big) }_{(S+T) T^{\prime}}                \bigg]     \bigg]^{\dagger}  \bigg[    \frac{1}{\sqrt{2}} \bigg\{  H^{\otimes 2n}   \bigg[       \ket{x_B + x_A} \bra{x_B}_{(B+A)B} + \ket{x_B + x_A} \\  \times  \bra{x_A }_{(B+A)A}      \bigg] H^{\otimes 2n}     \times        \bigg[     \ket{\mathcal{P}_2  x_A,  \mathcal{P}_2 x_A , \mathcal{P}_2  x_B,  \mathcal{P}_2 x_B , \mathcal{P}_2  x_A  ,  \mathcal{P}_2  x_B  }_{U_A U^{\prime}_A U_B U^{\prime}_B U^{\prime\prime}_B U^{\prime\prime}_A }    \\ +    \ket{\mathcal{P}_2  x_A,  \mathcal{P}_2 x_A , \mathcal{P}_2  x_B,  \mathcal{P}_2 x_B , \mathcal{P}_2 x_B   ,  \mathcal{P}_2 x_A }_{U_A U^{\prime}_A U_B U^{\prime}_B U^{\prime\prime}_B U^{\prime\prime}_A }        \bigg]  \bigg\}       \bigg]^{\dagger} \end{align*}

     \begin{align*}    \times  \bigg[          \frac{1}{\sqrt{2}} \bigg\{ H^{\otimes 2n}  \bigg[       \ket{x_B + x_A} \bra{x_B}_{(B+A)B} + \ket{x_B + x_A} \bra{x_A }_{(B+A)A}      \bigg] H^{\otimes 2n}   \\ \times   \bigg[     \ket{\mathcal{P}_2  x_A,  \mathcal{P}_2 x_A , \mathcal{P}_2  x_B,  \mathcal{P}_2 x_B ,  g_2 \big( \mathcal{P}_1 , \mathcal{P}_2 \big( x_A + x_B \big) \big)  ,  \mathcal{P}_2  x_A  }_{U_A U^{\prime}_A U_B U^{\prime}_B U^{\prime\prime}_B U^{\prime\prime}_A }    \\ +    \ket{\mathcal{P}_2  x_A,  \mathcal{P}_2 x_A , \mathcal{P}_2  x_B,  \mathcal{P}_2 x_B , \mathcal{P}_2 x_B   ,  g_2 \big( \mathcal{P}_1 , \mathcal{P}_2 \big( x_A + x_B \big) \big)         }_{U_A U^{\prime}_A U_B U^{\prime}_B U^{\prime\prime}_B U^{\prime\prime}_A }        \bigg]    \bigg\}            \bigg] \bigg\}     \bigg]   \end{align*}

     \begin{align*} \overset{(\mathrm{\textbf{Lemma}})}{=}     {\underset{\alpha^{\prime} \neq \beta^{\prime\prime} \in \textbf{F}^n_2}{\underset{L \in \textit{Mat}_{\textbf{F}^n_2} [ \textit{2} \times \textit{2}]}{\sum}}} \bigg[ p_L        \bigg\{    \bigg[      \underbrace{\underset{x_B \in \textbf{F}^n_2}{\sum}  H^{\otimes 2 n}  \bigg[  \ket{x_B + \alpha^{\prime} + \beta^{\prime\prime}}_B   \bigg[         \underset{x_A \in \textbf{F}^2_n}{\sum}    \ket{x_A , x_A + \alpha} \bra{x_A , x_A + \alpha }            \bigg]_{AA}}   \\ \times  \underbrace{\bra{x_B + \alpha^{\prime} + \beta^{\prime\prime}}_{B} \bigg]   H^{\otimes 2 n}}                                           \times   \frac{1}{\sqrt{2}} \bigg[   \ket{F ( \alpha^{\prime} + \beta^{\prime\prime} ) , F \big( \alpha^{\prime}  \big) }_{(S+T) S^{\prime}} \\    +   \ket{F ( \alpha^{\prime} + \beta^{\prime\prime} ) , F \big( \beta^{\prime\prime} \big) }_{(S+T) T^{\prime}}                \bigg]     \bigg]^{\dagger} \times  \bigg[     \frac{1}{\sqrt{2}} \bigg\{  H^{\otimes 2n}   \bigg[       \ket{x_B + x_A} \bra{x_B}_{(B+A)B} + \ket{x_B + x_A} \bra{x_A }_{(B+A)A}      \bigg] \\   \times H^{\otimes 2n}     \times        \bigg[     \ket{\mathcal{P}_2  x_A,  \mathcal{P}_2 x_A , \mathcal{P}_2  x_B,  \mathcal{P}_2 x_B , \mathcal{P}_2  x_A  ,  \mathcal{P}_2  x_B  }_{U_A U^{\prime}_A U_B U^{\prime}_B U^{\prime\prime}_B U^{\prime\prime}_A }    \\  +    \ket{\mathcal{P}_2  x_A,  \mathcal{P}_2 x_A , \mathcal{P}_2  x_B,  \mathcal{P}_2 x_B , \mathcal{P}_2 x_B   ,  \mathcal{P}_2 x_A }_{U_A U^{\prime}_A U_B U^{\prime}_B U^{\prime\prime}_B U^{\prime\prime}_A }        \bigg]  \bigg\}       \bigg]^{\dagger} \\   \times  \bigg[           \frac{1}{\sqrt{2}} \bigg\{ H^{\otimes 2n}  \bigg[       \ket{x_B + x_A} \bra{x_B}_{(B+A)B} + \ket{x_B + x_A} \bra{x_A }_{(B+A)A}      \bigg] H^{\otimes 2n}   \end{align*}

     \begin{align*}  \times   \bigg[     \ket{\mathcal{P}_2  x_A,  \mathcal{P}_2 x_A , \mathcal{P}_2  x_B,  \mathcal{P}_2 x_B ,  g_2 \big( \mathcal{P}_1 , \mathcal{P}_2 \big( x_A + x_B \big) \big)  ,  \mathcal{P}_2  x_A  }_{U_A U^{\prime}_A U_B U^{\prime}_B U^{\prime\prime}_B U^{\prime\prime}_A }    \\ +    \ket{\mathcal{P}_2  x_A,  \mathcal{P}_2 x_A , \mathcal{P}_2  x_B,  \mathcal{P}_2 x_B , \mathcal{P}_2 x_B   ,  g_2 \big( \mathcal{P}_1 , \mathcal{P}_2 \big( x_A + x_B \big) \big)         }_{U_A U^{\prime}_A U_B U^{\prime}_B U^{\prime\prime}_B U^{\prime\prime}_A }        \bigg]    \bigg\}            \bigg] \bigg\}    \bigg] \end{align*}

     \begin{align*}     =         \bigg(      \underset{x_A, x_B}{\underset{\alpha^{\prime} \neq \beta^{\prime\prime} \in \textbf{F}^n_2}{\underset{L \in \textit{Mat}_{\textbf{F}^n_2} [ \textit{2} \times \textit{2}]}{\sum}}}         p_L        \bigg\{    \bigg[      H^{\otimes 2 n}  \bigg[  \ket{x_B + \alpha^{\prime} + \beta^{\prime\prime}}_B   \bigg[          \ket{x_A , x_A + \alpha} \bra{x_A , x_A + \alpha }            \bigg]_{AA}   \bra{x_B + \alpha^{\prime} + \beta^{\prime\prime}}_{B} \bigg]   H^{\otimes 2 n}                                                                 \bigg\}   \\ \times   \frac{1}{\sqrt{2}}   \ket{F ( \alpha^{\prime} + \beta^{\prime\prime} ) , F \big( \alpha^{\prime}  \big) }^{\dagger}_{(S+T) S^{\prime}}    + \underset{x_A, x_B}{\underset{\alpha^{\prime} \neq \beta^{\prime\prime} \in \textbf{F}^n_2}{\underset{L \in \textit{Mat}_{\textbf{F}^n_2} [ \textit{2} \times \textit{2}]}{\sum}}}         p_L        \bigg\{    \bigg[     H^{\otimes 2 n}  \bigg[  \ket{x_B + \alpha^{\prime} + \beta^{\prime\prime}}_B   \bigg[          \ket{x_A , x_A + \alpha}  \\       \times   \bra{x_A , x_A + \alpha }            \bigg]_{AA}   \bra{x_B + \alpha^{\prime} + \beta^{\prime\prime}}_{B} \bigg]   H^{\otimes 2 n}                                                                 \bigg\} \frac{1}{\sqrt{2}}   \ket{F ( \alpha^{\prime} + \beta^{\prime\prime} ) , F \big( \beta^{\prime\prime} \big) }^{\dagger}_{(S+T) T^{\prime}}  \bigg)  \\ \times \bigg(  \bigg\{                \frac{1}{\sqrt{2}}  H^{\otimes 2n}   \bigg[       \ket{x_B + x_A} \bra{x_B}_{(B+A)B} + \ket{x_B + x_A} \bra{x_A }_{(B+A)A}      \bigg] H^{\otimes 2n}  \\ \times       \ket{\mathcal{P}_2  x_A,  \mathcal{P}_2 x_A , \mathcal{P}_2  x_B,  \mathcal{P}_2 x_B , \mathcal{P}_2  x_A  ,  \mathcal{P}_2  x_B  }^{\dagger}_{U_A U^{\prime}_A U_B U^{\prime}_B U^{\prime\prime}_B U^{\prime\prime}_A }  \\ \\  +    \frac{1}{\sqrt{2}}  H^{\otimes 2n}   \bigg[       \ket{x_B + x_A} \bra{x_B}_{(B+A)B} + \ket{x_B + x_A} \bra{x_A }_{(B+A)A}      \bigg] H^{\otimes 2n}      \\ \times   \ket{\mathcal{P}_2  x_A,  \mathcal{P}_2 x_A , \mathcal{P}_2  x_B,  \mathcal{P}_2 x_B , \mathcal{P}_2 x_B   ,  \mathcal{P}_2 x_A }^{\dagger}_{U_A U^{\prime}_A U_B U^{\prime}_B U^{\prime\prime}_B U^{\prime\prime}_A }                          \bigg\} \\ \times   \bigg\{              \frac{1}{\sqrt{2}} H^{\otimes 2n}  \bigg[       \ket{x_B + x_A} \bra{x_B}_{(B+A)B} + \ket{x_B + x_A} \bra{x_A }_{(B+A)A}      \bigg] H^{\otimes 2n}  \\ \times     \ket{\mathcal{P}_2  x_A,  \mathcal{P}_2 x_A , \mathcal{P}_2  x_B,  \mathcal{P}_2 x_B ,  g_2 \big( \mathcal{P}_1 , \mathcal{P}_2 \big( x_A + x_B \big) \big)  ,  \mathcal{P}_2  x_A  }_{U_A U^{\prime}_A U_B U^{\prime}_B U^{\prime\prime}_B U^{\prime\prime}_A }     \\   +         \frac{1}{\sqrt{2}} H^{\otimes 2n}  \bigg[       \ket{x_B + x_A} \bra{x_B}_{(B+A)B} + \ket{x_B + x_A} \bra{x_A }_{(B+A)A}      \bigg] H^{\otimes 2n}  \\ \times   \ket{\mathcal{P}_2  x_A,  \mathcal{P}_2 x_A , \mathcal{P}_2  x_B,  \mathcal{P}_2 x_B , \mathcal{P}_2 x_B   ,  g_2 \big( \mathcal{P}_1 , \mathcal{P}_2 \big( x_A + x_B \big) \big)         }_{U_A U^{\prime}_A U_B U^{\prime}_B U^{\prime\prime}_B U^{\prime\prime}_A }   \bigg\}     \bigg)  \end{align*}

     \begin{align*} =   \big\{ \mathscr{T}_1 + \mathscr{T}_2  \big\} \big\{  \mathscr{T}^{\prime}_1 + \mathscr{T}^{\prime}_2 \big\}  \big\{ \mathscr{T}^{\prime\prime}_1 + \mathscr{T}^{\prime\prime}_2 \big\}    .   \\ 
\end{align*} }

\noindent In the following, we perform computations from the terms identified in the above product expansion,

{\small

\begin{align*}
  \mathscr{T}_1 =     \underset{x_A, x_B}{\underset{\alpha^{\prime} \neq \beta^{\prime\prime} \in \textbf{F}^n_2}{\underset{L \in \textit{Mat}_{\textbf{F}^n_2} [ \textit{2} \times \textit{2}]}{\sum}}}         p_L        \bigg[      H^{\otimes 2 n}  \bigg[  \ket{x_B + \alpha^{\prime} + \beta^{\prime\prime}}_B   \bigg[          \ket{x_A , x_A + \alpha} \bra{x_A , x_A + \alpha }            \bigg]_{AA}   \bra{x_B + \alpha^{\prime} + \beta^{\prime\prime}}_{B} \bigg]   H^{\otimes 2 n}                                                               \end{align*}

     \begin{align*}  \times   \frac{1}{\sqrt{2}}   \ket{F ( \alpha^{\prime} + \beta^{\prime\prime} ) , F \big( \alpha^{\prime}  \big) }^{\dagger}_{(S+T) S^{\prime}}         , \\ \\ \mathscr{T}_2 =      \underset{x_A, x_B}{\underset{\alpha^{\prime} \neq \beta^{\prime\prime} \in \textbf{F}^n_2}{\underset{L \in \textit{Mat}_{\textbf{F}^n_2} [ \textit{2} \times \textit{2}]}{\sum}}}         p_L         \bigg[     H^{\otimes 2 n}  \bigg[  \ket{x_B + \alpha^{\prime} + \beta^{\prime\prime}}_B   \bigg[          \ket{x_A , x_A + \alpha}     \bra{x_A , x_A + \alpha }            \bigg]_{AA}    \bra{x_B + \alpha^{\prime} + \beta^{\prime\prime}}_{B} \bigg]   H^{\otimes 2 n}                             \\ \times                                      \frac{1}{\sqrt{2}}   \ket{F ( \alpha^{\prime} + \beta^{\prime\prime} ) , F \big( \beta^{\prime\prime} \big) }^{\dagger}_{(S+T) T^{\prime}}        , \end{align*}

  \begin{align*} \mathscr{T}^{\prime}_1 =     H^{\otimes 2n}   \bigg[       \ket{x_B + x_A} \bra{x_B}_{(B+A)B} + \ket{x_B + x_A} \bra{x_A }_{(B+A)A}      \bigg] H^{\otimes 2n}  \\ \times       \ket{\mathcal{P}_2  x_A,  \mathcal{P}_2 x_A , \mathcal{P}_2  x_B,  \mathcal{P}_2 x_B , \mathcal{P}_2  x_A  ,  \mathcal{P}_2  x_B  }^{\dagger}_{U_A U^{\prime}_A U_B U^{\prime}_B U^{\prime\prime}_B U^{\prime\prime}_A }          , \\ \\ \mathscr{T}^{\prime}_2  =      H^{\otimes 2n}   \bigg[       \ket{x_B + x_A} \bra{x_B}_{(B+A)B} + \ket{x_B + x_A} \bra{x_A }_{(B+A)A}      \bigg] H^{\otimes 2n}      \\ \times   \ket{\mathcal{P}_2  x_A,  \mathcal{P}_2 x_A , \mathcal{P}_2  x_B,  \mathcal{P}_2 x_B , \mathcal{P}_2 x_B   ,  \mathcal{P}_2 x_A }^{\dagger}_{U_A U^{\prime}_A U_B U^{\prime}_B U^{\prime\prime}_B U^{\prime\prime}_A }                               ,  \\ \\ \mathscr{T}^{\prime\prime}_1 =                 H^{\otimes 2n}  \bigg[       \ket{x_B + x_A} \bra{x_B}_{(B+A)B} + \ket{x_B + x_A} \bra{x_A }_{(B+A)A}      \bigg] H^{\otimes 2n}  \\ \times     \ket{\mathcal{P}_2  x_A,  \mathcal{P}_2 x_A , \mathcal{P}_2  x_B,  \mathcal{P}_2 x_B ,  g_2 \big( \mathcal{P}_1 , \mathcal{P}_2 \big( x_A + x_B \big) \big)  ,  \mathcal{P}_2  x_A  }_{U_A U^{\prime}_A U_B U^{\prime}_B U^{\prime\prime}_B U^{\prime\prime}_A }          , \\ \\ \mathscr{T}^{\prime\prime}_2 =  H^{\otimes 2n} \bigg[       \ket{x_B + x_A} \bra{x_B}_{(B+A)B} + \ket{x_B + x_A} \bra{x_A }_{(B+A)A}      \bigg] H^{\otimes 2n}  \\ \times   \ket{\mathcal{P}_2  x_A,  \mathcal{P}_2 x_A , \mathcal{P}_2  x_B,  \mathcal{P}_2 x_B , \mathcal{P}_2 x_B   ,  g_2 \big( \mathcal{P}_1 , \mathcal{P}_2 \big( x_A + x_B \big) \big)         }_{U_A U^{\prime}_A U_B U^{\prime}_B U^{\prime\prime}_B U^{\prime\prime}_A }        .  \\ 
\end{align*}

}

\noindent Further rearrangements of the above superposition, term by term, imply that the desired result for the $\mathscr{V}$ isometries, in place of the $\mathscr{U}$ isometries, hold. Straightforwardly, in place of $\mathscr{P}_1, \mathscr{P}_2, \mathscr{P}^{\prime}_1, \mathscr{P}^{\prime}_2, \mathscr{P}^{\prime\prime}_1, \mathscr{P}^{\prime\prime}_2$, one performs the same computations with with $\mathscr{T}_1, \mathscr{T}_2, \mathscr{T}^{\prime}_1, \mathscr{T}^{\prime}_2, \mathscr{T}^{\prime\prime}_1, \mathscr{T}^{\prime\prime}_2$. Crucially, in comparison to arguments for the previous result which claimed that,

{\tiny \begin{align*}
   {\underset{ \{ X^{\prime} = \mathcal{P}_1 z_A, Y^{\prime} = g_1 ( \mathcal{P}_1 , \mathcal{P}_1 ( z_A + z_B) ) \} }{\underset{ \{ X^{\prime} = \mathcal{P}_1 z_B, Y^{\prime} = g_1 ( \mathcal{P}_1 , \mathcal{P}_1 ( z_A + z_B) ) \} }{\mathrm{sup}}}}   \textbf{P}_L         \big[    F \big( \beta \big) = X^{\prime}  , F \big( \alpha + \beta^{\prime} \big) = Y^{\prime}    \big] =         \textbf{P}_L         \bigg[    {\underset{ \{ X^{\prime} = \mathcal{P}_1 z_A, Y^{\prime} = g_1 ( \mathcal{P}_1 , \mathcal{P}_1 ( z_A + z_B) ) \} }{\underset{ \{ X^{\prime} = \mathcal{P}_1 z_B, Y^{\prime} = g_1 ( \mathcal{P}_1 , \mathcal{P}_1 ( z_A + z_B) ) \} }{\mathrm{sup}}}}   \big\{   F \big( \beta \big) = X^{\prime} \\  , F \big( \alpha + \beta^{\prime} \big) = Y^{\prime}  \big\}   \bigg]    \\   \\ = 1 - \textbf{P}_L         \bigg[    {\underset{ \{ X^{\prime} = \mathcal{P}_1 z_A, Y^{\prime} = g_1 ( \mathcal{P}_1 , \mathcal{P}_1 ( z_A + z_B) ) \} }{\underset{ \{ X^{\prime} = \mathcal{P}_1 z_B, Y^{\prime} = g_1 ( \mathcal{P}_1 , \mathcal{P}_1 ( z_A + z_B) ) \} }{\mathrm{sup}}}}   \big\{   F \big( \beta \big) \neq  X^{\prime}  , F \big( \alpha + \beta^{\prime} \big) \neq  Y^{\prime}  \big\}   \bigg]   \\ \\ \overset{(\textbf{Corollary} \textit{1}), {\color{blue}[40]}}{>} 1 - 2^{-k + n h ( \frac{r}{n})}  ,
\end{align*}     }

\noindent one instead obtains the desired lower bound, up to constants, with,

{\small \begin{align*}
  \big[  2^{5-\frac{3}{2}} -    2^{-k + n h ( \frac{r}{n} ) + 5 \frac{3}{2}}   \big] \textbf{I}_{AB}  , 
\end{align*} }

\noindent from  the probability measure,

\begin{align*}
    \textbf{P}_{(L^{-1})^{\mathrm{T}}} \big[ \cdot \big] =  \textbf{P}_{M} \big[ \cdot \big]  .
\end{align*}

\noindent Hence,

{\small \begin{align*}
     \bra{\mathcal{L}} \bigg[ \underset{x_A, x_B}{\underset{\alpha^{\prime} \neq \beta^{\prime\prime} \in \textbf{F}^n_2}{\underset{L \in \textit{Mat}_{\textbf{F}^n_2} [ \textit{2} \times \textit{2}]}{\sum}}}     \mathscr{V}^{\dagger}_{\textit{Ideal}}   \mathscr{V}^{\dagger}_{\textit{Simulator}}  \mathscr{V}^{\prime}_{\textit{Real}}      \bigg] \ket{\mathcal{L}} \gtrsim   2^{5-\frac{3}{2}}   \big[ 1 -    2^{-k + n h ( \frac{r}{n} )}  \big] \textbf{I}_{AB}   =    \big[  2^{5-\frac{3}{2}} -    2^{-k + n h ( \frac{r}{n} ) + 5 \frac{3}{2}}   \big] \textbf{I}_{AB}  , 
\end{align*} }

\noindent from which we conclude the argument. \boxed{}

\subsection{$\textbf{Proposition}$ \textit{4}}

\subsubsection{Description of the anticommutation relation between products of $\mathscr{U}$ and $\mathscr{V}$ isometries}

\noindent To determine whether,

{\small \begin{align*}
  \underset{L \in \textit{Mat}_{\textbf{F}^n_2} [ \textit{2} \times \textit{2} ]}{\sum} \underset{\alpha \neq \beta^{\prime} \in \textbf{F}^n_2}{\sum}     \underset{\alpha^{\prime} \neq \beta^{\prime\prime} \in \textbf{F}^n_2}{\sum}  \bigg\{ \frac{\mathscr{U}_{\textit{Simulator}} \mathscr{V}^{\prime}_{\textit{Real}}}{\mathscr{V}^{\prime}_{\textit{Real}} \mathscr{U}_{\textit{Simulator}} } \bigg\} \lesssim 1  ,
\end{align*} }

\noindent observe that composing the projection operations,

{\small

\begin{align*}
   P  \bigg[    P \bigg[ \mathcal{P}_1  \big( \vec{\sigma_1}  \big)     , u^{\prime}  \bigg] ,  u   \bigg]  P  \bigg[       P \bigg[ \mathcal{P}_2  \big( \vec{\sigma_3}  \big)  , v^{\prime} \bigg] ,  v   \bigg]   , \\ 
\end{align*}

}

\noindent over $\vec{\sigma_1}$ and $\vec{\sigma_3}$, respectively, when normalized with the composition,

{\small

\begin{align*}
     P  \bigg[    P \bigg[ \mathcal{P}_2  \big( \vec{\sigma_3}  \big)     , u^{\prime}  \bigg] ,  u   \bigg]  P  \bigg[       P \bigg[  \mathcal{P}_1  \big( \vec{\sigma_1}  \big)  , v^{\prime} \bigg] ,  v   \bigg] , \\ 
\end{align*}

}

\noindent of projection operations over $\vec{\sigma_3}$ and $\vec{\sigma_1}$, respectively, implies,

{\small

\begin{align*}
   \underset{L \in \textit{Mat}_{\textbf{F}^n_2} [ \textit{2} \times \textit{2} ]}{\sum} \underset{\alpha \neq \beta^{\prime} \in \textbf{F}^n_2}{\sum}     \underset{\alpha^{\prime} \neq \beta^{\prime\prime} \in \textbf{F}^n_2}{\sum}  \bigg\{ \frac{\mathscr{U}_{\textit{Simulator}} \mathscr{V}^{\prime}_{\textit{Real}}}{\mathscr{V}^{\prime}_{\textit{Real}} \mathscr{U}_{\textit{Simulator}} }  \bigg(   \frac{  P  \bigg[    P \bigg[ \mathcal{P}_1  \big( \vec{\sigma_1}  \big)     , u^{\prime}  \bigg] ,  u   \bigg]  P  \bigg[       P \bigg[ \mathcal{P}_2  \big( \vec{\sigma_3}  \big)  , v^{\prime} \bigg] ,  v   \bigg]}{  P  \bigg[    P \bigg[ \mathcal{P}_2  \big( \vec{\sigma_3}  \big)     , u^{\prime}  \bigg] ,  u   \bigg]  P  \bigg[       P \bigg[  \mathcal{P}_1  \big( \vec{\sigma_1}  \big)  , v^{\prime} \bigg] ,  v   \bigg]}   \bigg)            \bigg\} \leq  1    , 
\end{align*}

}

\noindent for,

{\small

\begin{align*}
 \mathscr{F} =    \frac{  P  \bigg[    P \bigg[ \mathcal{P}_1  \big( \vec{\sigma_1}  \big)     , u^{\prime}  \bigg] ,  u   \bigg]  P  \bigg[       P \bigg[ \mathcal{P}_2  \big( \vec{\sigma_3}  \big)  , v^{\prime} \bigg] ,  v   \bigg]}{  P  \bigg[    P \bigg[ \mathcal{P}_2  \big( \vec{\sigma_3}  \big)     , u^{\prime}  \bigg] ,  u   \bigg]  P  \bigg[       P \bigg[  \mathcal{P}_1  \big( \vec{\sigma_1}  \big)  , v^{\prime} \bigg] ,  v   \bigg]}          . \\ 
\end{align*}

}

\subsubsection{Proof}

\noindent \textit{Proof of Proposition 4}. By direct computation, to demonstrate that,

{\small \begin{align*}
  \underset{L \in \textit{Mat}_{\textbf{F}^n_2} [ \textit{2} \times \textit{2} ]}{\sum} \underset{\alpha \neq \beta^{\prime} \in \textbf{F}^n_2}{\sum}     \underset{\alpha^{\prime} \neq \beta^{\prime\prime} \in \textbf{F}^n_2}{\sum}  \big\{ \mathscr{U}_{\textit{Simulator}} \mathscr{V}^{\prime}_{\textit{Real}} \big\}  = \mathscr{F} \bigg\{  \underset{L \in \textit{Mat}_{\textbf{F}^n_2} [ \textit{2} \times \textit{2} ]}{\sum} \underset{\alpha \neq \beta^{\prime} \in \textbf{F}^n_2}{\sum}     \underset{\alpha^{\prime} \neq \beta^{\prime\prime} \in \textbf{F}^n_2}{\sum}   \big\{ \mathscr{V}^{\prime}_{\textit{Real}} \\ \times \mathscr{U}_{\textit{Simulator}} \big\} \bigg\}  ,
\end{align*} }

\noindent observe that the LHS of the above equality,

\begin{align*}
\bigg[  \ket{L} \bra{L}_{\textbf{L}} \otimes \frac{1}{\sqrt{2}} \bigg\{  \bigg[       \ket{z_B + z_A} \bra{z_B}_{(B+A)B} + \ket{z_B + z_A} \bra{z_A }_{(B+A)A}      \bigg] \\ \otimes      \bigg[  \ket{\mathcal{P}_1  z_A,  \mathcal{P}_1 z_A , \mathcal{P}_1  z_B,  \mathcal{P}_1 z_B ,  \mathcal{P}_1  z_A     , \mathcal{P}_1 z_B     }_{U_A U^{\prime}_A U_B U^{\prime}_B U^{\prime\prime}_A U^{\prime\prime}_B }       \\ +   \ket{\mathcal{P}_1  z_A,  \mathcal{P}_1 z_A , \mathcal{P}_1  z_B,  \mathcal{P}_1 z_B , \mathcal{P}_1 z_B   ,  \mathcal{P}_1  z_A }_{U_A U^{\prime}_A U_B U^{\prime}_B S S^{\prime} }       \bigg] \bigg\}  \bigg] \end{align*}

\begin{align*} \times \bigg[   \ket{L} \bra{L}_{\textbf{L}} \otimes \frac{1}{\sqrt{2}} \bigg\{ H^{\otimes 2n}  \bigg[       \ket{x_B + x_A} \bra{x_B}_{(B+A)B} + \ket{x_B + x_A} \\ \times  \bra{x_A }_{(B+A)A}      \bigg] H^{\otimes 2n}   \\ \otimes   \bigg[     \ket{\mathcal{P}_2  x_A,  \mathcal{P}_2 x_A , \mathcal{P}_2  x_B,  \mathcal{P}_2 x_B ,  g_2 \big( \mathcal{P}_1 , \mathcal{P}_2 \big( x_A + x_B \big) \big)  ,  \mathcal{P}_2  x_A  }_{U_A U^{\prime}_A U_B U^{\prime}_B T U^{\prime\prime}_A }    \\ +    \ket{\mathcal{P}_2  x_A,  \mathcal{P}_2 x_A , \mathcal{P}_2  x_B,  \mathcal{P}_2 x_B , \mathcal{P}_2 x_B   ,  g_2 \big( \mathcal{P}_1 , \mathcal{P}_2 \big( x_A + x_B \big) \big)         }_{U_A U^{\prime}_A U_B U^{\prime}_B U^{\prime\prime}_B T^{\prime} }        \bigg]    \bigg\}        \bigg] ,
\end{align*}

\noindent equals,

\begin{align*}
 \ket{L} \bra{L}_{\textbf{L}} \otimes       \bigg[ \frac{1}{\sqrt{2}} \bigg]^2   \bigg\{  H^{\otimes 2n}   \bigg\{            \ket{z_B + z_A, x_B + x_A} \bra{z_B, x_B}_{(B+A)B}   + \ket{z_B + z_A, x_B + x_A} \\ \times   \bra{z_B , x_A}_{(B+A)A} +   \ket{z_B + z_A , x_B + x_A} \bra{z_A, x_B}_{(B+A)A} \\ +           \ket{z_B + z_A , x_B + x_A} \bra{z_A, x_A}_{(B+A)A}                \bigg\}    H^{\otimes 2n}              \\  +  \ket{\mathcal{P}_1 z_A , \mathcal{P}_2 x_A ,   \mathcal{P}_1 z_A , \mathcal{P}_2 x_A , \mathcal{P}_1 z_B , \mathcal{P}_2 x_B ,  \mathcal{P}_1 z_B , g_2 \big( \mathcal{P}_1 , \mathcal{P}_2 \big( x_A + x_B \big) \big)  ,  \mathcal{P}_1 z_A , \mathcal{P}_2 x_A }_{\mathscr{S}^{\prime}} \end{align*}

 \begin{align*} +   \ket{\mathcal{P}_1 z_A , \mathcal{P}_2 x_A ,      \mathcal{P}_1 z_A , \mathcal{P}_2 x_A ,        \mathcal{P}_1 z_B , \mathcal{P}_2 x_B , \mathcal{P}_1 z_B , \mathcal{P}_2 x_B ,             \mathcal{P}_1 z_A , g_2 \big( \mathcal{P}_1 , \mathcal{P}_2 \big( x_A + x_B \big)    }_{\mathscr{S}^{\prime\prime}}            \\ +  \ket{\mathcal{P}_1 z_A , \mathcal{P}_2 x_A , \mathcal{P}_1 z_A , \mathcal{P}_2 x_A , \mathcal{P}_1 z_B , \mathcal{P}_2 x_B ,         \mathcal{P}_1 z_B , g_2 \big( \mathcal{P}_1 , \mathcal{P}_2 \big( x_A + x_B \big)    , \mathcal{P}_2 x_A , \mathcal{P}_2 x_A                     }_{\mathscr{S}^{\prime\prime\prime}}            \\ +  \ket{\mathcal{P}_1 z_A , \mathcal{P}_2 x_A ,                  \mathcal{P}_1 z_A , \mathcal{P}_2 x_A ,  \mathcal{P}_1 z_B , \mathcal{P}_2  x_B , \mathcal{P}_1 z_B , \mathcal{P}_2 x_B ,        \mathcal{P}_2 x_A , g_2 \big( \mathcal{P}_1 , \mathcal{P}_2 \big( x_A + x_B \big) \big)          }_{\mathscr{S}^{\prime\prime\prime\prime}}        \bigg\}                   , 
\end{align*}

\noindent where,

{\small \begin{align*}
   \mathscr{S}^{\prime} =  U_A U^{\prime}_A U^{\prime\prime\prime}_A U^{\prime\prime\prime\prime}_A    U^{\prime\prime\prime\prime\prime}_A  U_B U^{\prime}_B U^{\prime\prime}_B  T  U^{\prime\prime\prime\prime\prime\prime}_A   U^{\prime\prime\prime\prime\prime\prime\prime}_A     , \end{align*}

   \begin{align*} \mathscr{S}^{\prime\prime} =              U_A U^{\prime}_A U^{\prime\prime}_A U^{\prime\prime}_A U_B U^{\prime}_B U^{\prime\prime}_B           U^{\prime\prime\prime}_B U^{\prime\prime}_A T  , \end{align*}

   \begin{align*} \mathscr{S}^{\prime\prime\prime} =      U_A U^{\prime}_A U^{\prime\prime}_A U^{\prime\prime\prime}_A U_B U^{\prime}_B U^{\prime\prime}_B T  U^{\prime\prime\prime\prime}_A     U^{\prime\prime\prime\prime}_A    ,  \end{align*}

   \begin{align*}  \mathscr{S}^{\prime\prime\prime\prime} =            U_A U^{\prime}_A U^{\prime\prime}_A U^{\prime\prime\prime}_A U_B U^{\prime}_B U^{\prime\prime}_B U^{\prime\prime\prime\prime}_A  T             . 
\end{align*} } 

\noindent The constant appearing in front of the summation of $\mathscr{V}^{\prime}_{\textit{Real}}$, and $\mathscr{U}_{\textit{Real}}$ arises from the fact that, unlike the commutation, 

\begin{align*}
   L_1 \big( \vec{\sigma_3} \big) M_2 \big( \vec{\sigma_1} \big) =   M_2 \big( \vec{\sigma_1} \big)      L_1 \big( \vec{\sigma_3} \big)  , 
\end{align*}

\noindent for the parity check matrices introduced in {\color{blue}[40]}, in which,

{\small \begin{align*}
  P \big( L_1 \big( \vec{\sigma_3} \big)  , u \big) P \big( M_2 \big( \vec{\sigma_1} \big)  , v \big) =  P \big( M_2 \big( \vec{\sigma_1} \big) , v \big)    P \big( L_1 \big( \vec{\sigma_3} \big)  , u \big)  , 
\end{align*} }

\noindent for the parity check matrices introduced in this work the product of projection operators, anticommutation with respect to the ratio of projections holds,

 \begin{align*} \frac{  P  \bigg[    P \bigg[ \mathcal{P}_1  \big( \vec{\sigma_1}  \big)     , u^{\prime}  \bigg] ,  u   \bigg]  P  \bigg[       P \bigg[ \mathcal{P}_2  \big( \vec{\sigma_3}  \big)  , v^{\prime} \bigg] ,  v   \bigg]}{  P  \bigg[    P \bigg[ \mathcal{P}_2  \big( \vec{\sigma_3}  \big)     , u^{\prime}  \bigg] ,  u   \bigg]  P  \bigg[       P \bigg[  \mathcal{P}_1  \big( \vec{\sigma_1}  \big)  , v^{\prime} \bigg] ,  v   \bigg]} , \end{align*}

\noindent from which we conclude the argument. \boxed{}

\subsection{$\textbf{Corollary}$}

\subsubsection{Description of the product of projection operators for demonstrating that the anticommutation prefactors take the desired closed-form representation}

\noindent To argue that,

{\small \[ \left\{\!\begin{array}{ll@{}l} 
   \mathscr{T}_1 \big(                  \vec{\sigma_1} , \vec{\sigma_3} ,               u , u^{\prime} , v  , v^{\prime}        \big)         \equiv \mathscr{T}_1   =   \frac{1}{1 + \underset{2 \leq k \leq 8}{\sum} \big[       \mathcal{T}^{\prime}_k  \big]^{-1}   }  , \\   \mathscr{T}_2 \big(                  \vec{\sigma_1} , \vec{\sigma_3} ,               u , u^{\prime} , v  , v^{\prime}        \big)  \equiv \mathscr{T}_2                =         \frac{1}{1 +  \big[ \mathcal{T}_1 \big]^{-1} + \underset{3 \leq k \leq 8}{\sum} \big[       \mathcal{T}^{\prime}_k  \big]^{-1}   }         ,   \\  \mathscr{T}_3 \big(                  \vec{\sigma_1} , \vec{\sigma_3} ,               u , u^{\prime} , v  , v^{\prime}        \big)  \equiv \mathscr{T}_3 =             \frac{1}{1 + \underset{1 \leq k \leq 2}{\sum} \big[       \mathcal{T}^{\prime}_k  \big]^{-1} + \underset{4 \leq k \leq 8}{\sum} \big[    \mathcal{T}^{\prime}_k    \big]^{-1}   }         , \\       \mathscr{T}_4 \big(                  \vec{\sigma_1} , \vec{\sigma_3} ,               u , u^{\prime} , v  , v^{\prime}        \big)  \equiv \mathscr{T}_4                   =              \frac{1}{1 + \underset{1 \leq k \leq 3}{\sum} \big[       \mathcal{T}^{\prime}_k  \big]^{-1} + \underset{5 \leq k \leq 8}{\sum} \big[    \mathcal{T}^{\prime}_k    \big]^{-1}   }      , \\              \mathscr{T}_5\big(                  \vec{\sigma_1} , \vec{\sigma_3} ,               u , u^{\prime} , v  , v^{\prime}        \big)  \equiv \mathscr{T}_5 =         \frac{1}{1+ \underset{1 \leq k \leq 4}{\sum} \big[           \mathcal{T}^{\prime}_k    \big]^{-1}  +    \underset{6 \leq k \leq 8}{\sum} \big[           \mathcal{T}^{\prime}_k    \big]^{-1}     }     , \\           \mathscr{T}_6\big(                  \vec{\sigma_1} , \vec{\sigma_3} ,               u , u^{\prime} , v  , v^{\prime}        \big)  \equiv \mathscr{T}_6 =  \frac{1}{1 + \underset{1 \leq k \leq 5}{\sum} \big[       \mathcal{T}^{\prime}_k  \big]^{-1} + \underset{7 \leq k \leq 8}{\sum} \big[    \mathcal{T}^{\prime}_k    \big]^{-1}   }   , \\       \mathscr{T}_7 \big(                  \vec{\sigma_1} , \vec{\sigma_3} ,               u , u^{\prime} , v  , v^{\prime}        \big)  \equiv \mathscr{T}_7 = \frac{1}{1 + \underset{1 \leq k \leq 6}{\sum} \big[ \mathcal{T}^{\prime}_k \big]^{-1} + \big[ \mathcal{T}^{\prime}_8  \big]^{-1} }                      ,  \\  \mathscr{T}_8 \big(                  \vec{\sigma_1} , \vec{\sigma_3} ,               u , u^{\prime} , v  , v^{\prime}        \big)  \equiv \mathscr{T}_8 =    \frac{1}{1 + \underset{1 \leq k \leq 7}{\sum}  \big[ \mathcal{T}^{\prime}_k \big]^{-1}  }, 
\end{array}\right. 
\]  }

\noindent can be used to obtain the anticommutation factor $\mathscr{F}$ provided in the previous result above one would like to show that,

{\small

\begin{align*}
  \frac{\mathscr{P} \big[       \mathcal{P}_1 \big( \vec{\sigma_1}  \big) , u   \big]   \mathscr{P} \big[       \mathcal{P}_2 \big( \vec{\sigma_3}  \big) ,  v \big]}{             2^{-2m} \textbf{I} +  \mathcal{T}_1 + \mathcal{T}_2 + \cdots + \mathcal{T}_8    } \lesssim 1   , \\ 
\end{align*}

}

\noindent where,

{\small \[ \left\{\!\begin{array}{ll@{}l} 
   \mathscr{T}_1 \big(                  \vec{\sigma_1} , \vec{\sigma_3} ,               u , u^{\prime} , v  , v^{\prime}        \big)         \equiv \mathscr{T}_1    , \\ \\   \mathscr{T}_2 \big(                  \vec{\sigma_1} , \vec{\sigma_3} ,               u , u^{\prime} , v  , v^{\prime}        \big)  \equiv \mathscr{T}_2                      ,   \\  \\ \mathscr{T}_3 \big(                  \vec{\sigma_1} , \vec{\sigma_3} ,               u , u^{\prime} , v  , v^{\prime}        \big)  \equiv \mathscr{T}_3    , \\  \\      \mathscr{T}_4 \big(                  \vec{\sigma_1} , \vec{\sigma_3} ,               u , u^{\prime} , v  , v^{\prime}        \big)  \equiv \mathscr{T}_4                      , \\    \\           \mathscr{T}_5\big(                  \vec{\sigma_1} , \vec{\sigma_3} ,               u , u^{\prime} , v  , v^{\prime}        \big)  \equiv \mathscr{T}_5    , \\ \\            \mathscr{T}_6\big(                  \vec{\sigma_1} , \vec{\sigma_3} ,               u , u^{\prime} , v  , v^{\prime}        \big)  \equiv \mathscr{T}_6  , \\ \\       \mathscr{T}_7 \big(                  \vec{\sigma_1} , \vec{\sigma_3} ,               u , u^{\prime} , v  , v^{\prime}        \big)  \equiv \mathscr{T}_7         , \\  \\  \mathscr{T}_8 \big(                  \vec{\sigma_1} , \vec{\sigma_3} ,               u , u^{\prime} , v  , v^{\prime}        \big)  \equiv \mathscr{T}_8   , 
\end{array}\right. 
\]  } 
 
\bigskip

\noindent The above up to constants estimate is sharpened to an inequality under the choice of a constant which equals,

{\small

\begin{align*}
     \underset{1 \leq i \leq 8 }{\sum} \mathscr{T}_i       . \\ 
\end{align*}

}

\subsubsection{Proof}

\noindent \textit{Proof of Corollary}. Fix iid $g_{1,j} \equiv g_j , g_{1,j^{\prime\prime}} \equiv g_{j^{\prime\prime}} \sim \mathcal{P}_1 \big( \vec{\sigma}_1 \big) $,$g_{1,j^{\prime}} \equiv g_{j^{\prime}}, g_{1,j^{\prime\prime}} \equiv g_{j^{\prime\prime\prime}} \sim \mathcal{P}_2 \big( \vec{\sigma_3} \big)$. By direct computation, observe,

{\small \begin{align*}
    \underline{\mathscr{P} \big[       \mathcal{P}_1 \big( \vec{\sigma_1}  \big) , u   \big]   \mathscr{P} \big[       \mathcal{P}_2 \big( \vec{\sigma_3}  \big) ,  v \big]     }  =     P  \bigg[    P \bigg[ \mathcal{P}_1  \big( \vec{\sigma_1}  \big)     , u^{\prime}  \bigg] ,  u   \bigg]  P  \bigg[       P \bigg[ \mathcal{P}_2  \big( \vec{\sigma_3}  \big)  , v^{\prime} \bigg] ,  v   \bigg]   \end{align*}

    \begin{align*} =        P \bigg[  \bigg\{           2^{-m} \underset{1 \leq j \leq m}{\prod } \big[ \textbf{I} +  \big( - 1 \big)^{u^{\prime}_j} g_j \big]     \big]          \bigg\}                             , u      \bigg]  P \bigg[          \bigg\{           2^{-m} \underset{1 \leq j^{\prime} \leq m}{\prod } \big[ \textbf{I} +  \big( - 1 \big)^{v^{\prime}_{j^{\prime}}} g_{j^{\prime}} \big]     \big]          \bigg\}          ,              v  \bigg]     \bigg]         \end{align*}

    \begin{align*} = 
\bigg\{ 2^{-m}     \underset{1 \leq j^{\prime\prime} \leq m}{\prod}    \bigg[ \textbf{I} +  \big( - 1 \big)^{u_j}           \bigg\{      \bigg\{  2^{-m} \underset{1 \leq j \leq m}{\prod } \big[ \textbf{I} +  \big( - 1 \big)^{u^{\prime}_j} g_j \big]     \big]  \bigg\} g_{j^{\prime\prime}} \bigg\}       \bigg]     \bigg\} \bigg\{   2^{-m} \underset{1 \leq j^{\prime\prime\prime} \leq m}{\prod}   \bigg[ \textbf{I} \\ +   \big( - 1 \big)^{v_j}  \bigg\{  \bigg\{  2^{-m}  \underset{1 \leq j^{\prime} \leq m}{\prod } \big[ \textbf{I}  +  \big( - 1 \big)^{v^{\prime}_{j^{\prime}}} g_{j^{\prime}} \big]     \big]    \bigg\} g_{j^{\prime\prime\prime}}  \bigg\}  \bigg]            \bigg\}      \end{align*}

    \begin{align*}  =  \bigg\{     \underset{1 \leq j^{\prime\prime} \leq m}{\prod}    \bigg[ 2^{-m}  \textbf{I} + 2^{-m}   \big( - 1 \big)^{u_j}           \bigg\{      \bigg\{  2^{-m} \underset{1 \leq j \leq m}{\prod } \big[ \textbf{I} +  \big( - 1 \big)^{u^{\prime}_j} g_j \big]     \big]  \bigg\} g_{j^{\prime\prime}} \bigg\}       \bigg]     \bigg\} \bigg\{        \underset{1 \leq j^{\prime\prime\prime} \leq m}{\prod}   \bigg[ 2^{-m} \textbf{I} \\ +  2^{-m}  \big( - 1 \big)^{v_j}  \bigg\{  \bigg\{  2^{-m}  \underset{1 \leq j^{\prime} \leq m}{\prod } \big[ \textbf{I}  +  \big( - 1 \big)^{v^{\prime}_{j^{\prime}}} g_{j^{\prime}} \big]     \big]    \bigg\} g_{j^{\prime\prime\prime}} \bigg\}  \bigg]            \bigg\}           \end{align*}

    \begin{align*} =   \underset{1 \leq j^{\prime\prime} \neq j^{\prime\prime\prime} \leq m}{\prod}  2^{-2m} \textbf{I} + \underset{1 \leq j^{\prime\prime} \neq j^{\prime\prime\prime} \leq m}{\prod}          2^{-2m} \textbf{I} \big( - 1 \big)^{v_j}  \bigg\{  \bigg\{  2^{-m}  \underset{1 \leq j^{\prime} \leq m}{\prod } \big[ \textbf{I}  +  \big( - 1 \big)^{v^{\prime}_{j^{\prime}}} g_{j^{\prime}} \big]     \big]    \bigg\} g_{j^{\prime\prime\prime}} \bigg\}               \\ +   \underset{1 \leq j^{\prime\prime} \neq j^{\prime\prime\prime} \leq m}{\prod}           \bigg\{  2^{-m}   \big( - 1 \big)^{u_j}           \bigg\{      \bigg\{  2^{-m} \underset{1 \leq j \leq m}{\prod } \big[ \textbf{I} +  \big( - 1 \big)^{u^{\prime}_j} g_j \big]     \big]  \bigg\} g_{j^{\prime\prime}} \bigg\}           \\ \times 2^{-m} \textbf{I}   \bigg\}    +   \underset{1 \leq j^{\prime\prime} \neq j^{\prime\prime\prime} \leq m}{\prod}           \bigg\{  2^{-m}   \big( - 1 \big)^{u_j}           \bigg\{      \bigg\{  2^{-m} \underset{1 \leq j \leq m}{\prod } \big[ \textbf{I} +  \big( - 1 \big)^{u^{\prime}_j} g_j \big]     \big]  \bigg\} \\ \times g_{j^{\prime\prime}}     \bigg\{   2^{-m}  \big( - 1 \big)^{v_j}  \bigg\{  \bigg\{  2^{-m}  \underset{1 \leq j^{\prime} \leq m}{\prod } \big[ \textbf{I}  +  \big( - 1 \big)^{v^{\prime}_{j^{\prime}}} g_{j^{\prime}} \big]     \big]    \bigg\} g_{j^{\prime\prime\prime}} \bigg\}           \bigg\}          \bigg\}             \end{align*}

    \begin{align*}  =              \underset{1 \leq j^{\prime\prime} \neq j^{\prime\prime\prime} \leq m}{\prod}  2^{-2m} \textbf{I} + \underset{1 \leq j^{\prime\prime} \neq j^{\prime\prime\prime} \leq m}{\prod}          2^{-2m}  \big( - 1 \big)^{v_j}  \bigg\{  \bigg\{  2^{-m}  \underset{1 \leq j^{\prime} \leq m}{\prod } \big[ \textbf{I}  +  \big( - 1 \big)^{v^{\prime}_{j^{\prime}}} g_{j^{\prime}} \big]     \big]    \bigg\} g_{j^{\prime\prime\prime}} \bigg\}               \\ +   \underset{1 \leq j^{\prime\prime} \neq j^{\prime\prime\prime} \leq m}{\prod}           \bigg\{  2^{-m}   \big( - 1 \big)^{u_j}           \bigg\{      \bigg\{  2^{-m} \underset{1 \leq j \leq m}{\prod } \big[ \textbf{I} +  \big( - 1 \big)^{u^{\prime}_j} g_j \big]     \big]  \bigg\} g_{j^{\prime\prime}} \bigg\}           \\ \times 2^{-m} \textbf{I}   \bigg\}    +   \underset{1 \leq j^{\prime\prime} \neq j^{\prime\prime\prime} \leq m}{\prod}           \bigg\{  2^{-m}   \big( - 1 \big)^{u_j}           \bigg\{      \bigg\{  2^{-m} \underset{1 \leq j \leq m}{\prod } \big[ \textbf{I} +  \big( - 1 \big)^{u^{\prime}_j} g_j \big]     \big]  \bigg\} \\ \times g_{j^{\prime\prime}}     \bigg\{   2^{-m}  \big( - 1 \big)^{v_j}  \bigg\{  \bigg\{  2^{-m}  \underset{1 \leq j^{\prime} \leq m}{\prod } \big[ \textbf{I}  +  \big( - 1 \big)^{v^{\prime}_{j^{\prime}}} g_{j^{\prime}} \big]     \big]    \bigg\} g_{j^{\prime\prime\prime}} \bigg\}           \bigg\}          \bigg\}    \bigg\}             \end{align*}

\begin{align*} =   2^{-2m} \textbf{I} +  \underset{1 \leq j^{\prime} \neq j^{\prime\prime} \neq j^{\prime\prime\prime} \leq m}{\prod}  \bigg\{   2^{-3m} \big( -1 \big)^{v_j } \textbf{I} g_{j^{\prime\prime\prime}}          +  2^{-2m} \big( - 1 \big)^{v_j} \big( - 1 \big)^{v^{\prime}_{j^{\prime}}}  g_{j^{\prime}} g_{j^{\prime\prime\prime}}   \bigg\}  \\ +       \underset{1 \leq j \neq j^{\prime\prime} \neq j^{\prime\prime\prime} \leq m}{\prod}          \bigg\{     \bigg\{ 2^{-2m} \big( - 1 \big)^{u_j} \textbf{I} g_{j^{\prime\prime}}   + 2^{-m} \big( - 1 \big)^{u_j} \big( - 1 \big)^{u^{\prime}_j} g_j g_{j^{\prime\prime}} \bigg\} 2^{-m} \textbf{I}   \bigg\}  \\ +  \underset{1 \leq j \neq j^{\prime} \neq j^{\prime\prime} \neq j^{\prime\prime\prime} \leq m}{\prod} \bigg\{   2^{-4m} \big( - 1 \big)^{u_j} \big( - 1 \big)^{v_j} \textbf{I}    g_{j^{\prime\prime}} g_{j^{\prime\prime\prime}}    +  2^{-4m} \big( - 1 \big)^{u_j}      \textbf{I} g_{j^{\prime\prime}}   \\ \times  \big( - 1 \big)^{v_j} \big( - 1 \big)^{v^{\prime}_{j^{\prime}}}   g_{j^{\prime}} g_{j^{\prime\prime\prime}}                      +   2^{-4m}                   \big( - 1 \big)^{u_j} \big( - 1 \big)^{u^{\prime}_j} g_j g_{j^{\prime\prime}}   \big( - 1 \big)^{v_j} \textbf{I} g_{j^{\prime\prime\prime}}   \\ +       2^{-4m}  \big( - 1 \big)^{u_j} \big( - 1 \big)^{u^{\prime}_j}  g_j g_{j^{\prime\prime}} \big( - 1 \big)^{v_j} \big( - 1 \big)^{v^{\prime}_{j^{\prime}}} g_{j^{\prime}} g_{j^{\prime\prime\prime}}               \bigg\} \\ \\  =                       2^{-2m} \textbf{I} +   \underset{\mathcal{T}_1}{\underbrace{\underset{1 \leq j^{\prime\prime\prime} \leq m}{\prod}   2^{-3m} \big( -1 \big)^{v_j } \textbf{I} g_{j^{\prime\prime\prime}}}}          + \underset{\mathcal{T}_2}{\underbrace{\underset{1 \leq j^{\prime} \neq j^{\prime\prime\prime} \leq m}{\prod}    2^{-2m} \big( - 1 \big)^{v_j} \big( - 1 \big)^{v^{\prime}_{j^{\prime}}}  g_{j^{\prime}} g_{j^{\prime\prime\prime}}}}    \\ +           \underset{\mathcal{T}_3}{\underbrace{\underset{1 \leq  j^{\prime\prime} \leq m}{\prod}      2^{-3m} \big( - 1 \big)^{u_j} \textbf{I} g_{j^{\prime\prime}}}}   +   \underset{\mathcal{T}_4}{\underbrace{\underset{1 \leq j \neq j^{\prime\prime}  \leq m}{\prod}      2^{-2m} \big( - 1 \big)^{u_j} \big( - 1 \big)^{u^{\prime}_j} g_j g_{j^{\prime\prime}}}}  \end{align*}

    \begin{align*}   +   \underset{\mathcal{T}_5}{\underbrace{\underset{1 \leq j^{\prime\prime} \neq j^{\prime\prime\prime} \leq m}{\prod}   2^{-4m} \big( - 1 \big)^{u_j} \big( - 1 \big)^{v_j} \textbf{I}    g_{j^{\prime\prime}} g_{j^{\prime\prime\prime}}}}  \\  +  \underset{\mathcal{T}_6}{\underbrace{\underset{1 \leq j^{\prime\prime}  \leq m}{\prod}  2^{-4m} \big( - 1 \big)^{u_j}      \textbf{I} g_{j^{\prime\prime}}   \big( - 1 \big)^{v_j} \big( - 1 \big)^{v^{\prime}_{j^{\prime}}}   g_{j^{\prime}} g_{j^{\prime\prime\prime}}}}                   \\    +      \underset{\mathcal{T}_7}{\underbrace{  \underset{1 \leq j \neq j^{\prime\prime} \neq j^{\prime\prime\prime} \leq m}{\prod}    2^{-4m}                   \big( - 1 \big)^{u_j} \big( - 1 \big)^{u^{\prime}_j} g_j g_{j^{\prime\prime}}   \big( - 1 \big)^{v_j} \textbf{I} g_{j^{\prime\prime\prime}}   }}  \\ +   \underset{\mathcal{T}_8}{\underbrace{\underset{1 \leq j \neq j^{\prime} \neq j^{\prime\prime} \neq j^{\prime\prime\prime} \leq m}{\prod}     2^{-4m}  \big( - 1 \big)^{u_j} \big( - 1 \big)^{u^{\prime}_j}  g_j g_{j^{\prime\prime}} \big( - 1 \big)^{v_j} \big( - 1 \big)^{v^{\prime}_{j^{\prime}}} g_{j^{\prime}} g_{j^{\prime\prime\prime}}}}              .                                   \end{align*} }

\noindent Straightforwardly,

{\small \begin{align*}
\mathcal{T}_1 \bigg/ \bigg\{   P  \bigg[    P \bigg[ \mathcal{P}_2  \big( \vec{\sigma_3}  \big)      , u^{\prime}  \bigg] ,  u   \bigg]  P  \bigg[       P \bigg[ \mathcal{P}_1  \big( \vec{\sigma_1} \big)     , v^{\prime} \bigg] ,  v   \bigg]  \bigg\}  \end{align*}

    \begin{align*} = \mathcal{T}_1 \bigg/ \bigg\{  P \bigg[  \bigg\{           2^{-m} \underset{1 \leq j \leq m}{\prod } \big[ \textbf{I} +  \big( - 1 \big)^{u^{\prime}_{3,j}} g_{3,j} \big]     \big]          \bigg\}                             , u      \bigg]  P \bigg[          \bigg\{           2^{-m} \underset{1 \leq j^{\prime} \leq m}{\prod } \big[ \textbf{I} +  \big( - 1 \big)^{v^{\prime}_{3,j^{\prime}}} g_{3,j^{\prime}} \big]     \big]          \bigg\}          ,              v  \bigg]     \bigg] \bigg\}  \end{align*}

    \begin{align*}         = \bigg\{  1 + \underset{2 \leq k \leq 8}{\sum} \big[       \mathcal{T}^{\prime}_k  \big]^{-1}    \bigg\}^{-1}     =                    \mathscr{T}_1                                                                                                                                                       ,  \end{align*}

\noindent corresponding to the first term normalized by a product of projection operators,

    \begin{align*} \mathcal{T}_2 \bigg/ \bigg\{  P  \bigg[    P \bigg[ \mathcal{P}_2  \big( \vec{\sigma_3}  \big)      , u^{\prime}  \bigg] ,  u   \bigg]  P  \bigg[       P \bigg[ \mathcal{P}_1  \big( \vec{\sigma_1} \big)     , v^{\prime} \bigg] ,  v   \bigg] \bigg\} \end{align*}

    \begin{align*}= \mathcal{T}_2 \bigg/ \bigg\{  P \bigg[  \bigg\{           2^{-m} \underset{1 \leq j \leq m}{\prod } \big[ \textbf{I} +  \big( - 1 \big)^{u^{\prime}_{3,j}} g_{3,j} \big]     \big]          \bigg\}                             , u      \bigg]  P \bigg[          \bigg\{           2^{-m} \underset{1 \leq j^{\prime} \leq m}{\prod } \big[ \textbf{I} +  \big( - 1 \big)^{v^{\prime}_{3,j^{\prime}}} g_{3,j^{\prime}} \big]     \big]          \bigg\}          ,              v  \bigg]     \bigg] \bigg\}  \end{align*}

    \begin{align*} =        \bigg\{ 1 +  \big[ \mathcal{T}_1 \big]^{-1} + \underset{3 \leq k \leq 8}{\sum} \big[       \mathcal{T}^{\prime}_k  \big]^{-1}   \bigg\}^{-1}                    =                    \mathscr{T}_2                                                                                      , \end{align*}

\noindent corresponding to the second term normalized by a product of projection operators, 
    
    \begin{align*}  \mathcal{T}_3 \bigg/ \bigg\{  P  \bigg[    P \bigg[ \mathcal{P}_2  \big( \vec{\sigma_3}  \big)      , u^{\prime}  \bigg] ,  u   \bigg]  P  \bigg[       P \bigg[ \mathcal{P}_1  \big( \vec{\sigma_1} \big)     , v^{\prime} \bigg] ,  v   \bigg] \bigg\}  \end{align*}

    \begin{align*}  = \mathcal{T}_3 \bigg/ \bigg\{  P \bigg[  \bigg\{           2^{-m} \underset{1 \leq j \leq m}{\prod } \big[ \textbf{I} +  \big( - 1 \big)^{u^{\prime}_{3,j}} g_{3,j} \big]     \big]          \bigg\}                             , u      \bigg]  P \bigg[          \bigg\{           2^{-m} \underset{1 \leq j^{\prime} \leq m}{\prod } \big[ \textbf{I} +  \big( - 1 \big)^{v^{\prime}_{3,j^{\prime}}} g_{3,j^{\prime}} \big]     \big]          \bigg\}          ,              v  \bigg]     \bigg] \bigg\} \end{align*}

    \begin{align*}  =      \bigg\{      1 + \underset{1 \leq k \leq 2}{\sum} \big[       \mathcal{T}^{\prime}_k  \big]^{-1} + \underset{4 \leq k \leq 8}{\sum} \big[    \mathcal{T}^{\prime}_k    \big]^{-1}    \bigg\}^{-1}                     =                    \mathscr{T}_3                                     , \end{align*}

\noindent corresponding to the third term normalized by a product of projection operators,

    \begin{align*}  \mathcal{T}_4 \bigg/ \bigg\{   P  \bigg[    P \bigg[ \mathcal{P}_2  \big( \vec{\sigma_3}  \big)      , u^{\prime}  \bigg] ,  u   \bigg]  P  \bigg[       P \bigg[ \mathcal{P}_1  \big( \vec{\sigma_1} \big)     , v^{\prime} \bigg] ,  v   \bigg]  \bigg\}  \end{align*}

    \begin{align*} = \mathcal{T}_4 \bigg/ \bigg\{  P \bigg[  \bigg\{           2^{-m} \underset{1 \leq j \leq m}{\prod } \big[ \textbf{I} +  \big( - 1 \big)^{u^{\prime}_{3,j}} g_{3,j} \big]     \big]          \bigg\}                             , u      \bigg]  P \bigg[          \bigg\{           2^{-m} \underset{1 \leq j^{\prime} \leq m}{\prod } \big[ \textbf{I} +  \big( - 1 \big)^{v^{\prime}_{3,j^{\prime}}} g_{3,j^{\prime}} \big]     \big]          \bigg\}          ,              v  \bigg]     \bigg]  \bigg\} \end{align*}

    \begin{align*}  =            \bigg\{     1 + \underset{1 \leq k \leq 3}{\sum} \big[       \mathcal{T}^{\prime}_k  \big]^{-1} + \underset{5 \leq k \leq 8}{\sum} \big[    \mathcal{T}^{\prime}_k    \big]^{-1}    \bigg\}^{-1}                       =                    \mathscr{T}_4                                                                , \end{align*}

\noindent corresponding to the fourth term normalized by a product of projection operators, 
    
    \begin{align*}   \mathcal{T}_5 \bigg/ \bigg\{  P  \bigg[    P \bigg[ \mathcal{P}_2  \big( \vec{\sigma_3}  \big)      , u^{\prime}  \bigg] ,  u   \bigg]  P  \bigg[       P \bigg[ \mathcal{P}_1  \big( \vec{\sigma_1} \big)     , v^{\prime} \bigg] ,  v   \bigg]  \bigg\} \end{align*}

    \begin{align*} =  \mathcal{T}_5 \bigg/ \bigg\{ P \bigg[  \bigg\{           2^{-m} \underset{1 \leq j \leq m}{\prod } \big[ \textbf{I} +  \big( - 1 \big)^{u^{\prime}_{3,j}} g_{3,j} \big]     \big]          \bigg\}                             , u      \bigg]  P \bigg[          \bigg\{           2^{-m} \underset{1 \leq j^{\prime} \leq m}{\prod } \big[ \textbf{I} +  \big( - 1 \big)^{v^{\prime}_{3,j^{\prime}}} g_{3,j^{\prime}} \big]     \big]          \bigg\}          ,              v  \bigg]     \bigg] \bigg\}  \end{align*}

  \begin{align*} =    \bigg\{ 1+ \underset{1 \leq k \leq 4}{\sum} \big[           \mathcal{T}^{\prime}_k    \big]^{-1}  +    \underset{6 \leq k \leq 8}{\sum} \big[           \mathcal{T}^{\prime}_k    \big]^{-1}     \bigg\}^{-1}   =                    \mathscr{T}_5    , \end{align*}

\noindent corresponding to the fifth term normalized by a product of projection operators, 
    
    \begin{align*}   \mathcal{T}_6 \bigg/ \bigg\{  P  \bigg[    P \bigg[ \mathcal{P}_2  \big( \vec{\sigma_3}  \big)      , u^{\prime}  \bigg] ,  u   \bigg]  P  \bigg[       P \bigg[ \mathcal{P}_1  \big( \vec{\sigma_1} \big)     , v^{\prime} \bigg] ,  v   \bigg] \bigg\}  \end{align*}

    \begin{align*}  = \mathcal{T}_6 \bigg/ \bigg\{ P \bigg[  \bigg\{           2^{-m} \underset{1 \leq j \leq m}{\prod } \big[ \textbf{I} +  \big( - 1 \big)^{u^{\prime}_{3,j}} g_{3,j} \big]     \big]          \bigg\}                             , u      \bigg]  P \bigg[          \bigg\{           2^{-m} \underset{1 \leq j^{\prime} \leq m}{\prod } \big[ \textbf{I} +  \big( - 1 \big)^{v^{\prime}_{3,j^{\prime}}} g_{3,j^{\prime}} \big]     \big]          \bigg\}          ,              v  \bigg]     \bigg] \bigg\}  \end{align*}

    \begin{align*} =    \bigg\{ 1 + \underset{1 \leq k \leq 5}{\sum} \big[       \mathcal{T}^{\prime}_k  \big]^{-1} + \underset{7 \leq k \leq 8}{\sum} \big[    \mathcal{T}^{\prime}_k    \big]^{-1}   \bigg\}^{-1}      =                    \mathscr{T}_6  ,  \end{align*}

\noindent corresponding to the sixth term normalized by a product of projection operators,

    \begin{align*}  \mathcal{T}_7 \bigg/ \bigg\{  P  \bigg[    P \bigg[ \mathcal{P}_2  \big( \vec{\sigma_3}  \big)      , u^{\prime}  \bigg] ,  u   \bigg]  P  \bigg[       P \bigg[ \mathcal{P}_1  \big( \vec{\sigma_1} \big)     , v^{\prime} \bigg] ,  v   \bigg] \bigg\} \end{align*}

    \begin{align*} = \mathcal{T}_7 \bigg/ \bigg\{ P \bigg[  \bigg\{           2^{-m} \underset{1 \leq j \leq m}{\prod } \big[ \textbf{I} +  \big( - 1 \big)^{u^{\prime}_{3,j}} g_{3,j} \big]     \big]          \bigg\}                             , u      \bigg]  P \bigg[          \bigg\{           2^{-m} \underset{1 \leq j^{\prime} \leq m}{\prod } \big[ \textbf{I} +  \big( - 1 \big)^{v^{\prime}_{3,j^{\prime}}} g_{3,j^{\prime}} \big]     \big]          \bigg\}          ,              v  \bigg]     \bigg] \bigg\} \end{align*}

    \begin{align*} =              \bigg\{ 1 + \underset{1 \leq k \leq 6}{\sum} \big[ \mathcal{T}^{\prime}_k \big]^{-1} + \big[ \mathcal{T}^{\prime}_8  \big]^{-1} \bigg\}^{-1}                          =                    \mathscr{T}_7                                          , \end{align*}

\noindent corresponding to the seventh term normalized by a product of projection operators, 
    
    \begin{align*} \mathcal{T}_8 \bigg/ \bigg\{  P  \bigg[    P \bigg[ \mathcal{P}_2  \big( \vec{\sigma_3}  \big)      , u^{\prime}  \bigg] ,  u   \bigg]  P  \bigg[       P \bigg[ \mathcal{P}_1  \big( \vec{\sigma_1} \big)     , v^{\prime} \bigg] ,  v   \bigg] \bigg\} \end{align*}

    \begin{align*} = \mathcal{T}_8 \bigg/ \bigg\{ P \bigg[  \bigg\{           2^{-m} \underset{1 \leq j \leq m}{\prod } \big[ \textbf{I} +  \big( - 1 \big)^{u^{\prime}_{3,j}} g_{3,j} \big]     \big]          \bigg\}                             , u      \bigg]  P \bigg[          \bigg\{           2^{-m} \underset{1 \leq j^{\prime} \leq m}{\prod } \big[ \textbf{I} +  \big( - 1 \big)^{v^{\prime}_{3,j^{\prime}}} g_{3,j^{\prime}} \big]     \big]          \bigg\}          ,              v  \bigg]     \bigg] \bigg\}  \end{align*}

    \begin{align*} =        \bigg\{1 + \underset{1 \leq k \leq 7}{\sum}  \big[ \mathcal{T}^{\prime}_k \big]^{-1}  \bigg\}^{-1}   =                    \mathscr{T}_8                                ,
\end{align*} }

\noindent corresponding to the eighth term normalized by a product of projection operators, from which we conclude the argument. \boxed{}

\subsection{$\textbf{Proposition}$ \textit{5}}

\subsubsection{Description of the up to constants upper bound for the fidelity}

\noindent To obtain the up to constants upper bound of,

{\small   \begin{align*} 1 -    2^{-k + n h ( \frac{r}{n} ) + 5-\frac{3}{2} }  , 
\end{align*} }

\noindent holds for the fidelity between the action of the $\mathscr{V}^{\prime}$ real and $\mathscr{U}$ simulator isometries, as opposed to that of the $\mathscr{V}^{\prime}$ real and $\mathscr{U}$ real isometries, observe that the multiplicative factor,

{\small
\begin{align*}
2^{ \frac{5}{2} ( 5 - \frac{3}{2} ) + \mathrm{log}_2 \sqrt{C}} , 
\end{align*}}

\noindent appearing in the security threshold of the QKD protocol $\pi^{\prime}$ implies that there exists a constant for which,

{\small \begin{align*}
 \frac{d_{\textit{Diamond}}  \big( \mathscr{E}_{\textit{Ideal}} ,  \mathscr{E}_{\textit{Real}} \big)}{\sqrt{2^{-k + n h ( \frac{r}{n} ) + 5-\frac{3}{2} }    }}  \lesssim  1 .
\end{align*} }

\noindent The above up to constants estimate for the diamond norm between ideal and real resources, normalized in the square root of the security threshold of $\pi^{\prime}$ without the $\mathrm{log}_2 \sqrt{C}$ factor, is readily obtained through a relation to the trace distance shown in the below computations.

\subsubsection{Proof}

\noindent \textit{Proof of Proposition 5}. First, observe,

{\small \begin{align*}
  \textit{Fidelity} \bigg[ \mathscr{V}^{\prime}_{\textit{Real}} \mathscr{U}_{\textit{Simulator}} \ket{\varphi} \ket{\mathcal{L}} ,   \mathscr{V}^{\prime}_{\textit{Real}} \mathscr{U}_{\textit{Real}} \ket{\varphi} \ket{\mathcal{L}} \bigg]  \lesssim    1 -    2^{-k + n h ( \frac{r}{n} ) + 5-\frac{3}{2} }  , 
\end{align*} }

\noindent where,

{\small \begin{align*}
  \textit{Fidelity} \bigg[ \mathscr{V}^{\prime}_{\textit{Real}} \mathscr{U}_{\textit{Simulator}} \ket{\varphi} \ket{\mathcal{L}} ,   \mathscr{V}^{\prime}_{\textit{Real}} \mathscr{U}_{\textit{Real}} \ket{\varphi} \ket{\mathcal{L}} \bigg] = \bigg|   \bigg| \big[ \mathscr{V}^{\prime}_{\textit{Real}} \mathscr{U}_{\textit{Simulator}} \ket{\varphi} \ket{\mathcal{L}} \big] \big[   \mathscr{V}^{\prime}_{\textit{Real}} \mathscr{U}_{\textit{Real}} \ket{\varphi}  \ket{\mathcal{L}}  \big] \bigg| \bigg|_1     
\end{align*} }

\noindent Next, apply the previous result, \textbf{Proposition} \textit{4}, particularly through the observation that,

{\small \begin{align*}
\sqrt{1 - \big( 1 -  2^{-k + n h ( \frac{r}{n} ) + 5-\frac{3}{2} }    \big) } =   \sqrt{2^{-k + n h ( \frac{r}{n} ) + 5-\frac{3}{2} } } .             \end{align*} }

\noindent Moreover, with respect to the trace norm,

{\small \begin{align*}
   \mathrm{Tr} \bigg[  \big[ \mathscr{U}_{\textit{Simulator}} \mathscr{V}^{\prime}_{\textit{Real}} \mathscr{U}_{\textit{Ideal}}  -          \mathscr{U}_{\textit{Simulator}}  \mathscr{V}_{\textit{Simulator}}          \mathscr{V}_{\textit{Ideal}} \mathscr{U}_{\textit{Ideal}}       \big] \ket{\varphi} \ket{\mathcal{L}}     \bigg] \gtrsim    1 -   2^{-k + n h ( \frac{r}{n} ) + 5-\frac{3}{2} }         , 
\end{align*} } 

\noindent hence implying that,

{\small \begin{align*}
   \mathrm{Tr} \bigg[    \big[ \mathscr{U}_{\textit{Simulator}} \mathscr{V}^{\prime}_{\textit{Real}} \mathscr{U}_{\textit{Ideal}}  -          \mathscr{U}_{\textit{Simulator}}  \mathscr{V}_{\textit{Simulator}}          \mathscr{V}_{\textit{Ideal}} \mathscr{U}_{\textit{Ideal}}       \big] \ket{\varphi} \ket{\mathcal{L}}      \bigg]  \lesssim   2^{-k + n h ( \frac{r}{n} ) +  5-\frac{3}{2} }   \\  \Updownarrow \\    \mathrm{Tr} \bigg[    \mathscr{E}_{\textit{Real}} \big( \rho \big) - \mathscr{E}_{\textit{Ideal}} \big( \rho \big)    \bigg]  \lesssim   2^{-k + n h ( \frac{r}{n} ) +  3 (5-\frac{3}{2}) }   . 
\end{align*} }

\noindent As a result, one has that,

{\small \begin{align*}
  d_{\textit{Diamond}}  \big( \mathscr{E}_{\textit{Ideal}} ,  \mathscr{E}_{\textit{Real}} \big)  \lesssim  \sqrt{2^{-k + n h ( \frac{r}{n} ) + 5-\frac{3}{2} }    } , 
\end{align*} }

\noindent from which we conclude the argument. \boxed{}

\subsection{$\textbf{Proposition}$ \textit{6}}

\subsubsection{Description of computations for the purified state shared by Alice, Bob and Eve}

\noindent To show that,

{\small \begin{align*} \big[ \ket{\psi_{\alpha \beta^{\prime}}}_{AB^{\prime}} \big[ \ket{\psi_{\alpha^{\prime} \beta^{\prime\prime}}}_{A^{\prime} B^{\prime\prime}} \big]  \big] \otimes \big[ \ket{\gamma_{\alpha \beta^{\prime}}}_{E E^{\prime}}  \ket{\gamma_{\alpha^{\prime} \beta^{\prime\prime}}}_{E^{\prime} E^{\prime\prime}} \big]
, 
\end{align*} }

\noindent corresponding to the state shared by Alice, Bob and Eve, $\ket{\varphi}$, sums to the identity, it suffices to compute,

{\small

\begin{align*}
    \underset{\alpha \neq \beta^{\prime} \in \textbf{F}^n_2}{\sum}     \underset{\alpha^{\prime} \neq \beta^{\prime\prime} \in \textbf{F}^n_2}{\sum}  \big\{   \mathscr{V}_{\textit{Ideal}} \mathscr{U}_{\textit{Ideal}} \big\}  , \end{align*}

    \begin{align*} \underset{L \in \textit{Mat}_{\textbf{F}^n_2} [ \textit{2} \times \textit{2} ]}{\sum} \underset{\alpha \neq \beta^{\prime} \in \textbf{F}^n_2}{\sum}     \underset{\alpha^{\prime} \neq \beta^{\prime\prime} \in \textbf{F}^n_2}{\sum}   \big\{  \mathscr{W} \mathscr{V}_{\textit{Simulator} } \mathscr{U}_{\textit{Simulator} }   \big\}    , \\ 
\end{align*}

}

\noindent which then implies the desired result on the joint state between Alice, Bob and Eve, from rearrangement of the terms appearing in,

{\small

\begin{align*}
    \bigg\{  \underset{L \in \textit{Mat}_{\textbf{F}^n_2} [ \textit{2} \times \textit{2} ]}{\sum} \underset{\alpha \neq \beta^{\prime} \in \textbf{F}^n_2}{\sum}     \underset{\alpha^{\prime} \neq \beta^{\prime\prime} \in \textbf{F}^n_2}{\sum}   \big\{  \mathscr{W} \mathscr{V}_{\textit{Simulator} } \mathscr{U}_{\textit{Simulator} }  \big\}  \bigg\}   \ket{\varphi} \ket{\mathcal{L}}    . \\ 
\end{align*}

}

\noindent As a result with respect to the trace operation,

{\small \begin{align*} \mathrm{Tr}_{AB \textbf{L}^{\prime} S^{\prime} T^{\prime} U^{\prime}_A U^{\prime}_B  V^{\prime}_A V^{\prime}_B  W^{\prime}_A W^{\prime}_B       }     \big[ \ket{\cdot } \bra{\cdot }   \big]  .
\end{align*}
}

\noindent one computes, 

{\small \begin{align*} \mathrm{Tr}_{AB \textbf{L}^{\prime} S^{\prime} T^{\prime} U^{\prime}_A U^{\prime}_B  V^{\prime}_A V^{\prime}_B  W^{\prime}_A W^{\prime}_B       }     \big[ \ket{\tau_{\textit{Reject}}} \bra{\tau_{\textit{Reject}}}   \big]  .
\end{align*}
}

\subsubsection{Proof}

\noindent \textit{Proof of Proposition 6}. To argue that

\begin{align*}
       \mathscr{E}_{\textit{Ideal}} \big[        \ket{\varphi} \bra{\varphi}    \big]  , \end{align*}

\noindent takes the desired form, observe, 

{\small \begin{align*}
  \underline{\underset{\alpha \neq \beta^{\prime} \in \textbf{F}^n_2}{\sum}     \underset{\alpha^{\prime} \neq \beta^{\prime\prime} \in \textbf{F}^n_2}{\sum}  \big\{   \mathscr{V}_{\textit{Ideal}} \mathscr{U}_{\textit{Ideal}}} \big\}  =     \bigg\{  \underset{\alpha^{\prime} \neq \beta^{\prime\prime} \in \textbf{F}^n_2}{\sum}  \ket{\psi_{\alpha^{\prime}  \beta^{\prime\prime} }} \bra{\psi_{\alpha^{\prime} \beta^{\prime\prime} }}   \otimes   \frac{1}{\sqrt{2}} \bigg[   \ket{F ( \alpha^{\prime} + \beta^{\prime\prime} ) , F \big( \alpha^{\prime}  \big) }_{(S+T) S^{\prime}} \\ +   \ket{F ( \alpha^{\prime} + \beta^{\prime\prime} ) , F \big( \beta^{\prime\prime} \big) }_{(S+T) T^{\prime}}                \bigg]  \bigg\} \\ \times \bigg\{     \underset{\alpha \neq \beta^{\prime} \in \textbf{F}^n_2}{\sum}      \ket{\psi_{\alpha \beta^{\prime} }} \bra{\psi_{\alpha \beta^{\prime} }}  \otimes    \frac{1}{\sqrt{2}} \bigg[  \ket{F \big( \alpha \big) , F \big( \alpha + \beta^{\prime} \big)}_{S(S^{\prime} + T^{\prime})} +    \ket{F \big( \beta  \big) , F \big( \alpha + \beta^{\prime} \big)}_{T(S^{\prime} + T^{\prime})}    \bigg]           \bigg\} \end{align*}
  
  \begin{align*} =  \bigg[  \frac{1}{\sqrt{2}}    \bigg]^2              \underset{\alpha \neq \beta^{\prime} \in \textbf{F}^n_2 }{\underset{\alpha^{\prime} \neq \beta^{\prime\prime} \in \textbf{F}^n_2}{\sum}}  \bigg\{        \ket{\psi_{\alpha^{\prime} \beta^{\prime\prime}}}  \big[    \ket{\psi_{\alpha \beta^{\prime}}} \bra{\psi_{\alpha \beta^{\prime}}}   \big]    \bra{\psi_{\alpha^{\prime} \beta^{\prime\prime}}} \\   \otimes \bigg[ \ket{F \big( \alpha^{\prime} + \beta^{\prime\prime} \big) , F \big( \alpha^{\prime} \big) , F \big( \alpha \big) , F \big( \alpha + \beta^{\prime} \big) }_{(S+T) S S^{\prime} ( S^{\prime} + T^{\prime}) } \\ +      \ket{F \big( \alpha^{\prime} + \beta^{\prime} \big) , F \big( \alpha^{\prime} \big) ,  F \big( \beta \big) , F \big( \alpha + \beta^{\prime} \big)     }_{(S+T) T S^{\prime} ( S^{\prime} + T^{\prime}) } \\ + \ket{F \big( \alpha^{\prime} + \beta^{\prime\prime} \big) , F \big( \beta^{\prime\prime} \big) ,   F \big( \alpha \big) , F \big( \alpha + \beta^{\prime} \big)       }_{(S+T) S T^{\prime} (S+T^{\prime})}    \\ +   \ket{F \big( \alpha^{\prime} + \beta^{\prime\prime} \big) , F \big( \beta^{\prime\prime} \big) ,    F \big( \beta \big) , F \big( \alpha+ \beta^{\prime}    }_{(S+T)T T^{\prime} (S^{\prime} + T^{\prime}) }    \bigg]          \bigg\}            ,  \end{align*}

  \begin{align*} \underline{\underset{L \in \textit{Mat}_{\textbf{F}^n_2} [ \textit{2} \times \textit{2} ]}{\sum} \underset{\alpha \neq \beta^{\prime} \in \textbf{F}^n_2}{\sum}     \underset{\alpha^{\prime} \neq \beta^{\prime\prime} \in \textbf{F}^n_2}{\sum}   \big\{  \mathscr{W} \mathscr{V}_{\textit{Simulator} } \mathscr{U}_{\textit{Simulator} }   \big\}      }  =     \underset{x_A, x_B}{\underset{\alpha^{\prime} \neq \beta^{\prime\prime} \in \textbf{F}^n_2}{\underset{L \in \textit{Mat}_{\textbf{F}^n_2} [ \textit{2} \times \textit{2}]}{\sum}}}    \bigg\{   \bigg\{      \ket{L} \bra{L}_{\textbf{L}}  \otimes  \frac{1}{\sqrt{2}}  H^{\otimes 2n} \bigg[  \ket{x_A + x_B} \\ \times \bra{x_A }   +    \ket{x_A + x_B }    \bra{x_B}             \bigg]_{(A+B)A} H^{\otimes 2n}  \otimes       \ket{\mathscr{M} x_A,  \mathscr{M} x_A , \mathscr{M} x_B \mathscr{M} x_B     }_{W_A W^{\prime}_A W_B W^{\prime}_B}      \bigg\} \\ \times \bigg\{  \ket{L} \bra{L}_{\textbf{L}} \otimes \frac{1}{\sqrt{2}} \bigg\{  H^{\otimes 2n}   \bigg[       \ket{x_B + x_A} \bra{x_B}_{(B+A)B} + \ket{x_B + x_A} \bra{x_A }_{(B+A)A}      \bigg] H^{\otimes 2n}    \\ \otimes        \bigg[     \ket{\mathcal{P}_2  x_A,  \mathcal{P}_2 x_A , \mathcal{P}_2  x_B,  \mathcal{P}_2 x_B , \mathcal{P}_2  x_A  ,  \mathcal{P}_2  x_B  }_{U_A U^{\prime}_A U_B U^{\prime}_B U^{\prime\prime}_B U^{\prime\prime}_A }    \\ +    \ket{\mathcal{P}_2  x_A,  \mathcal{P}_2 x_A , \mathcal{P}_2  x_B,  \mathcal{P}_2 x_B , \mathcal{P}_2 x_B   ,  \mathcal{P}_2 x_A }_{U_A U^{\prime}_A U_B U^{\prime}_B U^{\prime\prime}_B U^{\prime\prime}_A }        \bigg]  \bigg\}                     \bigg\} \\  \times \bigg\{              \ket{L} \bra{L}_{\textbf{L}} \otimes \frac{1}{\sqrt{2}} \bigg\{  \bigg[       \ket{z_B + z_A} \bra{z_B}_{(B+A)B} + \ket{z_B + z_A} \bra{z_A }_{(B+A)A}      \bigg]  \\  \otimes      \bigg[  \ket{\mathcal{P}_1  z_A,  \mathcal{P}_1 z_A , \mathcal{P}_1  z_B,  \mathcal{P}_1 z_B ,  \mathcal{P}_1  z_A     , \mathcal{P}_1 z_B     }_{U_A U^{\prime}_A U_B U^{\prime}_B S S^{\prime} }   \\ +   \ket{\mathcal{P}_1  z_A,  \mathcal{P}_1 z_A , \mathcal{P}_1  z_B,  \mathcal{P}_1 z_B , \mathcal{P}_1 z_B   ,  \mathcal{P}_1  z_A }_{U_A U^{\prime}_A U_B U^{\prime}_B S S^{\prime} }       \bigg] \bigg\}               \bigg\} \bigg\}  \end{align*}

  \begin{align*} = \bigg[ \frac{1}{\sqrt{2}} \bigg]^3   \underset{x_A, x_B}{\underset{\alpha^{\prime} \neq \beta^{\prime\prime} \in \textbf{F}^n_2}{\underset{L \in \textit{Mat}_{\textbf{F}^n_2} [ \textit{2} \times \textit{2}]}{\sum}}}           \bigg\{     \ket{L} \bra{L}_{\textbf{L}}   \otimes          \bigg\{     \bigg[      \mathscr{P} \bigg[  \bigg[   \mathcal{P}_1 \vec{\sigma_3}    ,          \mathcal{P}_3 \vec{\sigma_1}     ,   L \bigg|_{\textit{Third column}}  \vec{\sigma_1}  \bigg]^{\mathrm{T}} ,      \big[ u_A, v_A, w_A \big]^{\mathrm{T}}                                         \bigg]     \bigg]_A \\ \otimes \bigg[ \mathscr{P} \bigg[   \bigg[   \mathcal{P}_1 \vec{\sigma_3}    ,          \mathcal{P}_3 \vec{\sigma_1}     ,   L \bigg|_{\textit{Third column}}  \vec{\sigma_1}  \bigg]^{\mathrm{T}} ,                     \big[ u_B , v_B , w_B \big]^{\mathrm{T}}                           \bigg]    \bigg]_B    \bigg\}                         \bigg\}  \\ \otimes \ket{u_A , u_A , u_B , u_B, v_A , v_A , v_B , v_B , w_A , w_A , w_B ,  w_B }_{U_A U^{\prime}_A U_B U^{\prime}_B V_A V^{\prime}_A V_B V^{\prime}_B   W_A W^{\prime}_A W_B W^{\prime}_B }            .
\end{align*} }

\noindent From the above expressions, to compute the product of,

{\small \begin{align*} \underset{L \in \textit{Mat}_{\textbf{F}^n_2} [ \textit{2} \times \textit{2} ]}{\sum} \underset{\alpha \neq \beta^{\prime} \in \textbf{F}^n_2}{\sum}     \underset{\alpha^{\prime} \neq \beta^{\prime\prime} \in \textbf{F}^n_2}{\sum}  \big\{   \mathscr{W} \mathscr{V}_{\textit{Simulator} } \mathscr{U}_{\textit{Simulator} } \big\}  , 
\end{align*} }

\noindent with,

{\small \begin{align*}
   \ket{\varphi} \ket{\mathcal{L}} ,
\end{align*} }

\noindent observe,

{\small \begin{align*}
     \bigg\{  \underset{L \in \textit{Mat}_{\textbf{F}^n_2} [ \textit{2} \times \textit{2} ]}{\sum} \underset{\alpha \neq \beta^{\prime} \in \textbf{F}^n_2}{\sum}     \underset{\alpha^{\prime} \neq \beta^{\prime\prime} \in \textbf{F}^n_2}{\sum}   \big\{  \mathscr{W} \mathscr{V}_{\textit{Simulator} } \mathscr{U}_{\textit{Simulator} }  \big\}  \bigg\}   \ket{\varphi} \ket{\mathcal{L}}  \equiv   \underset{L \in \textit{Mat}_{\textbf{F}^n_2} [ \textit{2} \times \textit{2} ]}{\sum}  \bigg\{   \underset{\alpha \neq \beta^{\prime} \in \textbf{F}^n_2}{\sum}     \underset{\alpha^{\prime} \neq \beta^{\prime\prime} \in \textbf{F}^n_2}{\sum}   \big\{  \mathscr{W} \end{align*}

     \begin{align*}  \times \mathscr{V}_{\textit{Simulator} } \mathscr{U}_{\textit{Simulator} }   \ket{\varphi} \ket{\mathcal{L}} \big\}   \bigg\} \\ \\    \equiv   \underset{L \in \textit{Mat}_{\textbf{F}^n_2} [ \textit{2} \times \textit{2} ]}{\sum}  \underset{\alpha \neq \beta^{\prime} \in \textbf{F}^n_2}{\sum}      \bigg\{   \underset{\alpha^{\prime} \neq \beta^{\prime\prime} \in \textbf{F}^n_2}{\sum}  \big\{   \mathscr{W} \mathscr{V}_{\textit{Simulator} } \mathscr{U}_{\textit{Simulator} }   \ket{\varphi} \ket{\mathcal{L}} \big\}   \bigg\}       , 
    \end{align*} }

    \noindent equals,

{\small \begin{align*}
  \bigg\{ \bigg[ \frac{1}{\sqrt{2}} \bigg]^3   \underset{x_A, x_B}{\underset{\alpha^{\prime} \neq \beta^{\prime\prime} \in \textbf{F}^n_2}{\underset{L \in \textit{Mat}_{\textbf{F}^n_2} [ \textit{2} \times \textit{2}]}{\sum}}}           \bigg\{     \ket{L} \bra{L}_{\textbf{L}}   \otimes          \bigg\{     \bigg[      \mathscr{P} \bigg[  \bigg[   \mathcal{P}_1 \vec{\sigma_3}    ,          \mathcal{P}_3 \vec{\sigma_1}     ,   L \bigg|_{\textit{Third column}}  \vec{\sigma_1}  \bigg]^{\mathrm{T}} ,      \big[ u_A, v_A, w_A \big]^{\mathrm{T}}                                         \bigg]     \bigg]_A \\ \otimes \bigg[ \mathscr{P} \bigg[   \bigg[   \mathcal{P}_1 \vec{\sigma_3}    ,          \mathcal{P}_3 \vec{\sigma_1}     ,   L \bigg|_{\textit{Third column}}  \vec{\sigma_1}  \bigg]^{\mathrm{T}} ,                     \big[ u_B , v_B , w_B \big]^{\mathrm{T}}                           \bigg]    \bigg]_B    \bigg\}                         \bigg\}  \end{align*}

          \begin{align*}  \otimes \ket{u_A , u_A , u_B , u_B, v_A , v_A , v_B , v_B , w_A , w_A , w_B ,  w_B }_{U_A U^{\prime}_A U_B U^{\prime}_B V_A V^{\prime}_A V_B V^{\prime}_B   W_A W^{\prime}_A W_B W^{\prime}_B }           \bigg\} \\ \times \ket{\varphi} \ket{\mathcal{L}} \\ \\ =   \bigg[ \frac{1}{\sqrt{2}} \bigg]^3   \underset{x_A, x_B}{\underset{\alpha^{\prime} \neq \beta^{\prime\prime} \in \textbf{F}^n_2}{\underset{L \in \textit{Mat}_{\textbf{F}^n_2} [ \textit{2} \times \textit{2}]}{\sum}}}  \bigg\{ \sqrt{p_L} \ket{L} \bra{L}_{\textbf{L}}  \otimes \mathscr{P} \bigg[      \bigg[   \mathcal{P}_1 \vec{\sigma_3}    ,          \mathcal{P}_3 \vec{\sigma_1}     ,   L \bigg|_{\textit{Third column}}  \vec{\sigma_1}  \bigg]^{\mathrm{T}}     ,   \big[ u_A, v_A, w_A \big]^{\mathrm{T}}      \bigg]          \\ \times    \bigg[ \big[ \ket{\gamma_{\alpha^{\prime} \beta^{\prime\prime}}} \big]_{EE^{\prime}} \big[   \ket{\gamma_{\alpha \beta^{\prime}}} \bra{\gamma_{\alpha \beta^{\prime}}} \big]_{AB} \big[ \bra{\gamma_{\alpha^{\prime} \beta^{\prime\prime}}} \big]_{EE^{\prime}} \bigg]                 \\     \otimes   \ket{u_A , u_A ,  u_A + \mathcal{P}_1  \alpha , u_A + \mathcal{P}_1    \alpha  }_{U_A U^{\prime}_A U_B U^{\prime}_B}       \otimes      \ket{v_A , v_A ,  v_A + \mathcal{P}_1  \alpha , v_A + \mathcal{P}_1    \alpha  }_{V_A V^{\prime}_A V_B V^{\prime}_B} \\ \otimes \ket{w_A , w_A ,  w_A + \mathcal{P}_1  \alpha , w_A + \mathcal{P}_1    \alpha  }_{W_A W^{\prime}_A W_B W^{\prime}_B}           \otimes    \frac{1}{\sqrt{2}} \bigg\{  \bigg[  \ket{F \big( \alpha \big) , F \big( \alpha + \beta^{\prime} \big)}_{S(S^{\prime} + T^{\prime})} \\  +    \ket{F \big( \beta  \big) , F \big( \alpha + \beta^{\prime} \big)}_{T(S^{\prime} + T^{\prime})}    \bigg]      \bigg[   \ket{F ( \alpha^{\prime} + \beta^{\prime\prime} ) , F \big( \alpha^{\prime}  \big) }_{(S+T) S^{\prime}} \\ +   \ket{F ( \alpha^{\prime} + \beta^{\prime\prime} ) , F \big( \beta^{\prime\prime} \big) }_{(S+T) T^{\prime}}                \bigg]  \bigg\}    . \tag{\textit{Tensor Product Purification}}
\end{align*}}

\noindent Moreover,

{\small \begin{align*}
  (\textit{Tensor Product Purification})  =  \bigg[ \frac{1}{\sqrt{2}} \bigg]^3   \underset{x_A, x_B}{\underset{\alpha^{\prime} \neq \beta^{\prime\prime} \in \textbf{F}^n_2}{\underset{L \in \textit{Mat}_{\textbf{F}^n_2} [ \textit{2} \times \textit{2}]}{\sum}}}  \bigg\{ \sqrt{p_L} \ket{L} \bra{L}_{\textbf{L}}  \otimes \mathscr{P} \bigg[      \bigg[   \mathcal{P}_1 \vec{\sigma_3}    ,          \mathcal{P}_3 \vec{\sigma_1}     \\ ,   L \bigg|_{\textit{Third column}}  \vec{\sigma_1}  \bigg]^{\mathrm{T}}      ,   \big[ u_A, v_A, w_A \big]^{\mathrm{T}}      \bigg]           \bigg[ \big[ \ket{\gamma_{\alpha^{\prime} \beta^{\prime\prime}}} \big]_{EE^{\prime}} \big[   \ket{\gamma_{\alpha \beta^{\prime}}} \bra{\gamma_{\alpha \beta^{\prime}}} \big]_{AB} \big[ \bra{\gamma_{\alpha^{\prime} \beta^{\prime\prime}}} \big]_{EE^{\prime}} \bigg]       \end{align*}

     \begin{align*}   \otimes   \ket{u_A , u_A ,  u_A + \mathcal{P}_1  \alpha , u_A + \mathcal{P}_1    \alpha  }_{U_A U^{\prime}_A U_B U^{\prime}_B}       \otimes      \ket{v_A , v_A ,  v_A + \mathcal{P}_1  \alpha , v_A + \mathcal{P}_1    \alpha  }_{V_A V^{\prime}_A V_B V^{\prime}_B} \\ \otimes \ket{w_A , w_A ,  w_A + \mathcal{P}_1  \alpha , w_A + \mathcal{P}_1    \alpha  }_{W_A W^{\prime}_A W_B W^{\prime}_B}              \\  \otimes    \frac{1}{\sqrt{2}} \bigg\{   \ket{F \big( \alpha \big) , F \big( \alpha + \beta^{\prime} \big) , F \big( \alpha^{\prime} + \beta^{\prime\prime} \big) , F \big( \alpha^{\prime} \big) }_{S(S+T)(S^{\prime} + T^{\prime})S^{\prime}} \\     +    \ket{F \big( \alpha \big) , F \big( \alpha + \beta^{\prime} \big) , F \big( \alpha^{\prime} + \beta^{\prime\prime } \big) ,     F \big( \beta^{\prime\prime} \big) }_{S(S+T)(S^{\prime} + T^{\prime}) T^{\prime}} \\     + \ket{F \big( \beta \big) , F \big( \alpha + \beta^{\prime} \big) , F \big( \alpha^{\prime} + \beta^{\prime\prime} \big) , F \big( \alpha + \beta^{\prime} \big) }_{T(S +T)(S^{\prime} + T^{\prime})S^{\prime}}    \end{align*}

  \begin{align*}      +  \ket{F \big( \beta \big) , F \big( \alpha + \beta^{\prime} \big) ,   F \big( \alpha^{\prime} + \beta^{\prime\prime} \big)          ,       F \big( \beta^{\prime\prime} \big)        }_{T(S+T)(S^{\prime} + T^{\prime} ) T^{\prime}}             \bigg\}     \bigg\} \\ \\ =              \bigg[ \frac{1}{\sqrt{2}} \bigg]^3   \underset{x_A, x_B}{\underset{\alpha^{\prime} \neq \beta^{\prime\prime} \in \textbf{F}^n_2}{\underset{L \in \textit{Mat}_{\textbf{F}^n_2} [ \textit{2} \times \textit{2}]}{\sum}}}  \bigg\{ \sqrt{p_L} \ket{L} \bra{L}_{\textbf{L}}  \otimes \mathscr{P} \bigg[      \bigg[   \mathcal{P}_1 \vec{\sigma_3}    ,          \mathcal{P}_3 \vec{\sigma_1}      ,   L \bigg|_{\textit{Third column}}  \vec{\sigma_1}  \bigg]^{\mathrm{T}}      ,   \big[ u_A , v_A, w_A \big]^{\mathrm{T}}      \bigg]  \\ \times          \bigg[ \big[ \ket{\gamma_{\alpha^{\prime} \beta^{\prime\prime}}} \big]_{EE^{\prime}} \big[   \ket{\gamma_{\alpha \beta^{\prime}}} \bra{\gamma_{\alpha \beta^{\prime}}} \big]_{AB} \big[ \bra{\gamma_{\alpha^{\prime} \beta^{\prime\prime}}} \big]_{EE^{\prime}} \bigg]                 \\     \otimes   \ket{u_A , u_A ,  u_A + \mathcal{P}_1  \alpha , u_A + \mathcal{P}_1    \alpha  }_{U_A U^{\prime}_A U_B U^{\prime}_B}       \otimes      \ket{v_A , v_A ,  v_A + \mathcal{P}_1  \alpha , v_A + \mathcal{P}_1    \alpha  }_{V_A V^{\prime}_A V_B V^{\prime}_B} \\ \otimes \ket{w_A , w_A ,  w_A + \mathcal{P}_1  \alpha , w_A + \mathcal{P}_1    \alpha  }_{W_A W^{\prime}_A W_B W^{\prime}_B}          \\  \otimes    \frac{1}{\sqrt{2}}  \ket{F \big( \alpha \big) , F \big( \alpha + \beta^{\prime} \big) , F \big( \alpha^{\prime} + \beta^{\prime\prime } \big) ,     F \big( \beta^{\prime\prime} \big) }_{S(S+T)(S^{\prime} + T^{\prime}) T^{\prime}}       \\ +  \bigg[ \frac{1}{\sqrt{2}} \bigg]^3   \underset{x_A, x_B}{\underset{\alpha^{\prime} \neq \beta^{\prime\prime} \in \textbf{F}^n_2}{\underset{L \in \textit{Mat}_{\textbf{F}^n_2} [ \textit{2} \times \textit{2}]}{\sum}}}  \bigg\{ \sqrt{p_L} \ket{L} \bra{L}_{\textbf{L}}  \otimes \mathscr{P} \bigg[      \bigg[   \mathcal{P}_1 \vec{\sigma_3}    ,          \mathcal{P}_3 \vec{\sigma_1}    \\  ,   L \bigg|_{\textit{Third column}}  \vec{\sigma_1}  \bigg]^{\mathrm{T}}      ,   \big[ u_A, v_A, w_A \big]^{\mathrm{T}}      \bigg]           \bigg[ \big[ \ket{\gamma_{\alpha^{\prime} \beta^{\prime\prime}}} \big]_{EE^{\prime}} \big[   \ket{\gamma_{\alpha \beta^{\prime}}} \bra{\gamma_{\alpha \beta^{\prime}}} \big]_{AB} \big[ \bra{\gamma_{\alpha^{\prime} \beta^{\prime\prime}}} \big]_{EE^{\prime}} \bigg]                 \\     \otimes   \ket{u_A , u_A ,  u_A + \mathcal{P}_1  \alpha , u_A + \mathcal{P}_1    \alpha  }_{U_A U^{\prime}_A U_B U^{\prime}_B}       \otimes      \ket{v_A , v_A ,  v_A + \mathcal{P}_1  \alpha , v_A + \mathcal{P}_1    \alpha  }_{V_A V^{\prime}_A V_B V^{\prime}_B} \\ \otimes \ket{w_A , w_A ,  w_A + \mathcal{P}_1  \alpha , w_A + \mathcal{P}_1    \alpha  }_{W_A W^{\prime}_A W_B W^{\prime}_B}          \\  \otimes    \frac{1}{\sqrt{2}}       \ket{F \big( \alpha \big) , F \big( \alpha + \beta^{\prime} \big) , F \big( \alpha^{\prime} + \beta^{\prime\prime } \big) ,     F \big( \beta^{\prime\prime} \big) }_{S(S+T)(S^{\prime} + T^{\prime}) T^{\prime}}       \\     +  \bigg[ \frac{1}{\sqrt{2}} \bigg]^3   \underset{x_A, x_B}{\underset{\alpha^{\prime} \neq \beta^{\prime\prime} \in \textbf{F}^n_2}{\underset{L \in \textit{Mat}_{\textbf{F}^n_2} [ \textit{2} \times \textit{2}]}{\sum}}}  \bigg\{ \sqrt{p_L} \ket{L} \bra{L}_{\textbf{L}}  \otimes \mathscr{P} \bigg[      \bigg[   \mathcal{P}_1 \vec{\sigma_3}    ,          \mathcal{P}_3 \vec{\sigma_1}      ,   L \bigg|_{\textit{Third column}}  \vec{\sigma_1}  \bigg]^{\mathrm{T}}      ,   \big[ u_A, v_A, w_A \big]^{\mathrm{T}}      \bigg] \\ \times           \bigg[ \big[ \ket{\gamma_{\alpha^{\prime} \beta^{\prime\prime}}} \big]_{EE^{\prime}} \big[   \ket{\gamma_{\alpha \beta^{\prime}}} \bra{\gamma_{\alpha \beta^{\prime}}} \big]_{AB} \big[ \bra{\gamma_{\alpha^{\prime} \beta^{\prime\prime}}} \big]_{EE^{\prime}} \bigg]      \\ \otimes   \ket{u_A , u_A ,  u_A + \mathcal{P}_1  \alpha , u_A + \mathcal{P}_1    \alpha  }_{U_A U^{\prime}_A U_B U^{\prime}_B}       \otimes      \ket{v_A , v_A ,  v_A + \mathcal{P}_1  \alpha , v_A + \mathcal{P}_1    \alpha  }_{V_A V^{\prime}_A V_B V^{\prime}_B} \end{align*}

  \begin{align*}    \otimes \ket{w_A , w_A ,  w_A + \mathcal{P}_1  \alpha , w_A + \mathcal{P}_1    \alpha  }_{W_A W^{\prime}_A W_B W^{\prime}_B}        \\  \otimes    \frac{1}{\sqrt{2}}  \ket{F \big( \beta \big) , F \big( \alpha + \beta^{\prime} \big) , F \big( \alpha^{\prime} + \beta^{\prime\prime} \big) , F \big( \alpha + \beta^{\prime} \big) }_{T(S +T)(S^{\prime} + T^{\prime})S^{\prime}} \\     \\ +  \bigg[ \frac{1}{\sqrt{2}} \bigg]^3   \underset{x_A, x_B}{\underset{\alpha^{\prime} \neq \beta^{\prime\prime} \in \textbf{F}^n_2}{\underset{L \in \textit{Mat}_{\textbf{F}^n_2} [ \textit{2} \times \textit{2}]}{\sum}}}  \bigg\{ \sqrt{p_L} \ket{L} \bra{L}_{\textbf{L}}  \otimes \mathscr{P} \bigg[      \bigg[   \mathcal{P}_1 \vec{\sigma_3}    ,          \mathcal{P}_3 \vec{\sigma_1}      ,   L \bigg|_{\textit{Third column}}  \vec{\sigma_1}  \bigg]^{\mathrm{T}}      ,   \big[ u_A, v_A, w_A \big]^{\mathrm{T}}      \bigg]                    \\     \times    \bigg[ \big[ \ket{\gamma_{\alpha^{\prime} \beta^{\prime\prime}}} \big]_{EE^{\prime}} \big[   \ket{\gamma_{\alpha \beta^{\prime}}} \bra{\gamma_{\alpha \beta^{\prime}}} \big]_{AB} \big[ \bra{\gamma_{\alpha^{\prime} \beta^{\prime\prime}}} \big]_{EE^{\prime}} \bigg]                 \\     \otimes   \ket{u_A , u_A ,  u_A + \mathcal{P}_1  \alpha , u_A + \mathcal{P}_1    \alpha  }_{U_A U^{\prime}_A U_B U^{\prime}_B}    \\    \otimes      \ket{v_A , v_A ,  v_A + \mathcal{P}_1  \alpha , v_A + \mathcal{P}_1    \alpha  }_{V_A V^{\prime}_A V_B V^{\prime}_B} \\ \otimes \ket{w_A , w_A ,  w_A + \mathcal{P}_1  \alpha , w_A + \mathcal{P}_1    \alpha  }_{W_A W^{\prime}_A W_B W^{\prime}_B}          \\  \otimes    \frac{1}{\sqrt{2}}       \ket{F \big( \beta \big) , F \big( \alpha + \beta^{\prime} \big) ,   F \big( \alpha^{\prime} + \beta^{\prime\prime} \big)          ,       F \big( \beta^{\prime\prime} \big)        }_{T(S+T)(S^{\prime} + T^{\prime} ) T^{\prime}}     \bigg\}  \\ \tag{\textit{Normalized Purification}} .
\end{align*} }

\noindent Altogether, the desired expression for,

\begin{align*}
       \mathscr{E}_{\textit{Ideal}} \big[        \ket{\varphi} \bra{\varphi}    \big]  , \end{align*}

\noindent follows from the fact that, for,

{\small \begin{align*}
  \sigma^{\textit{Reject}}_{LEE^{\prime} E^{\prime} E^{\prime\prime}S T U_A V_A W_A U_B V_B W_B   } = \mathrm{Tr}_{AB \textbf{L}^{\prime} S^{\prime} T^{\prime} U^{\prime}_A U^{\prime}_B  V^{\prime}_A V^{\prime}_B  W^{\prime}_A W^{\prime}_B       }     \big[ \ket{\tau_{\textit{Reject}}} \bra{\tau_{\textit{Reject}}}   \big]          ,  \end{align*}

\begin{align*} \mathscr{E}_{\textit{Ideal}} \big[        \ket{\varphi} \bra{\varphi}    \big] =   \mathrm{Tr}_{AB \textbf{L}^{\prime} S^{\prime} T^{\prime} U^{\prime}_A U^{\prime}_B  V^{\prime}_A V^{\prime}_B  W^{\prime}_A W^{\prime}_B       }     \big[ \ket{\tau_{\textit{Reject}}} \bra{\tau_{\textit{Reject}}}          \big]  + \mathrm{Tr}_{AB \textbf{L}^{\prime} S^{\prime} T^{\prime} U^{\prime}_A U^{\prime}_B  V^{\prime}_A V^{\prime}_B  W^{\prime}_A W^{\prime}_B       }     \big[ \ket{\tau_{\textit{Accept}}} \\ \times \bra{\tau_{\textit{Accept}}}         \big]        ,
\end{align*} }

\noindent where $\ket{\tau_{\textit{Accept}}}$, and $\ket{\tau_{\textit{Reject}}}$, are two vectors, the remaining desired expression for the outer product follows from expanding,

{\small \begin{align*}
     \mathrm{Tr}_{AB \textbf{L}^{\prime} S^{\prime} T^{\prime} U^{\prime}_A U^{\prime}_B  V^{\prime}_A V^{\prime}_B  W^{\prime}_A W^{\prime}_B       }     \big[ \ket{\tau_{\textit{Accept}}} \bra{\tau_{\textit{Accept}}}         \big]       , \\ 
\end{align*} }

\noindent by substituting in the superposition previously obtained in the normalized state after purification (\textit{Normalized Purification} expressed above), from which we conclude the argument. \boxed{}

\subsection{$\textbf{Theorem}$ \textit{1}}

\subsubsection{Proof} 

\noindent \textit{Proof of Theorem 1}. A straightforward application of a probabilistic union bound, along with $\Theta$ provided, which is a counterpart to $\epsilon$ in \textbf{Definition} \textit{4} of {\color{blue}[40]} implies the desired result, from which we conclude the argument. \boxed{}

\subsection{$\textbf{Theorem}$ \textit{2}}

\subsubsection{Description of the computations associated with the Hamming distance, specifically through the number of bit flip, and phase flip, errors over another joint state shared by Alice, Bob and Eve}

\noindent To demonstrate that the probability that $\pi^{\prime}$ accepts $\rho_{ABE}$, one demonstrates that, with respect to the probability measures over $L$,

{\small

\begin{align*}
\textbf{P}_L \big[ \cdot \big]     , 
\end{align*}

}

\noindent one has that,

{\small

\begin{align*}
 \textbf{P}_L \bigg[  \textit{The number of bit flip and phase flip errors between } \rho_{ABE}    \textit{ and } \underset{\alpha^{\prime} \neq \beta^{\prime\prime} \in \textbf{F}^n_2}{\underset{\alpha \neq \beta^{\prime} \in \textbf{F}^n_2}{\sum}}  \ket{\gamma_{\alpha^{\prime} \beta^{\prime\prime}}} \big[   \ket{\gamma_{\alpha \beta^{\prime}}} \\ \times   \bra{\gamma_{\alpha \beta^{\prime}}} \big] \bra{\gamma_{\alpha^{\prime} \beta^{\prime\prime}}}    \textit{ equals } 2^{-\frac{k}{2} + n \frac{h}{2} ( \frac{r}{n} ) + \frac{3}{2} ( 5 - \frac{3}{2} )+ \mathrm{log}_2 C^{\frac{1}{4}} }  \bigg] . 
\end{align*}
}

\noindent The fact that the above probability occuring is dependent upon the factor of $ 2^{-\frac{k}{2} + n \frac{h}{2} ( \frac{r}{n} ) + \frac{3}{2} ( 5 - \frac{3}{2} )+ \mathrm{log}_2 C^{\frac{1}{4}} }$ is important to keep in mind as a comparison to the security threshold $ 2^{-\frac{k}{2} + n \frac{h}{2} ( \frac{r}{n} ) + \frac{5}{2} ( 5 - \frac{3}{2} ) + \mathrm{log}_2 \sqrt{C}} $. Indeed, the ratio,

{\small

\begin{align*}
\frac{2^{-\frac{k}{2} + n \frac{h}{2} ( \frac{r}{n} ) + \frac{5}{2} ( 5 - \frac{3}{2} ) + \mathrm{log}_2 \sqrt{C}}}{2^{-\frac{k}{2} + n \frac{h}{2} ( \frac{r}{n} ) + \frac{3}{2} ( 5 - \frac{3}{2} )+ \mathrm{log}_2 C^{\frac{1}{4}} }}   , 
\end{align*}

}

\noindent demonstrates that the difference between $\mathrm{log}_2 \sqrt{C}$ and $\mathrm{log}_2 \sqrt{\sqrt{C}}$, respectively, demonstrates how the previously obtained expansions for,

 {\small

 \begin{align*} \underset{\alpha \neq \beta^{\prime} \in \textbf{F}^n_2}{\sum}        \ket{\psi_{\alpha \beta^{\prime} }} \bra{\psi_{\alpha \beta^{\prime} }}  , \\  \underset{\alpha^{\prime} \neq \beta^{\prime\prime} \in \textbf{F}^n_2}{\sum}    \ket{\psi_{\alpha^{\prime}  \beta^{\prime\prime} }} \bra{\psi_{\alpha^{\prime} \beta^{\prime\prime} }}   , 
\end{align*}

}

\noindent corresponding to taking the entangled outer products in the Bell basis are related to the action of a suitably defined projection operator, $\Pi_{n,r}$.

\subsubsection{Proof}

\noindent \textit{Proof of Theorem 2}. For Eve's vectors which satisfy,

{\small \begin{align*}
\underset{\alpha^{\prime} \neq \beta^{\prime\prime} \in \textbf{F}^n_2}{\underset{\alpha \neq \beta^{\prime} \in \textbf{F}^n_2}{\sum}}  \bigg[   \big[ \ket{\gamma_{\alpha^{\prime} \beta^{\prime\prime}}} \big]_{EE^{\prime}} \big[   \ket{\gamma_{\alpha \beta^{\prime}}} \bra{\gamma_{\alpha \beta^{\prime}}} \big]_{AB} \big[ \bra{\gamma_{\alpha^{\prime} \beta^{\prime\prime}}} \big]_{EE^{\prime}} \bigg]    = \textbf{I} ,
\end{align*} }

\noindent the final desired result, the $ 2^{-\frac{k}{2} + n \frac{h}{2} ( \frac{r}{n} ) + \frac{5}{2} ( 5 - \frac{3}{2} ) + \mathrm{log}_2 \sqrt{C}} $ security, follows from the observation that, given an input state $\rho_{ABE}$, which takes the form,

\begin{align*}
  \rho_{ABE} = \mathrm{Tr}_{E^{\prime}}      \bigg[ \big[ \ket{\varphi} \big[ \ket{\varphi} \bra{\varphi}_{ABEE^{\prime}}   \big]  \bra{\varphi} \big]_{ABEE^{\prime}}      \bigg]      , 
\end{align*}

\noindent the probability,

{\small \begin{align*}
 \textbf{P}_L \big[   \pi^{\prime} \big( n , k ,r \big) \textit{ accepts } \rho_{ABE}    \big]  =  \textbf{P}_L \big[  \textit{The Hamming distance between } \rho_{ABE}   \textit{ and } \Pi_{n,r} \\ \times  \big[ \Pi_{n,r} \rho_{AB} \Pi_{n,r} \big] \Pi_{n,r} \textit{ equals } 2^{-\frac{k}{2} + n \frac{h}{2} ( \frac{r}{n} ) + \frac{3}{2} ( 5 - \frac{3}{2} ) + \mathrm{log}_2 C^{\frac{1}{4}} }  \big] \end{align*}

 \begin{align*} =  \textbf{P}_L \bigg[  \textit{The Hamming distance between } \rho_{ABE}   \textit{ and } \underset{\alpha^{\prime} \neq \beta^{\prime\prime} \in \textbf{F}^n_2}{\underset{\alpha \neq \beta^{\prime} \in \textbf{F}^n_2}{\sum}}  \ket{\gamma_{\alpha^{\prime} \beta^{\prime\prime}}} \big[   \ket{\gamma_{\alpha \beta^{\prime}}} \bra{\gamma_{\alpha \beta^{\prime}}} \big] \bra{\gamma_{\alpha^{\prime} \beta^{\prime\prime}}}   \textit{ equals } \\ 2^{-\frac{k}{2} + n \frac{h}{2} ( \frac{r}{n} ) + \frac{3}{2} ( 5 - \frac{3}{2} )+ \mathrm{log}_2 C^{\frac{1}{4}} }  \bigg] \end{align*}

 \begin{align*}  =  \textbf{P}_L \bigg[  \textit{The number of bit flip and phase flip errors between } \rho_{ABE}    \textit{ and } \underset{\alpha^{\prime} \neq \beta^{\prime\prime} \in \textbf{F}^n_2}{\underset{\alpha \neq \beta^{\prime} \in \textbf{F}^n_2}{\sum}}  \ket{\gamma_{\alpha^{\prime} \beta^{\prime\prime}}} \big[   \ket{\gamma_{\alpha \beta^{\prime}}} \\ \times   \bra{\gamma_{\alpha \beta^{\prime}}} \big] \bra{\gamma_{\alpha^{\prime} \beta^{\prime\prime}}}    \textit{ equals } 2^{-\frac{k}{2} + n \frac{h}{2} ( \frac{r}{n} ) + \frac{3}{2} ( 5 - \frac{3}{2} )+ \mathrm{log}_2 C^{\frac{1}{4}} }  \bigg] \end{align*}

 \begin{align*}   =   2^{-\frac{k}{2} + n \frac{h}{2} ( \frac{r}{n} ) + \frac{5}{2} ( 5 - \frac{3}{2} )+ \mathrm{log}_2 \sqrt{C}  }   , 
\end{align*}} 

\noindent for the projection $\Pi_{n,r}$,

\begin{align*}
  \textit{Projection onto the systems AB with at most r bit flip, and r phase flip, errors} = \Pi_{n,r} , 
\end{align*}

\noindent from which we conclude the argument. \boxed{}

\section{Conclusion}

In this work we demonstrated how a novel class of two-universal QKD hashing protocols is $2^{ \frac{5}{2} ( 5 - \frac{3}{2} ) + \mathrm{log}_2 \sqrt{C}}$ less secure, for some strictly positive constant $C$. A strictly positive factor in the security parameter is expected; the computational complexity of functions $g_1$ and $g_2$ of $\mathcal{P}_1$ and $\mathcal{P}_2$, respectively, increases the probability that Eve can determine qubits of the secret key. While many papers within the Quantum Information processing literature describe unconditional security guarantees, characterizing conditional security guarantees can be easier to formalize. The first component of the factor appearing in the exponent of $2$ is dependent upon terms introduced in Ideal, Real, and Simulator isometries. As transformations which preserve length, parts of the isometry were introduced to replace idealized values of functions of the set of errors which were first computed in {\color{blue}[40]}. As such, computations associated isometries introduced in this work are related to the collection of states,

\begin{align*}
 \ket{\mathcal{P}_1  z_A,  \mathcal{P}_1 z_A , \mathcal{P}_1  z_B,  \mathcal{P}_1 z_B ,  g_1 \big( \mathcal{P}_1 , \mathcal{P}_1 \big( z_A + z_B \big)  \big)    , \mathcal{P}_1 z_B     }_{U_A U^{\prime}_A U_B U^{\prime}_B S U^{\prime\prime}_B}      , \\ \\    \ket{\mathcal{P}_1  z_A,  \mathcal{P}_1 z_A , \mathcal{P}_1  z_B,  \mathcal{P}_1 z_B , \mathcal{P}_1 z_B   ,  g_1 \big( \mathcal{P}_1 , \mathcal{P}_1 \big( z_A + z_B \big) \big)              }_{U_A U^{\prime}_A U_B U^{\prime}_B  U_B U^{\prime\prime}_B  S }       , \\ \\    \ket{\mathcal{P}_1  z_A,  \mathcal{P}_1 z_A , \mathcal{P}_1  z_B,  \mathcal{P}_1 z_B ,  \mathcal{P}_1  z_A     , \mathcal{P}_1 z_B     }_{U_A U^{\prime}_A U_B U^{\prime}_B U^{\prime\prime}_A U^{\prime\prime}_B}          ,  \\ \\   \ket{\mathcal{P}_1  z_A,  \mathcal{P}_1 z_A , \mathcal{P}_1  z_B,  \mathcal{P}_1 z_B , \mathcal{P}_1 z_B   ,  \mathcal{P}_1  z_A }_{U_A U^{\prime}_A U_B U^{\prime}_B U^{\prime\prime}_B U^{\prime\prime}_A  }        , \\ \\ \ket{\mathcal{P}_2  x_A,  \mathcal{P}_2 x_A , \mathcal{P}_2  x_B,  \mathcal{P}_2 x_B ,  g_2 \big( \mathcal{P}_1 , \mathcal{P}_2 \big( x_A + x_B \big) \big)  ,  \mathcal{P}_2  x_A  }_{U_A U^{\prime}_A U_B U^{\prime}_B T U^{\prime\prime}_A }       , \\ \\   \ket{\mathcal{P}_2  x_A,  \mathcal{P}_2 x_A , \mathcal{P}_2  x_B,  \mathcal{P}_2 x_B , \mathcal{P}_2 x_B   ,  g_2 \big( \mathcal{P}_1 , \mathcal{P}_2 \big( x_A + x_B \big) \big)         }_{U_A U^{\prime}_A U_B     U^{\prime}_B U^{\prime\prime}_A T     }    , \\ \\  \ket{\mathcal{P}_2  x_A,  \mathcal{P}_2 x_A , \mathcal{P}_2  x_B,  \mathcal{P}_2 x_B , \mathcal{P}_2  x_A  ,  \mathcal{P}_2  x_B  }_{U_A U^{\prime}_A U_B U^{\prime}_B U^{\prime\prime}_A U^{\prime\prime}_B }  , \\ \\   \ket{\mathcal{P}_2  x_A,  \mathcal{P}_2 x_A , \mathcal{P}_2  x_B,  \mathcal{P}_2 x_B , \mathcal{P}_2 x_B   ,  \mathcal{P}_2 x_A }_{U_A U^{\prime}_A U_B U^{\prime}_B U^{\prime\prime}_B U^{\prime\prime\prime}_B U^{\prime\prime}_A }        . 
\end{align*}

\noindent The second factor appearing in the power of $2$ is determined by factors in the isometry which do not cancel out, which would give more copies of $\textbf{I}_{AB}$. Computations associated with taking products of $\mathscr{U}$ and $\mathscr{V}$ isometries demonstrate how the efficient computation of functions of parity check matrices comes at a cost. To this end, it would be interesting to determine additional implications of the $2^{ \frac{5}{2} ( 5 - \frac{3}{2} ) + \mathrm{log}_2 \sqrt{C}}$ security gap. 

Besides the security result, the probability that Alice, or Bob, decides to abort the hashing protocol draws our attention to whether the protocol is robust. Considering whether additional steps of the protocol can be included to increase robustness is of related interest to explore.

\section{Declarations}

\subsection{Ethics approval and consent to participate}

The author consents to participate in the peer review process.

\subsection{Consent for publication}

The author consents to submit the following work for publication.

\subsection{Availability of data and materials}

Not applicable

\subsection{Competing interests}

Not applicable

\subsection{Funding}

Not applicable

\subsection{Acknowledgments}

Not applicable

\subsection{Author's contributions}

PR wrote the manuscript, and performed rounds of editing for submitting the work.

\section{References}

\noindent [1] Amr, A., Villanueva, I.: Quantum one way vs. classical two way communication in XOR games. Quantum Information Processing 20(79) (2021).

\bigskip

\noindent [2] Bannik, T., al.: Bounding quantum-classical separations for classes of nonlocal games. STACS 12, 1–12 (2019). https://doi.org/10.4230/LIPIcs.STACS.2019.12.

\bigskip

\noindent [3] Briet, J., Buhrman, H., Toner, B.: A generalized grothendieck inequality and entanglement in XOR games. Comm. Math. Phys. 305, 827–843 (2011) https://doi.org/10.1007/s00220-011-1280-3.

\bigskip

\noindent [4] Broadbent, A., Methot, A.A.: On the power of non-local boxes. Theoretical
Computer Science 358, 3–14 (2006) https://doi.org/10.1016/j.tcs.2005.08.035.

\bigskip

\noindent [5] Brassard, G., Broadbent, A., Tapp, A.: Quantum pseudo-telepathy. Found. Phys. 35, 1877–1907 (2005) https://doi.org/https://philpapers.org/rec/BRAQP.

\bigskip

\noindent [6] Benedetti, M., Coyle, B., Fiorentini, M., Lubasch, M., Rosenkranz, M.: Variational Inference with a Quantum Computer. Phys Rev Applied 16(044057) (2021). https://doi.org/10.1103/PhysRevApplied.16.044057.

\bigskip

\noindent [7] Bittel, L., Kliesch, M.: Training variational quantum algorithms is np-hard. Physical Review Letters 127(120502) (2021) https://doi.org/10.1103/PhysRevLett.127.120502.

\bigskip

\noindent [8] Catani, L., Faleiro, R., Emeriau, P.E., Mansfield, S., Pappa, A.: Connecting XOR and XOR* games. Phys. Rev. A 109(012427) (2024) https://doi.org/10.1103/PhysRevA.109.012427.

\bigskip

\noindent [9] Chen, H., Vives, M., Metcalf, M.: Parametric amplification of an optomechanical Quantum interconnect. Physical Review Research 4(043119) (2022) https://doi.org/10.1103/PhysRevResearch.4.043119.

\bigskip

\noindent [10] Cong, I., Duan, L.: Quantum discriminant analysis for dimensionality reduction and classification. New Journal of Physics 18(073011) (2016) https://doi.org/10.1088/1367-2630/18/7/073011.

\bigskip

\noindent [11] Cleve, R., Hoyer, P., Toner, B., Watrous, J.: Consequences and limits of non-local strategies. 19th IEEE Annual Conference on Computational Complexity Proceedings, 236–249 (2004) https://doi.org/10.1109/CCC.2004.1313847.

\bigskip

\noindent [12] Culf, E., Mousavi, H., Spirig, T.: Approximation algorithms for noncommutative csps. IEEE 65th Annual Symposium on Foundations of Computer Science
(FOCS), 920–929 (2024) https://doi.org/10.1109/FOCS61266.2024.00061.

\bigskip

\noindent [13] Cui, D., Malavolta, G., Mehta, A., Natarajan, A., Paddock, C., Schmidt, S., Walter, M., Zhang, T.: A computational tsireslson’s theorem for the value of compiled xor games. arXiv: 2402.17301 (2024).

\bigskip

\noindent [14] Doherty, A.C., Liang, Y.C., Toner, B., Wehner, S.: The quantum moment problem and bounds on entangled multi-prover games. 23rd Annual IEEE Conference on Computational Complexity 8 (2018).

\bigskip

\noindent [15] Drmota, P., Main, D., Ainley, E.M., Agrawal, A., Araneda, G., Nadlinger, R. Srinivas, Cabello, A., al.: Experimental quantum advantage in the odd-cycle game. Phys. Rev. Lett. 134(070201) (2025) https://doi.org/10.1103/PhysRevLett.134.070201.

\bigskip

\noindent [16] Ewe, W.-B., Koh, D.E., Goh, S.T., Chu, H.-S., Png, C.E.: Variational quantum-based simulation of waveguide modes. IEEE Transactions on Microwave Theory and Techniques 70(5), 2517–2525 (2022) https://doi.org/10.1109/TMTT.2022.3151510.

\bigskip

\noindent [17] Pierre-Emmanuel Emeriau, P.-E., Howard, M., Mansfield, S.: Quantum advantage in information retrieval. PRX Quantum 3(020307) (2022) https://doi.org/
10.1103/PRXQuantum.3.02030.

\bigskip

\noindent [18] Faleiro, R.: Quantum strategies for simple 2-player xor games. Quantum Inf
Process 19(229) (2020) https://doi.org/10.1007/s11128-020-02717-.

\bigskip

\noindent [19] Garg, D., Ikbal, S., Srivastava, S.K., Vishwakarma, H., Karanam, H., Subramaniam, L.V.: Quantum embedding of knowledge for reasoning. Advance in Neural Information Processing Systems 32 (2019).

\bigskip

\noindent [20] Genoni, M.G., Tufarelli, T.: Non-orthogonal bases for quantum metrology. Journal of Physics A: Mathematical and Theoretical 52(43) (2019) https://doi.org/10.1088/1751-8121/ab3fe0.

\bigskip

\noindent [21] Gidi, J.A., Candia, B., Munoz-Moller, A.D., Rojas, A., Pereira, L., Munoz, M., Zambrano, L., Delgado, A.: Stochastic optimization algorithms for quantum applications. Phys.Rev.A 108(032409) (2023) https://doi.org/10.1103/
PhysRevA.108.032409.

\bigskip

\noindent [22] Givi, P., Daley, A.J., Mavriplis, D., Malik, M.: Quantum speedup for aeroscience and engineering. AIAA 58(8) (2020).

\bigskip

\noindent [23] Helton, J.W., Mousavi, H., Nezhadi, S.S., al.: Synchronous values of
games. Ann. Henri Poincar´e 25, 4357–4397 (2024) https://doi.org/10.1007/s00023-024-01426-1.
\bigskip

\noindent [24] Hadiashar, S.B., Nayak, A., Sinha, P.: Optimal lower bounds for quantum learning via information theory. IEEE Transactions on Information Theory 70(3), 1876–1896 (2024) https://doi.org/10.1109/TIT.2023.3324527.

\bigskip

\noindent [25] Hur, T., Kim, L., Park, D.K.: Quantum convolutional neural network for classical data classification. Quantum Machine Intelligence 4(3) (2022) https://doi.org/10.1007/s42484-021-00061-x.

\bigskip

\noindent [26] Holmes, Z., Coble, N.J., Sornborger, A.T., Subasi, Y.: On nonlinear transformations in quantum computation. Phys. Rev. Research 5(013105) (2023) https:
//doi.org/10.1103/PhysRevResearch.5.013105.

\bigskip

\noindent [27] Jing, H., Wang, Y., Li, Y.: Data-driven quantum approximate optimization algorithm for cyber-physical power systems. arXiv: 2204.00738 (2022) https://doi.org/10.48550/arXiv.2204.00738.

\bigskip

\noindent [28] Junge, M., Palazuelos, C.: On the power of quantum entanglement in multipartite Quantum xor games. Journal of the London Mathematical Society 110(5) (2024).

\bigskip

\noindent [29] Kubo, K., Nakagawa, Y.O., Endo, S., Nagayama, S.: Variational quantum simulations of stochastic differential equations. Physical Review A 103(052425) (2021). https://doi.org/10.1103/PhysRevA.103.052425.

\bigskip

\noindent [30] Kribs, D.W.: A quantum computing primer for operator theorists. Linear Algebra and its Applications 400, 147–167 (2005) https://doi.org/10.48550/arXiv.math/0404553.

\bigskip

\noindent [31] Li, R.Y., Di Felice, R., Rohs, R., Lidar, D.A.: Quantum annealing versus classical machine learning applied to a simplied computational biology problem. npj Quantum Information 4(14) (2008) https://doi.org/10.1038/s41534-018-0060-8.

\bigskip

\noindent [32] Mahdian, M., Yeganeh, H.D.: Toward a quantum computing algorithm to quantify classical and quantum correlation of system states. Quantum Information
Processing 20(393) (2021) https://doi.org/10.1007/s11128-021-03331-6.

\bigskip

\noindent [33] Maldonado, T.J., Flick, J., Krastanov, S., Galda, A.: Error rate reduction of single-qubit gates via noise-aware decomposition into native gates. Scientific Reports 12(6379) (2022) https://doi.org/10.1038/s41598-022-10339-0.

\bigskip

\noindent [34] Manby, F.R., Stella, M., Goodpaster, J.D., Miller, T.F.: A simple, exact density-functional-theory embedding scheme. Journal of Chemical Theory and
Computation 8(8), 2564–2568 (2012) https://doi.org/10.1021/ct300544e.

\bigskip

\noindent [35] Maurer, U.: Perfect cryptographic security from partially independent channels. Proc. 23rd ACM Symposium on Theory of Computing — STOC, 561–572 (1991). https://doi.org/https://crypto.ethz.ch/publications/Maurer91b.html.

\bigskip

\noindent [36] Mensa, S., Sahin, E., Tacchino, F., Barkoutsos, P.K., Tavernelli, I.: Quantum machine learning framework for virtual screening in drug discovery: a prospective Quantum advantage. Mach. Learn.: Sci. Technol. 4(015023) (2023) https://doi.org/10.1088/2632-2153/acb900.

\bigskip

\noindent [37] Nan Sheng, H.M., Govono, M., Galli, G.: Quantum embedding theory for
strongly-correlated states in materials. J. Chem. Theory Comput. 17(4), 2116–
2125 (2021) https://doi.org/10.1021/acs.jctc.0c01258.

\bigskip

\noindent [38] Ostrev, D.: The structure of nearly-optimal quantum strategies for the non-local XOR games. Quantum Information and Computation 16(13-14), 1191–1211 (2016)
https://doi.org/10.26421/QIC16.13-14-6.

\bigskip

\noindent [39] Ostrev, D.: Composable, unconditionally secure message authentication without
any secret key. IEEE International Symposium on Information Theory 10(1109),
622–626 (2019) https://doi.org/10.1109/ISIT.2019.8849510.

\bigskip

\noindent [40] Ostrev, D.: Qkd parameter estimation by two-universal hashing. Quantum 7, 894
(2023) https://doi.org/10.22331/q-2023-01-13-894.

\bigskip

\noindent [41] Paine, A.E., Elfving, V.E., Kyriienko, O.: Quantum kernel methods for solving
differential equations. Physical Review A 107(032428) (2023) https://doi.org/10.
1103/PhysRevA.107.032428.

\bigskip

\noindent [42] Paudel, H.P., Syamlal, M., Crawford, S.E., Lee, Y.-L., Shugayev, R.A., Lu, P.,
Ohodnicki, P.R., Mollot, D., Duan, Y.: Quantum computing and simulations for
energy applications: Review and perspective. ACS Eng. Au 3, 151–196 (2022)
https://doi.org/10.1021/acsengineeringau.1c00033.

\bigskip

\noindent [43] Przhiyalkovskiy, Y.V.: Quantum process in probability representation of quan-
tum mechanics. Journal of Physics A: Mathematical and Theoretical 55(085301)
(2022) https://doi.org/10.1088/1751-8121/ac4b15.

\bigskip

\noindent [44] Perc, M.: Statistical physics of human cooperation. Physics Reports 687, 1–
51 (2017) https://doi.org/https://papers.ssrn.com/sol3/papers.cfm?abstract id=
2972841.

\bigskip

\noindent [45] Ravishankar Ramanathan, R., Augusiak, R., Murta, G.: Generalized XOR games
with d outcomes and the task of nonlocal computation. Phys. Rev. A 93(022333)
(2016) https://doi.org/10.1103/PhysRevA.93.022333.

\bigskip

\noindent [46] Rigas, P.: Optimal, and approximately optimal, quantum strategies for XOR* and FFL games. arXiv: 2311.12887 (submitted) (2023).

\bigskip

\noindent [47] Rigas, P.: Variational quantum algorithm for measurement extraction from the navier-stokes, einstein, maxwell, b-type, lin-tsien, camassa-holm, dsw, h-s, kdv-b, non-homogeneous kdv, generalized kdv, kdv, translational kdv, skdv, b-l and airy equations. arXiv: 2209.07714 (submitted) (2025) https://doi.org/10.48550/
arXiv.2209.07714.

\bigskip

\noindent [48] Rigas, P.: Quantum error bounds, optimality, and duality gaps for multiplayer XOR, XOR*, compiled XOR, XOR*, and strong parallel repetition of XOR, XOR*, and FFL games. arXiv: 2209.07714 (submitted) (2025). https://doi.org/10.48550/arXiv.2209.07714.

\bigskip

\noindent [49] Rigas, P.: Error correction, authentication, and false acceptance, probabilities for communication over noisy quantum channels: converse upper bounds on the bit transmission rate. arXiv: 2507.03035 (submitted) (2025) https://doi.org/10.48550/arXiv.2507.03035.

\bigskip

\noindent [50] Rigas, P.: Parallel repetition of expanded, and multiplayer, quantum games: anchoring, optimal values, generalized error bounds, dependency-breaking as symmetry-breaking. arXiv: 2508.09380 (submitted) (2025) https://doi.org/10.48550/arXiv.2508.09380.

\bigskip

\noindent [51] Roscika, M., Mazurek, P., Grudka, A., Horodecki, M.: Generalized XOR non-locality games with graph description on a square lattice. Journal of Phys A: Math. Theor. 53(265302) (2020) https://doi.org/10.1088/1751-8121/ab8f3e.

\bigskip

\noindent [52] Slofstra, W.: Lower bounds on the entanglement needed to play xor non-local games. Journal of Mathematical Physics 52(10), 102202 (2011) https://doi.org/10.1063/1.3652924.

\bigskip

\noindent [53] Dam, W., Sasaki, Y.: Quantum algorithms for problems in number theory, algebraic geometry, and group theory. Diversities in Quantum Computation and Quantum Information, 79–105 (2012) https://doi.org/10.1142/97898144259880003.

\bigskip

\noindent [54] Wang, Y., Krstic, P.S.: Multistate transition dynamics by strong time-dependent perturbation in nisq era. J. Phys. Commun. 7(075004) (2023) https://doi.org/10.1088/2399-6528/ace67a.

\bigskip

\noindent [55] Zhao, L., Zhao, Z., Rebentrost, P., Fitzsimons, J.: Compiling basic linear algebra subroutines for quantum computers. Quantum Machine Intelligence 3(21) (2021). https://doi.org/10.1007/s42484-021-00048-8.

\end{document}